\documentclass[journal,twocolumn]{IEEEtran}

\usepackage[utf8]{inputenc}
\usepackage[usenames, dvipsnames]{color}

\usepackage{mathpazo}
\usepackage{amssymb}
\usepackage{amsmath}
\usepackage{mathtools}
\usepackage{relsize}
\usepackage{hyperref}
\usepackage{amsthm}
\usepackage{physics}
\usepackage{mathrsfs}
\usepackage{comment}
\usepackage[most]{tcolorbox}

\usepackage{verbatim}
\DeclareFontFamily{U}{mathx}{\hyphenchar\font45}
\DeclareFontShape{U}{mathx}{m}{n}{
      <5> <6> <7> <8> <9> <10>
      <10.95> <12> <14.4> <17.28> <20.74> <24.88>
      mathx10
      }{}
\DeclareSymbolFont{mathx}{U}{mathx}{m}{n}
\DeclareFontSubstitution{U}{mathx}{m}{n}
\DeclareMathAccent{\widecheck}{0}{mathx}{"71}

\usepackage{float} 
\allowdisplaybreaks 
\usepackage{hyperref} 
\usepackage{mwe}
\usepackage{listing}
\usepackage[en-US]{datetime2}
\usepackage{placeins}
\usepackage{capt-of}
\usepackage{pgf,tikz}
\usetikzlibrary{arrows}
\usetikzlibrary{shapes}
\usetikzlibrary{patterns}
\usepackage{enumitem}
\usepackage{subcaption}
\usepackage{longtable,lscape}
\usepackage{supertabular}
\usepackage{booktabs}

\newcommand{\wt}{\widetilde}

\newcommand{\ve}{\varepsilon}
\newcommand{\mrm}{\mathrm}
\newcommand{\mbb}{\mathbb}
\newcommand{\ol}{\overline}

\newcommand{\Bve}{\mathscr{B}^{\ve}}
\newcommand{\Bvec}[1]{\mathscr{B}^{#1}}
\newcommand{\QMC}{\mrm{QMC}}
\newcommand{\upuparrow}{\uparrow\uparrow}
\newcommand{\spec}{\mrm{spec}}
\newcommand{\mcol}{{\,:\,}}
\newcommand{\TD}{\mrm{TD}}
\newcommand{\Lone}{\mrm{L}1}

\newcommand{\opvec}{\operatorname{vec}}

\renewcommand{\tr}{\operatorname{Tr}}

\newcommand{\cB}{\mathcal{B}}
\newcommand{\cC}{\mathcal{C}}

\newcommand{\cE}{\mathcal{E}}
\newcommand{\cF}{\mathcal{F}}
\newcommand{\cG}{\mathcal{G}}

\newcommand{\cJ}{\mathcal{J}}
\newcommand{\cK}{\mathcal{K}}

\newcommand{\cP}{\mathcal{P}}

\newcommand{\cR}{\mathcal{R}}

\newcommand{\cU}{\mathcal{U}}
\newcommand{\cV}{\mathcal{V}}
\newcommand{\cW}{\mathcal{W}}
\newcommand{\cX}{\mathcal{X}}
\newcommand{\cY}{\mathcal{Y}}

\newcommand{\mc}{\mathcal}

\newcommand{\Lin}{\mathrm{L}}

\newcommand{\Pos}{\mathrm{Pos}}
\newcommand{\Herm}{\mathrm{Herm}}
\newcommand{\Channel}{\mathrm{C}}

\newcommand{\Density}{\mathrm{D}}

\newcommand{\Sep}{\mathrm{Sep}}

\newcommand{\id}{\textrm{id}}

\newenvironment{mylist}[1]{\begin{list}{}{
	\setlength{\leftmargin}{#1}
	\setlength{\rightmargin}{0mm}
	\setlength{\labelsep}{2mm}
	\setlength{\labelwidth}{8mm}
	\setlength{\itemsep}{0mm}}}
	{\end{list}}

\newcounter{problemcounter}

\theoremstyle{definition}
\newtheorem{theorem}{Theorem}
\newtheorem*{theorem*}{Theorem}
\newtheorem{lemma}[theorem]{Lemma}

\newtheorem{definition}[theorem]{Definition}

\newtheorem{corollary}[theorem]{Corollary}
\newtheorem{conjecture}[theorem]{Conjecture}

\newtheorem{example}[theorem]{Example}
\newtheorem{remark}[theorem]{Remark}
\newtheorem{proposition}[theorem]{Proposition}

\newtheorem*{proposition*}{Proposition}

\usepackage{ulem}

\usepackage{soul,xcolor}
\setstcolor{red}

\definecolor{cool_green}{rgb}{0.0, 0.5, 0.0}

\newcommand{\edit}[1]{{\color{black}#1}}

\newcommand{\seedto}[1]{%
\begin{tikzpicture}[#1]%
\draw (0,0.25ex) -- (1ex,0.8ex);%
\draw (0,0) -- (1.4ex,0ex);%
\draw (0,-0.25ex) -- (1ex,-0.8ex);
\end{tikzpicture}%
}

\begin{document}

\title{One-Shot Distributed Source Simulation: As Quantum as it Can Get}
\author{Ian George, Min-Hsiu Hsieh, and Eric Chitambar%
\thanks{Ian George is with the Centre for Quantum Technologies, National University of Singapore, Singapore, email: \edit{qit.george@gmail.com}.}
\thanks{Eric Chitambar is with the Department of Electrical and Computer Engineering, University of Illinois at Urbana-Champaign, Urbana, Illinois, 61801, USA.}%
\thanks{Min-Hsiu Hsieh is with Hon Hai (Foxconn) Research Institute, Taipei, Taiwan, email: min-hsiu.hsieh@foxconn.com.}
\thanks{A weaker version of these results was presented in part at the 2023 IEEE International Symposium on Information Theory.}}
\maketitle
\begin{abstract}
    Distributed source simulation is the task where two (or more) parties share some correlated randomness and use local operations and no communication to convert this into some target correlation. Wyner's seminal result showed that asymptotically the rate of uniform shared randomness needed for this task is given by a mutual information induced measure, now referred to as Wyner's common information. This asymptotic result was extended by Hayashi in the quantum setting to separable states, the largest class of states for which this task can be performed to vanishing error. In this work we characterize this task in a near-tight manner in the one-shot setting using the smooth entropy framework. We do this by introducing one-shot operational quantities and correlation measures that characterize them. We establish asymptotic equipartition properties for our correlation measures thereby recovering the previous vanishing-error asymptotic results. In doing so, we consider technical points in one-shot network information theory and provide methods for cardinality bounds in the smooth entropy calculus. We also introduce entangled state versions of the distributed source simulation task and determine bounds in this setting via quantum embezzling. This provides a strong characterization of this network task in the one-shot, quantum regime.
\end{abstract}

\section{Introduction}
Correlation is at the heart of information theory. Indeed, one may view Shannon's seminal results \cite{Shannon-48} of source coding and channel coding as the study of establishing perfect correlation between a sender and a receiver. This notion of correlation as the key idea in information theory has deepened with the advent of quantum information theory and its new form of correlation, quantum entanglement. In quantum information theory, a complementary pair of fundamental problems are entanglement distillation \cite{Bennett-1996a,Bennett-1996b,Devetak-2005a,Horodecki-2009a, Regula-2019a} and dilution \cite{Bennett-1996a,Buscemi-2011a,Theurer-2023a} between spatially separated parties under various communication models. Both Shannon's original problem \cite{Shannon-48} and the tasks of entanglement distillation/dilution belong to the general problem of \textit{correlation conversions under locality constraints}.  For entanglement, the processing is typically restricted to local quantum operations and classical communication (LOCC) between two or more parties.  If instead of entanglement we consider the analogous task of manipulating classical correlations, different types of locality constraints naturally arise. In an adversarial setting, one can distinguish between public correlations and private correlations and how these correlations transform under public and private communication \cite{Ahlswede-1993a,Ahlswede-1998a}.  An alternative approach historically taken is to remove communication all together and consider the processing of joint classical distributions under just local processing \cite{Gacs-1973a,Witsenhausen-1975a,Wyner-1975a}.  This has motivated the notion of common information between two or more random variables, and recently there has been a resurgence of interest in studying different common information measures (see \cite{Yu-2022a} for a recent summary).

\begin{figure}[ht]
    \begin{center}
        \begin{tikzpicture}
            \tikzstyle{porte} = [draw=black!50, fill=black!20]
                \draw
                (0,0) node (qy) {$p_{X}$}
                ++(1.5,0) node[porte] (copy) {Copy}
                ++(2,0.75) node[porte] (Alice-Proc) {$\Phi_{X \to A}$}
                ++(1.75,0) node (A) {$A$}
                ++(0,-1.5) node (C) {$C$}
                ++(-1.75,0) node[porte] (Bob-Proc) {$\Psi_{X' \to C}$}
                ++(-2,-0.25) node (X') {$X'$}
                ++(0,2) node (X) {$X$}
                ++(4.25,-1) node (approx) {$\approx_{\ve}$}
                ++(0.8,0) node (pxz) {$\rho_{AC}$}
                ;
                \draw
                (2.25,-1.5) node (chi) {$\chi_{XX'}^{|p}$}
                ;
                \draw
                (4.75,-1.5) node (rhotilde) {$\wt{\rho}_{AC}$}
                ;
                \draw [draw=blue!50, thick , densely dashed ] (2.25,1.2) -- (2.25,-1.2)
                ;
                \draw [draw=blue!120, thick , densely dashed ] (4.75,1.2) -- (4.75,-1.2)
                ;
                \path[draw = black, ->] (qy) -- (copy); 
                \path[draw=black, ->] (copy) |- (Alice-Proc);
                \path[draw=black, ->] (Alice-Proc) -- (A);
                \path[draw=black, ->] (copy) |- (Bob-Proc);
                \path[draw=black, ->] (Bob-Proc) -- (C);
        \end{tikzpicture}
    \end{center}
    \caption{Diagram of distributed source simulation of a quantum state from a classical `seed'. After the copying procedure (at the light blue line), the two parties share a perfectly correlated state $\chi^{|p}_{XX'}$. After their local processing (at the dark blue line), the parties share a state $\wt{\rho}_{AC}$ which should be approximately the target state $\rho_{AC}$. We distinguish between the scenarios of uniform and non-uniform random seed $p_X$.}
    \label{fig:DSS-fig-intro}
\end{figure} 
In this work, we study the canonical problem introduced by Wyner of distributed source simulation \cite{Wyner-1975a}, where two (or more) non-communicating receivers try to generate a target joint distribution $p_{XZ}$ using perfect classical correlation distributed by a hub. Wyner established that the vanishing-error rate (necessary number of bits of correlation to send the receivers per copy of target state $p_{XZ}$ such that the error goes to zero) is given by what is now known as the Wyner common information:
\begin{align}
	R = \min_{X-Y-Z} I(XZ:Y)_{p'}. 
\end{align}
Here, the minimization is over all Markov chains $p'_{X-Y-Z}$ such that $p'_{XZ} = p_{XZ}$, and the error can be measured in terms of the normalized relative entropy (as in Wyner's work) or more strongly in terms of the variational distance \cite{Cuff-2013a} or the unnormalized relative entropy \cite{Cuff-2015a}. This result was generalized by Hayashi in \cite{Hayashi-2006b} to the simulation of separable quantum states $\rho_{AC}$, which are convex combinations of product states, $\rho_{AC}=\sum_{i}p_i\sigma_A\otimes\tau_C$, and lacking any entanglement. Using the trace distance to quantify the error in simulation, Hayashi showed the optimal rate to be
\begin{align}\label{eq:Wyner-common-information}
	R = \min_{A-Y-C} I(AC:Y)_{\rho'} =: C(A\mcol C)_{\rho}, 
\end{align}
where the minimization is taken over all short quantum Markov chains $\rho_{A-\edit{Y}-C}'$ such that $\rho_{AC}' = \rho_{AC}$.

This work approaches the classical and quantum source simulation problem in the one-shot setting using the smooth entropy calculus \cite{Tomamichel-2015a}, where we do not require a multi-copy structure on the target state and the tolerated error may be non-zero (see Fig.~\ref{fig:DSS-fig-intro} for depiction).  The bulk of our work focuses on the simulation of states using a classical seed.  As argued below, there is still a gap in our understanding of simulating general quantum states with a classical seed and nonzero error.  Our use of smooth entropies allow us to close this.  Furthermore, we believe this methodology is suitable for tackling more general problems in quantum network information theory, of which one-shot mutual information quantities and smooth entropy calculus remain nascent.

It is particularly noteworthy that a single classical seed is universal for the generation of all separable states.  The situation is much different when considering the distributed simulation of fully entangled states.  First, it is no longer possible to build an entangled state using a classical seed; some type of quantum correlations (i.e. entanglement) is needed.  Second, even if we allow `quantum seeds' of this type, there is no universal seed that could be used to generate every target quantum state at some nonzero rate, as there is when restricted to separable states.  Thus, it is not entirely clear how one should proceed in generalizing the distributed source simulation problem to the full quantum setting.  One approach is to allow a sublinear amount of communication between the receivers, and this scenario has been linked to a correlation measure known as entanglement of purification \cite{Terhal-2002a}.  Another approach, followed in this work, is to keep the model of zero communication between the receivers and instead allow for some $\ve>0$ error in the simulated state.  In this case, it is possible to recover the Fig. \ref{fig:DSS-fig-intro} for arbitrary quantum states $\rho_{AC}$ by replacing the classical seed by a universal class of using \textit{quantum embezzling} states \cite{vanDam-2003a, Cleve-2017a}. 
 We formalize this idea by developing a one-shot fully quantum theory of distributed source simulation using embezzling states.

\paragraph{Brief Summary of Results} Our primary result (Theorem \ref{thm:one-shot-DSS-main-theorem}) is provably near-tight one-shot bounds on one-shot distributed source simulation from a common classical seed in terms of new smooth correlation measures. Our results hold for both classical and quantum target states.  While only separable states can be generated using just classical shared randomness and local processing, as  in Fig. \ref{fig:DSS-fig-intro}, for one-shot simulations some error tolerance is allowed.  In this case, one can consider the minimal rate of shared classical randomness that enables the simulation of a general state $\rho_{AC}$ within the allowed error $\ve$.  Of course these results are of greatest relevance for states lying near the set of separable states, for then $\ve$ can be taken to be small.  But it is easy to construct such states (see Example \ref{example:Werner-states}), and any complete solution to the distributed source simulation problem must handle states lying near the boundary of the separable set. Our results offer such a solution that is `near tight' in the sense that our correlation measures capture the size of shared randomness needed to simulate $n$ copies of the state up to correction terms scaling as $O(\log n)$. \edit{This means if the second-order expansion of the quantity scales as $O(\sqrt{n})$, as is common due to the Berry-Esseen theorem (see \cite{Tan-2014a}), then our bounds are tight to second-order. This aligns with other recent work in quantum information theory \cite{Shen-2022a, Cheng-2023b} that have been deriving tight to second-order results by relying upon already known second-order expansions and establishing bounds such that the correction term only scales as $O(\log n)$, which is what we achieve.}
 
A technical contribution of this work is the identification of the appropriate method for getting both near-tight and general bounds, which does not follow from the recent soft-covering lemma of Shen et al. \cite{Shen-2022a}. Unfortunately, we are not able to derive a second-order expansion due to challenges in working with the auxiliary random variable in the Markov chain, an issue that is known even classically \cite{Tan-2014a}. We are however able to recover the result of Hayashi by establishing a vanishing error asymptotic equipartition property for our correlation measures. We also establish cardinality bounds for our one-shot quantities by both generalizing standard classical network information theory methods and providing a new method; both of which may be useful elsewhere in quantum network information theory. 

We also expand the scope of the classical distributed source simulation problem in different directions to highlight some key structural details.  We first note a nuanced distinction in the one-shot setting between uniform and non-uniform seeds (e.g. the box labeled `copy' in Fig.\ref{fig:DSS-fig-intro}), something that has not been previously considered.
Our results are near-tight in both settings. By proposing different variants to the distributed source simulation problem, we clarify a technical distinction between establishing a near-tight soft-cover lemma (as in Shen et al. \cite{Shen-2022a}) and a near-tight distributed source simulation rate in the one-shot setting. All of these results extend easily to scenarios with more than two receivers. Finally, we consider the direct quantum version of Wyner's original problem by replacing the classical seed with entangled states that are then used to generate arbitrary entangled states by local processing. Most of our results in this direction make use of embezzling states \cite{van-2003a} and structurally behave different than standard classical information theory results, e.g. there exist near-tight upper and lower bounds that diverge.

\paragraph{Contribution in Relation to Previous Work}
We now provide a more detailed (but inevitably incomplete) history of distributed source simulation, focusing on the aspects most relevant to this work: the development of refined soft-cover lemmas, the study of common information quantities, and the broader theory of correlation conversions. The soft-cover lemma was introduced by Wyner for the achievability of distributed source simulation, \cite{Wyner-1975a}, though he used the regularized relative entropy to measure error. Following this original work, there has been much work on classical soft-covering. All with respect to error in trace distance, Han and Verd\'{u} studied the strong converse and non-memoryless setting through their notion of resolvability \cite{Han-1993a}, Hayashi established a bound on memoryless channels \cite{Hayashi-2006a}, which was later improved by Cuff \cite{Cuff-2013a}. Yu and Tan studied soft-covering under R\'{e}nyi divergence error \cite{Yu-2019a}. In the quantum setting, the first soft-covering lemma was established by  Ahlswede and Winter for channel resolvability \cite{Ahlswede-2002a}, and Hayashi provided a one-shot bound \cite{Hayashi-2006b} a few years later. Since then, there have been further advancements, resulting in error exponents \cite{Cheng-2023a, Cheng-2023b} and optimal to second-order rates \cite{Shen-2022a}. However, a major observation of this work is that distributed source simulation is captured by a `max mutual information'-induced quantity thereby separating it from a direct application of \cite{Shen-2022a}, which works with the hypothesis testing mutual information.

Since Wyner's initial paper, which was inspired by prior work of G\'{a}cs-K\"{o}rner \cite{Gacs-1973a} and Witsenhausen \cite{Witsenhausen-1975a}, many variations of common information and refinements of distributed source simulation have been considered. Liu et al. extended distributed source simulation to multipartite joint distributions \cite{Liu-2010a}. Yu and Tan considered R\'{e}nyi divergences and total variation as measures of error, which in particular enabled them to establish a strong converse for distributed source simulation under the total variation measure \cite{Yu-2018a,Yu-corrections-2018a}. Cuff established a general tradeoff region between Wyner common information and the classical reverse Shannon theorem when simulating a classical channel \cite{Cuff-2008a}. Reverse Shannon theorems have found use in other coding schemes as well \cite{Hsieh-2016a}. There is also the related problem of exact common information introduced by Kumar et al., which require the target state $p_{XY}^{\otimes n}$ to be \textit{exactly} constructed using variable-length codes \cite{Kumar-2014a}. It was established by Yu and Tan that the exact common information corresponds to the common information with the error measured in terms of the max-divergence \cite{Yu-2020a}. Winter extended to the problem to the adversarial setting and drew relations to key distillation \cite{Winter-2005a}. Chitambar et al.\ compared this adversarial setting to the collaborative alternative \cite{Chitambar-2016a}. Moreover, Chitambar et al.\ related the adversarial setting to quantum entanglement manipulations \cite{Chitambar-2018a,Chitambar-2015b} at one point using the G\'{a}cs-K\"{o}rner common information \cite{Gacs-1973a}, which also is relevant in round complexity of state transformations \cite{Chitambar-2017a}. In terms of common information, our major contribution is, to the best of our knowledge, the first study of such quantities in the smooth entropy calculus and establishing the tools necessary to do this.

Finally, as noted, we can view the problem of distributed source simulation as a specific instance of the general set of problems of correlation conversions under locality constraints. In the classical setting perhaps the seminal works of interest are the works of Ahlswede and Csisz\`{a}r \cite{Ahlswede-1993a,Ahlswede-1998a}, which study the distillation of commmon randomness with and without privacy constraints. Both of these works were in effect extended to quantum states by Devetak and Winter \cite{Devetak-2005a,Devetak-2004a} and have had some renewed interest \cite{Salek-2021a}. The related problems of entanglement distillation and dilution have been long studied both asymptotically and in the one-shot setting (see \cite{Bennett-1996a,Bennett-1996b,Devetak-2005a,Horodecki-2009a} and \cite{Hayashi-2006c,Buscemi-2010a,Buscemi-2011a,Regula-2019a, Theurer-2023a} respectively and references therein). With regards to this class of work, our major contribution is resolving a gap in the existing literature for the scenario of distributed state preparation with no communication, which has been previously considered  in \cite{Hayden-2003a,pure-state-revisited}.  We also hope that the tools developed here, such as cardinality bounds, properties of R\'{e}nyi mutual information quantities, etc., may be useful in related problems of converting correlations under locality constraints.  

\paragraph{Structure of Paper}
In Section \ref{sec:technical-background}, we provide the technical background necessary to read the rest of the paper in-depth. For clarity, in Section \ref{sec:main-theorems} we provide a more dectailed summary of our results and discussion on the technical barriers in the work. In Section \ref{sec:operational-quantities}, we introduce one-shot distributed source simulation and provide a particularly direct way of relating separability, the notion of a Markov chain extension, and our network structure. In Section \ref{sec:one-shot-corr-meas} we introduce our one-shot correlation measures, max Wyner common information quantities, and establish their properties. In Section \ref{sec:asymptotics}, we prove the asymptotic equipartition property of our measures and discuss technical barriers and open problems to showing a strong converse or second-order expansion of our correlation measures. In Section \ref{sec:multi-receiver}, we discuss how our methods extend to more receivers. In Section \ref{sec:ent-source-sim}, we consider `fully quantum' analogues of distributed source simulation. In Section \ref{sec:conclusion} we re-capitulate what we have established and discuss open problems. In the appendix, we provide technical results as well as our findings on variants of one-shot distributed source simulation.

\section{Technical Background \& Notation}\label{sec:technical-background}
In this section we provide sufficient technical background for this paper. We largely follow notation from standard works \cite{Wilde-2011a,Watrous-Book,Tomamichel-2015a,Khatri-2020a} to which we refer the reader to further background. We provide a small look-up table of common notation in the main text for the convenience of the reader.

\begin{table}
\begin{tabular}{p{0.38\columnwidth}|p{0.55\columnwidth}} 
Notation & Meaning \\
\hline
$[n]$ & The set $\{1,...,n\}$ \\
$\cX,\cY,\cdots$ & Finite alphabets \\
$A,B,C,\cdots$ & Hilbert spaces \\
$|A|,|B|,|C|,\cdots$ & Dimension of Hilbert spaces\\
$A^n$ & Simplification of $A_1 \otimes A_2 \otimes \cdots \otimes A_n$\\
$A-B-C$ & (Short) Markov Chain \\
$\QMC(\rho)$ & Set of Markov chain extensions of $\rho_{AC}$ \\
$\pi_{A}$ & Maximally mixed state or uniform distribution\\
$\chi_{XX'}^{|p}$ & Perfectly correlated state according to distribution $p$ \eqref{eq:perf-corr-rand} \\
$D_{\max}(\rho||\sigma)$ & Max divergence \eqref{eq:Dmax} \\
$I^{z}_{\max}(A\mcol B)_{\rho}, \, z \in  \{\upuparrow,\uparrow,\downarrow\}$ & Max mutual informations \eqref{eq:Iupupmax-defn}-\eqref{eq:Idownmax-defn} \\
$I^{z,\ve}_{\max}(A \mcol B)_{\rho} \, , z \in  \{\upuparrow,\uparrow,\downarrow\}$ & Smooth Max Mutual Informations \eqref{eq:smooth-max-mutual-infos} \\
$C^{\ve}_{F}(A\mcol C)_{\rho},C^{\ve}_{U,F}(A\mcol C)_{\rho}$ & Correlation of Formation \eqref{eq:corr-form-defn} and uniform variation \eqref{eq:corr-form-unif-defn}\\
$C_{\max}(A\mcol B)_{\rho},C^{\ve}_{\max}(A\mcol B)_{\rho}$ & Max Wyner common information \eqref{eq:max-common-information} and smooth variant \eqref{eq:SMCI} \\
\label{tab: notations}
\end{tabular}
\caption{\small Major notational conventions and where to find the corresponding definitions.}
\end{table}

\paragraph{General Notation} Finite alphabets are denoted by caligraphic Roman letters at the end of the alphabet, e.g. $\cX,\cY,$ etc, and random variables over these alphabets are denoted by the corresponding capital Roman letters $X$, $Y$, etc. We denote the space of probability distributions on an alphabet $\cX$ by $\cP(\cX)$ and a specific probability distribution $p_X \in \cP(\cX)$ will have its elements denoted  $p_X(x)$. When the underlying random variable is clear, the subscript $X$ will be omitted on these objects. Hilbert spaces are labeled with capital roman letters at the start of the alphabet, $A,B,C,$ etc. The space of linear maps from Hilbert space $A$ to Hilbert space $B$ is denoted $\Lin(A,B)$ and the space of endomorphisms on Hilbert space is simplified to $\Lin(A) := \Lin(A,A)$. We denote the identity on $A$ by $\mbb{1}_{A}$. The tensor product space $A_{1} \otimes A_{2} \otimes \hdots \otimes A_{n}$ will sometimes be simplified to $A^{n}$  to mirror common notation, $\cX^{n} := \times_{i \in [n]} \cX$.  We denote the space of positive operators on a Hilbert space $A$ by $\Pos(A)$. We generally denote generic positive operators as $P,Q \in \Pos(A)$. The space of density matrices on a Hilbert space $A$ is denoted $\Density(A) := \{\rho: \Tr[\rho]= 1\}$ where $\Tr[\cdot]$ is the trace functional. We denote the space of subnormalized density matrices $\Density_{\leq}(A) := \{\rho: \Tr[\rho] \leq 1\}$. We often will place subscripts on endomorphisms to specify what space they act on, e.g. $\rho_{AB} \in \Density(A \edit{\otimes} B)$. The space of separable operators on $A \otimes B$ is denoted $\Sep(A\mcol B)$ and the set of separable density matrices is denoted $\Sep\Density(A\mcol B)$ for clarity. The space of quantum channels (completely-positive, trace-preserving maps) from $\Lin(A)$ to $\Lin(B)$ is denoted $\Channel(A,B)$. We denote the identity channel by $\text{id}_{A}$.

\paragraph{Divergence-Induced Quantities} 
 Recall the definition of the Umegaki divergence, a.k.a. relative entropy:
\begin{align}
	D(\rho \Vert \sigma) := \Tr[\rho \log(\rho) - \rho \log(\sigma))] \ .
\end{align}
This quantity induces the mutual information of a bipartite operator $P_{AB}$
\begin{align}
	I(A \mcol B)_{\rho} := D(\rho_{AB} \Vert \rho_{A} \otimes \rho_{B}) \ .
\end{align}
A key property of the mutual information is that it admits multiple equivalent variational forms:
\begin{equation}\label{eq:variational-mutual-information}
\begin{aligned}
	I(A:B)_{\rho} &= \min_{\sigma_{B} \in \Density(B)} D(\rho_{AB} \Vert \rho_{A} \otimes \sigma_{B}) \\
	 &= \min_{\tau_{A} \in \Density(A)} D(\rho_{AB} \Vert \tau_{A} \otimes \rho_{B}) \\
	&= \min_{\tau_{A} \in \Density(A), \sigma_{B} \in \Density(B)} D(\rho_{AB} \Vert \tau_{A} \otimes \sigma_{B}) \ ,
\end{aligned}
\end{equation}
which follows from the chain rule 
$D(\rho_{AB}\Vert \sigma_{A} \otimes \tau_{B}) = D(\rho_{AB} \Vert \rho_{A} \otimes \rho_{B}) + D(\rho_{A} \Vert \sigma_{A}) + D(\rho_{B} \Vert \tau_{B})$ and the non-negativity of Umegaki divergence when both arguments have the same trace.

In the one-shot scenario, a commonly used alternative to the Umegaki divergence is the \textit{max-divergence}
\begin{align}\label{eq:Dmax}
	D_{\max}(\rho \Vert \sigma) := \inf\{\lambda : \rho \leq \exp(\lambda) \sigma \} \ .
\end{align}
This quantity then induces the max mutual information quantities.
\begin{definition}\label{def:max-mutual-information-quantities}
 Let $\rho \in \Density_{\leq}(A\edit{\otimes}B)$.  Then we define the doubly up arrow max mutual information, the max mutual information, and the down arrow max mutual information respectively as
\begin{align}
	I^{\upuparrow}_{\max}(A\mcol B)_{\rho} &:= D_{\max}(\rho_{AB} \Vert \rho_{A} \otimes \rho_{B}) \label{eq:Iupupmax-defn} \\
	I^{\uparrow}_{\max}(A\mcol B)_{\rho} &:= \min_{\sigma_{B} \in \Density(B)} D_{\max}(\rho_{AB}\Vert\rho_{A} \otimes \sigma_{B}) \label{eq:Iupmax-defn} \\
	I^{\downarrow}_{\max}(A\mcol B)_{\rho} &:= \min_{\substack{\tau_{A} \in \Density(A) \\ \sigma_{B} \in \Density(B)}} D_{\max}(\rho_{AB}\Vert\tau_{A} \otimes \sigma_{B}) \ ,  \label{eq:Idownmax-defn}
\end{align}
\end{definition}
We make a few remarks about these definitions. First, the multiple definitions are motivated by \eqref{eq:variational-mutual-information}, but we no longer guarantee they all take the same value. Second, \eqref{eq:Iupmax-defn} was introduced in \cite{Berta-2010a} while \eqref{eq:Iupupmax-defn} and \eqref{eq:Idownmax-defn} were introduced in \cite{Ciganovic-2013a}. Our notation differs from both of these to make the relation $I^{\upuparrow}_{\max}(A\mcol B) \geq I^{\uparrow}_{\max}(A\mcol B) \geq I^{\downarrow}_{\max}(A\mcol B) $ explicit and also to align with the notation of more recent work, namely \cite{Mckinlay-2020a}. Beyond this, to simplify the notation of measuring a number of bits, we will use the \textit{Hartley entropy}
\begin{align}\label{eq:Hartley-entropy}
	H_{0}(A)_{\rho} := \log(\text{rank}(\rho)) \ .
\end{align}
We will also at times make use of the min-entropy:
\begin{align}
	H_{\min}(A|B)_{\rho} := \max_{\sigma_{B} \in \Density(B)} - D_{\max}(\rho_{AB} \Vert \mbb{1}_{A} \otimes \sigma_{B}) \ .
\end{align}

\paragraph{Smooth Entropy Calculus and Asymptotic Equipartition Properties} To handle error in the one-shot setting and bridge the gap between the one-shot and asymptotic settings, one may use the smooth entropy calculus, which involves a smoothing ball. In this work, we will use the following form.
\begin{definition}
Let $\rho \in \Density_{\leq}(A)$. Then,
\begin{align}
	\Bve(\rho) := \{\wt{\rho} \in \Density_{\leq}(A) : P(\rho,\wt{\rho}) \leq \ve \} \ ,
\end{align}
where $P(\cdot,\cdot)$ is the purified distance, which is defined in \eqref{eq:purif-dist-def}.
\end{definition}
A key property that we will use at times in this work is the following.
\begin{lemma}\cite{Ciganovic-2013a}\label{lem:re-norm-in-ball}
Let $\rho_{AB} \in \Density(A\edit{\otimes}B)$. Then for $\ve \geq 0$, if $\wt{\rho}_{AB} \in \Bve(\rho)$, then $\Bve(\rho) \ni \widehat{\rho} := \wt{\rho}/\Tr[\wt{\rho}]$.
\end{lemma}

Using the smoothing ball, one defines smooth entropic quantities. Of particular relevance for this work are the smooth max mutual information quantities:
\begin{align}\label{eq:smooth-max-mutual-infos}
	\hspace{-2mm} I^{z,\ve}_{\max}(A \mcol B)_{\rho} := \min_{\wt{\rho} \in \Bve(\rho)} I^{z}_{\max}(A \mcol B)_{\wt{\rho}} \quad z \in \{\upuparrow,\uparrow,\downarrow\} \ .
\end{align}
We note that these smooth quantities can be related up to some smoothing correction terms as established in \cite{Ciganovic-2013a}. We will also at times make reference to the smooth min-entropy:
\begin{align}
	H^{\ve}_{\min}(A \vert B)_{\rho} := \max_{\wt{\rho} \in \Bve(\rho)} H_{\min}(A \vert B)_{\rho} \ .
\end{align}

The remaining aspect of the smooth entropy calculus is the establishment of asymptotic equipartition properties, which recover asymptotic results from the one-shot bounds. For example, \cite{Berta-2011a} establishes the following `weak' asymptotic equipartition property for $\rho_{AB} \in \Density(A\edit{\otimes}B)$, 
\begin{align*}
	\lim_{\ve \to 0} \lim_{n \to \infty} \left[\frac{1}{n} I^{\uparrow,\ve}_{\max}(A^{n} \mcol B^{n})_{\rho^{\otimes n}} = I(A\mcol B)_{\rho} \right] \ .
\end{align*}
We call this `weak' as it `only' establishes the rate for vanishing-error, i.e. it does not establish a strong converse. In contrast, if the smoothing parameter is not required to go to zero, we call this a `strong' asymptotic equipartition property. Beyond such asymptotic equipartition properties, there are also \textit{second-order expansions} of some smooth quantities. For example, \cite{Tomamichel-2013a} established that the smooth max-divergence, $D^{\ve}_{\max}(\rho \Vert \sigma) := \min_{\wt{\rho} \in \Bve(\rho)} D_{\max}(\wt{\rho} \Vert \sigma)$, satisfies the following second order expansion
\begin{equation} 
\begin{aligned}
	D^{\ve}_{\max}(\rho^{\otimes n} \Vert \sigma^{\otimes n}) &= nD(\rho \Vert \sigma) - \sqrt{nV(\rho \Vert \sigma)}\Phi^{-1}(\ve^{2}) \\
	& \hspace{25mm} + O(\log(n)) \ ,
\end{aligned}
\end{equation}
where $V(\rho \Vert \sigma)$ is the quantum information variance and $\Phi^{-1}$ is the inverse of the cumulative distribution function of the normal distribution (see \cite{Tomamichel-2013a} for more details in the quantum setting and \cite{Tan-2014a} for the classical setting). The key point of this result for this work is that the `second-order' term, the varentropy term, scales as $\sqrt{n}$. We will later show that our quantities have correction terms that scale as $O(\log(n))$, and they are thus tight to any second-order term that also scales as $\sqrt{n}$.

\paragraph{Distance and Entanglement Measures} In this work we use three distance measures. We will at times use the trace distance $\text{TD}(\rho,\sigma) := \frac{1}{2}\Vert \rho - \sigma \Vert_{1}$ (and the $L_{1}$-norm that induces it). We also use the fidelity between positive semidefinite operators $P,Q$:\footnote{We use the definition of fidelity with a square to align with \cite{van-2003a}, which we appeal to later in the paper.}
\begin{align}
	F(P,Q) := \Vert \sqrt{P} \sqrt{Q} \Vert_{1}^{2} \ .
\end{align}
As we already saw in the definition of the smoothing ball, we will also make use of the purified distance, which is well defined for $\rho,\sigma \in \Density_{\leq}(A)$, \begin{align}\label{eq:purif-dist-def}
P(\rho,\sigma) \coloneq \sqrt{1-F_{\ast}(\rho,\sigma)} \ , 
\end{align} where $F_{\ast}(\rho,\sigma)$ is the ``generalized fidelity",
\begin{align}
	F_{\ast}(\rho,\sigma) \coloneq \left( \sqrt{F(\rho,\sigma)} + \sqrt{(1-\Tr[\rho])(1-\Tr[\sigma])} \right)^{2} \ .
\end{align}

We will at times use the relationship between trace distance and purified distance when one of the states is normalized:
\begin{align}\label{eq:Purified-Dist-FvdG}
	\mrm{TD}(\rho,\sigma) \leq P(\rho,\sigma) \leq \sqrt{2\mrm{TD}(\rho,\sigma)} \ .
\end{align}  
To simplify our notation, we introduce the following notation for similarity under given distance measures:
\begin{align}\label{eq:TD-PD-relation}
P(\rho,\sigma) \leq \varepsilon & \Leftrightarrow \rho \approx_{\ve} \sigma \\  
2\mrm{TD}(\rho,\sigma) \leq \varepsilon & \Leftrightarrow \rho \approx_{\varepsilon}^{\Lone} \sigma \\
F(\rho,\sigma) \geq 1- \varepsilon & \Leftrightarrow \rho \approx_{\ve}^{F} \sigma \ . 
\end{align}

As the above are dissimilarity measures, we can use them to quantify the distance between a state and the set of separable states, which will be useful later.
\begin{definition}
For $\rho_{AB} \in \Density(A \otimes B)$, the trace distance of entanglement is defined as
$$ E_{T}(A\mcol B)_{\rho} := \inf_{\sigma_{AB} \in \Sep\Density(A\mcol B)} \mrm{TD}(\rho,\sigma) \ , $$
and the purified distance of entanglement, $E_{P}(A\mcol \edit{B})_{\rho}$, is defined identically with $\mrm{TD}(\cdot,\cdot)$ replaced with the purified distance $P(\cdot,\cdot)$.
\end{definition}
All further technical notation will be introduced as it arises.

\section{Main Results and Summary of Technical Barriers}\label{sec:main-theorems}
In this section we state the main results of our work. This allows us to highlight various technical nuances that crop up in the following sections and how they are addressed.

\subsection{Distributed Source Simulation with a Classical Seed}

We begin with the central theorem of this work, which are one-shot bounds on non-uniform and uniform distributed source simulation (i.e. Fig.~\ref{fig:DSS-fig-intro}), represented by operational quantities that we call `correlations of formations,' $C^{\ve}_{F}(A \mcol C)_{\rho}$ and $C^{\ve}_{U,F}(A \mcol C)_{\rho}$ respectively. The bounds are presented using correlation measures that are variations of Wyner's common information induced by max mutual information quantities. The operational quantities are discussed in detail in Section \ref{sec:operational-quantities}, the correlation measures are discussed in detail in Section \ref{sec:one-shot-corr-meas}, and the one-shot rates themselves are derived in Section \ref{sec:one-shot-rate}.
\begin{tcolorbox}[width=\linewidth, sharp corners=all, colback=white!95!black, boxrule=0pt,frame hidden]
\begin{theorem}\label{thm:one-shot-DSS-main-theorem}
Let $\ve_{1},\ve_{2},\ve \in (0,1)$ such that $\ve_{1}+\ve_{2} \leq \ve$. Let $\rho_{AC} \in \Density(A\edit{\otimes}C)$ such that $2E_{T}(A\mcol C)_{\rho} \leq \ve_{1}$. Then,
\begin{equation}
\begin{aligned}
	C^{\sqrt{\ve}}_{\max}(A \mcol C)_{\rho} &\leq  C^{\ve}_{F}(A \mcol C)_{\rho} \\
	 &\leq C^{\sqrt{\ve_{1}}}_{\max}(A\mcol C)_{\rho}  + \text{corr}(\ve_{2})  \label{eq:non-uniform-DSS-one-shot-bounds}
\end{aligned}
\end{equation}
\begin{equation}
\begin{aligned}
	 C^{\upuparrow,\sqrt{\ve}}_{\max}(A \mcol C)_{\rho} &\leq  C^{\ve}_{U,F}(A \mcol C)_{\rho} \\
	 &\leq C^{\upuparrow,\sqrt{\ve_{1}}}_{\max}(A\mcol C)_{\rho}  +\text{corr}(\ve_{2}) \ , \label{eq:uniform-DSS-one-shot-bounds}
\end{aligned}
\end{equation}
where the operational quantities $C^{\ve}_{F}(A \mcol C)_{\rho}$, $C^{\ve}_{U,F}(A \mcol C)_{\rho}$ are given in Definition \ref{def:corr-of-form}, the correlation measures $C^{\ve}_{F}(A \mcol C)_{\rho}$, $C^{\upuparrow,\ve}_{F}(A \mcol C)_{\rho}$ are given in Definition \ref{def:max-Wyner-CIs}, and $\text{corr}(\ve_{2}) = \frac{5}{2}\log\left(\frac{5}{\ve_{2}}\right) + \frac{1}{2}$. \\

In particular this implies that when $\lim_{n \to \infty} 2E_{T}(A^{n} \mcol C^{n})_{\rho^{\otimes n}} < \ve$, the second-order asymptotics of $C_{F}^\ve$ and $C_{U,F}^{\ve}$ correspond to the second order terms in the expansion of our quantities $C_{\max}^{\sqrt{\ve}}$ and $C_{\max}^{\upuparrow,\sqrt{\ve}}$, respectively, provided they scale proportional to $\sqrt{n}$. This especially holds for all separable states as $2E_{T}(A^{n} \mcol C^{n}) = 0$ for all $n \in \mbb{N}$.
\end{theorem}
\end{tcolorbox}
\begin{proof}
	The achievability bounds follow from Lemma \ref{lem:achievability-bounds-for-one-shot-DSS}. The converse bounds follow from Lemma \ref{lem:one-shot-formation-converse}.

	For the `in particular' statement, note that if we let $\ve_{1,n} := (1-\frac{1}{n})\ve$ and $\ve_{2,n}:= \frac{1}{n}\ve$, then $\frac{5}{2}\log\left(\frac{5}{\ve_{2,n}}\right) = \frac{5}{2}\log\left(\frac{5n}{\ve}\right) = O(\log(n))$. Thus, so long as the second-order expansion scales as $\sqrt{n}$, we know the correction terms of the size $O(\log(n))$ are irrelevant for establishing second-order asymptotics by letting $\ve_{1,n}$ and $\ve_{2,n}$ be constructed as given.
\end{proof}
\begin{remark}
	The achievable bound in \eqref{eq:non-uniform-DSS-one-shot-bounds} is in fact also achievable for $C^{\ve}_{U,F}(A \mcol C)_{\rho}$. However, as the one-shot converse is in terms of a stronger measure, we know higher order asymptotics are limited by the converse measure, which is why we have expressed it in this form. This suggests $C^{\sqrt{\ve}}_{\max}$ and $C^{\upuparrow,\sqrt{\ve_1}}_{\max}$ may agree to second-order in their expansions.
\end{remark}

\paragraph{Desiderata and Technical Difficulties}
We now highlight the appeal of Theorem \ref{thm:one-shot-DSS-main-theorem} and discuss the technical difficulties in establishing it. As is standard in information theory, the ultimate goal is to prove bounds on different tasks in terms of entropic quantities. Here, the goal is to bound the correlation cost in the task of Fig.~\ref{fig:DSS-fig-intro} with entropic quantities $\mbb{C}^{?}$ that satisfy the following desiderata:
\begin{enumerate}
\renewcommand\labelitemi{}
\setlength\itemsep{2.5pt}
	\item \textbf{Generality:} The one-shot rate applies under all feasible situations of the task,
	\item \textbf{Approximately Sharp:} the correlation measures can be established as approximately tight under some condition,
	\item \textbf{Obtainable:} the correlation measure's solution is obtained, i.e. all optimizations are minimizations rather than infimizations,\footnote{We note this does not imply the quantity is efficient to compute.} and
	\item \textbf{Asymptotically Appropriate:} the correlation measures provably recover Wyner's common information in the asymptotic limit, i.e. on i.i.d. inputs it should satisfy 
	$$ \lim_{n \to \infty} \left[ \frac{1}{n} \mbb{C}^{?}(A^{n}\mcol C^{n})_{\rho^{\otimes n}} \right] = \min_{A-X-C} I(AC\mcol X)_{\rho}  \ , $$
where we note the right hand side is Wyner's common information as per \eqref{eq:Wyner-common-information}.
\end{enumerate}

We stress that Item 1 holds as Theorem \ref{thm:one-shot-DSS-main-theorem} applies even for the simulation of non-separable states that are sufficiently close to the separable set, which has not been considered previously. To highlight this nuance, we provide the following example.
\begin{example}[One-Shot DSS of Entangled State]\label{example:Werner-states}
	For any local dimension of size $d \geq 2$, consider the family of Werner states which may be parameterized as
\begin{align}
    \rho^{\lambda}_{AC} &:= \frac{2\lambda}{d(d+1)}\Pi_{+} + \frac{2(1-\lambda)}{d(d-1)}\Pi_{-} \, , \lambda \in [0,1] \label{eq:WH-defn-lambda} \ ,
\end{align}
where $\Pi_{+}, \Pi_{-}$ are the projectors onto the symmetric and anti-symmetric spaces respectively. It is known that $\rho^{\lambda}$ is entangled for $\lambda \in [0,1/2)$ \cite{Watrous-Book}. Consider the Werner state $\rho^{1/2-\delta}$ where $\delta \in (0,1/2)$. This state is entangled, but by letting $\delta$ being close to zero, this state may be made sufficiently close to a separable state to simulate. For example, a direct calculation will determine $F(\rho^{1/2-\delta},\rho^{1/2}) = \left( \sqrt{(1/2-\delta)/2} + \sqrt{(1/2+\delta)/2} \right)^{2}$, which means $F(\rho^{1/2},\rho^{1/2-\delta}) > 0.99$ for $\delta \leq 0.0994$. It follows that it would be reasonable to use distributed source simulation to simulate close-to-separable states, and as one-shot distributed source simulation only measures the error in terms of the output state $\rho_{AC}$ (see Definition \ref{def:corr-of-form}), nothing precludes this approximate simulation from an operational or formal standpoint.
\end{example}
Not only does our result cover certain entangled states like the above example, but it in principle extends to the asymptotic setting in the following sense. Theorem \ref{thm:one-shot-DSS-main-theorem} shows that our one-shot rate is tight enough such that if $\rho_{AC}^{\otimes n}$ stays `sufficiently close' to the space of separable states relative to the tolerated simulation $\ve\in(0,1)$ in $L_1$-norm, then our result still applies even if the underlying state $\rho_{AC}$ is entangled. This possibility has not been previously considered, and we leave it as a challenging open problem to construct an explicit example of this phenomenon.\footnote{If there exists an entangled state $\rho$ and parameters $0 < \ve_{1} < \ve < 1$ such that for all sufficiently large $n$,  $2E_{T}(A^{n} \mcol C^{n})_{\rho^{\otimes n}} \leq \ve_{1}$ so that $C^{\ve}_{\max}(A^{n} \mcol C^{n})_{\rho^{\otimes n}}$ is finite, then the first-order term of its expansion at this point cannot be $C(A \mcol C)_{\rho}$ since only separable states admit a Markov chain extension (See Section \ref{sec:operational-quantities}). This may suggest such a phenomenon does not exist. Moreover, using Pinsker's inequality and the Generalized Quantum Stein's Lemma \cite{Hayashi-2024a,Lami-2024a} guarantees this is not a phenomenon that generically arises for $\ve$ close to unity.} The key technical difficulty in establishing these one-shot rates is determining methods that apply even for target entangled states, as otherwise we could use previous results, in particular \cite{Shen-2022a}. See Section \ref{sec:one-shot-rate} for details.

Item 2 of our desiderata holds as we have reduced the problem of second-order asymptotics to second-order expansions of our correlation measures \edit{in the sense given} in Theorem \ref{thm:one-shot-DSS-main-theorem}, which shows the results are quite tight. Item 3 is important as the correlation measures optimize over a Markov chain extension and thus involve a classical auxiliary random variable (See Definition \ref{def:max-Wyner-CIs}). To address Item 3, we establish cardinality bounds for each correlation measure in Section \ref{subsec:cardinality-bounds}. While Theorem \ref{thm:one-shot-DSS-main-theorem} in effect already tells us our measure is asymptotically meaningful, \edit{likely} even to second-order, it turns out it is difficult to actually establish this fact even to first order. This difficulty again arises from optimizing over the auxiliary random variable $X$ in a Markov chain extension. In particular, the auxiliary random variable  need not admit an i.i.d.~form. Consequently, to establish our result we have needed to alter from standard approaches and instead appeal to strong conditional typicality of quantum states  (see \cite{Wilde-2011a} for a review on quantum conditional typicality). This move, however, only allows us to establish an AEP for our measures that require the smoothing parameter $\ve$ to tend to zero.
\begin{tcolorbox}[width=\linewidth, sharp corners=all, colback=white!95!black, boxrule=0pt,frame hidden]
\begin{theorem*}
	Let $\rho_{AB} \in \Sep\Density(A\edit{:}B)$. Then,
	\begin{align}
		\edit{\lim_{\ve \to 0}} \lim_{n \to \infty} \frac{1}{n} C^{\ve}_{\max}(A^{n} \mcol \edit{B}^{n})_{\rho^{\otimes n}} = C(A \mcol \edit{B})_{\rho} \ ,
	\end{align}
\end{theorem*}
where the right hand side is the Wyner common information given in \eqref{eq:Wyner-common-information}.
\end{tcolorbox}
As explained in Section \ref{sec:technical-background}, we call this result a `weak' asymptotic equipartition property as it is not strong enough to establish a strong converse for our problem. The result itself is proven in Section \ref{sec:asymptotics} where we also discuss barriers to establishing a strong form or second-order expansion of our correlation measures. As partial progress in this respect, we provide the first proof of a strong converse for a certain definition of smooth max mutual information:
\begin{tcolorbox}[width=\linewidth, sharp corners=all, colback=white!95!black, boxrule=0pt,frame hidden]
$\forall \rho \in \Density(A\edit{\otimes}B) \, , \, \ve \in (0,1),$
\begin{align}
 \lim_{n \to \infty} \left[ \frac{1}{n} I^{\ve}_{\max}(A^{n}\mcol B^{n})_{\rho^{\otimes n}} \right] = I(A\mcol \edit{B})_{\rho} \ . 
\end{align} 
\end{tcolorbox}
\noindent This result has independent value for establishing strong converses through the smooth entropy calculus.

\subsection{Extensions and Variations of Distributed Source Simulation}
Beyond the above main result, we consider other operationally or technically motivated variations to the distributed source simulation problem. Namely, we consider (i) the extension to more than two receivers, (ii) the extension to seed states that are fully quantum, and (iii) other subtle variants that naturally emerge when analyzing the relationship between distributed source simulation and the soft-covering lemma. We summarize these results and insights here. 
\paragraph{On Multi-Receiver and Fully Quantum Extensions}
\begin{figure}
\centering
\begin{subfigure}[b]{\columnwidth}
   \begin{center}
        \begin{tikzpicture}
            \tikzstyle{porte} = [draw=black!50, fill=black!20]
            		\draw ++(-2.5,0) node (sigma) {$\ket{\mu(n)}_{A'C'} \otimes \chi^{\vert p}_{XX'}$};
                \draw ++(-1.25,0.8) node (A') {$A'X$};
                \draw ++(-1.25,-0.8) node (C') {$C'X'$};
                \draw ++(0.25,1) node[porte] (A-prep) {$\Phi_{A'X \to A}$};
                \draw ++(0.25,-1) node[porte] (C-prep) {$\Phi_{C'X' \to C}$};
                \draw ++(2,1) node (A) {$A$};
                \draw ++(2,0.25) node (C) {$C$};
                \draw ++(3.5,0.6) node (rhoout) {$\rho_{AC}$};
                \draw ++(2.75,0.6) node (approx) {$\approx_{\ve}$};
                \draw[->] (A-prep) -- (A);
                \draw[->] (C-prep) to[out=+0,in=+180] (C);
                \draw[->] (sigma) to[out=+0,in=+180] (A-prep);
                \draw[->] (sigma) to[out=+0,in=+180] (C-prep);
        \end{tikzpicture}
    \end{center}
   \caption{Embezzling Entangled Source Simulation}
\end{subfigure}
\\[5mm]
\begin{subfigure}[b]{\columnwidth}
\begin{center}
	     \begin{tikzpicture}
            \tikzstyle{porte} = [draw=black!50, fill=black!20]
            		\draw ++(0,0) node (sigma) {$\sigma_{A'C'}$};
                \draw ++(0.5,0.8) node (A') {$A'$};
                \draw ++(0.5,-0.8) node (C') {$C'$};
                \draw ++(1.75,1) node[porte] (A-prep) {$\Phi_{A' \to AA'}$};
                \draw ++(1.75,-1) node[porte] (C-prep) {$\Phi_{C' \to CC'}$};
                \draw ++(3.75,1) node (A) {$A$};
                \draw ++(3.75,0.25) node (C) {$C$};
                \draw ++(3.75,-0.25) node (Apout) {$A'$};
                \draw ++(3.75,-1) node (Cpout) {$C'$};
                \draw ++(5.25,0.75) node (rhoout) {$\rho_{AC}$};
                \draw ++(4.5,0) node (approx) {$\approx_{\ve}$};
                \draw ++(5.25,0) node (otimes) {$\otimes$};
                \draw ++(5.25,-0.75) node (sigmaout) {$\sigma_{A'C'}$};
                \draw[->] (A-prep) -- (A);
                \draw[->] (A-prep) to[out=+0,in=+180] (Apout);
                \draw[->] (C-prep) to[out=+0,in=+180] (C);
                \draw[->] (C-prep) -- (Cpout);
                \draw[->] (sigma) to[out=+0,in=+180] (A-prep);
                \draw[->] (sigma) to[out=+0,in=+180] (C-prep);
        \end{tikzpicture}
    \end{center}
   \caption{Catalytic Entangled Source Simulation}
\end{subfigure}
\caption{Primary versions of entangled state source simulation considered. (a) Target state $\rho_{AC}$ is prepared using local operations on an embezzling state $\ket{\mu(n)}$ and arbitrary classical correlation. (b) Target state $\rho_{AC}$ is constructed using local operations with $\sigma_{A'C'}$ as a catalysis. This means the $A'$ and $C'$ registers are relevant to the error criterion rather than implicitly ignored as in (a).}
\label{fig:EDSS-intro}
\end{figure}
In the classical asymptotic scenario, the extension to multiple receivers has been studied \cite{Liu-2010a}. Similarly, in the one-shot setting the extension to multiple receivers is rather straightforward, and we provide details on this extension in Section \ref{sec:multi-receiver}. With regards to the simulation of quantum states, there is some question as to the appropriate way in which Wyner's original problem should be generalized into the quantum setting. As argued in the introduction, one can recover the notion of a universal quantum seed by replacing the classical seed of Fig. \ref{fig:DSS-fig-intro} with a source of entangled embezzling states. We refer to this case as embezzling entangled source simulation. This identification of embezzling entangled source simulation as the appropriate analogue of distributed source simulation is further justified by our definition and analysis of other forms of entangled source simulation, including
catalytic entangled source simulation, where the input quantum correlations must be approximately independent of thesim ulated state (see Fig. \ref{fig:EDSS-intro}). For catalytic entangled source simulation, when using embezzling states, we show that for fixed approximation error, the one-shot rate is additive over number of copies (Proposition \ref{prop:incompressibility}), i.e. the needed entanglement under this strategy is incompressible. For embezzling entangled source simulation, we establish upper and lower bounds on the one-shot rate, $C^{\ve}_{\mrm{Emb-EoS}}(A\mcol C)_{\rho}$, in terms of an entanglement measure, $\mrm{Ent}^{\ve}_{A:C}(\rho)$, that are near-optimal yet diverge as the error goes to zero.
\begin{tcolorbox}[width=\linewidth, sharp corners=all, colback=white!95!black, boxrule=0pt,frame hidden]
\begin{theorem*}
Let $\rho \in \Density(A \otimes C)$ and $\ve \in [0,1)$. Then 
\begin{align*}
 \log(\mrm{Ent}^{\ve}_{A:C}(\rho)) &\leq C^{\ve}_{\mrm{Emb-EoS}}(A\mcol C)_{\rho} \\
 &\leq \frac{1}{\ve} \log(\mrm{Ent}_{A:C}(\rho)) \ .
\end{align*}
Moreover, there exist distinct cases where both bounds are near-optimal, although they don't match.
\end{theorem*}
\end{tcolorbox}
\noindent These results complement related recent work on converting entangled states with zero communication by a subset of the authors of this work \cite{pure-state-revisited}.

\paragraph{Variations of Distributed Source Simulation} 
As noted in the introduction, prior to this work, distributed source simulation has been studied for classical states or separable states with vanishing error \cite{Hayashi-2006a}. In these settings, distributed source simulation is nearly immediate from any soft-covering lemma. In the one-shot setting, to have the most general result that includes close-to-separable target states this is no longer the case. In particular, we cannot apply the recent tight to second-order soft-covering lemma of Shen et al. \cite{Shen-2022a} without either:
\begin{enumerate}[itemsep=0pt]
	\item Adding a secondary smoothing ball to the already-smoothed hypothesis testing mutual information quantity which introduces a technical issue with the correction term, or
	\item Restricting to simulating separable states.
\end{enumerate}
Given this, it is reasonable to ask if there is a task that is more directly captured by one-shot soft-covering. In the appendix, we introduce variants of distributed source simulation where the system is required to simulate the whole Markov chain structure rather than just the output. The achievability of these variants are more directly addressed by \cite{Shen-2022a}, which highlights the difference between analyzing one-shot soft-covering and one-shot distributed source simulation. In one of these variants, we then allow the hub to send entangled states, which we call `entanglement-assisted' distributed source simulation and denote its operational quantity by $\wt{C}^{\ve}_{F}(A\mcol C)_{\rho}$. We show that when the target state is separable, the entanglement-assistance is not helpful even in the one-shot setting (Lemma \ref{lem:characterization-of-variant-of-DSS}) and establish the following one-shot bounds.
\begin{proposition}\label{prop:one-shot-rate-for-EA-DSS}
Let $\rho_{AC}$ be a separable quantum state and tolerated error be $\ve \in (0,1)$. Then for $\delta \in (0, \ve/2)$,
\begin{equation}
\begin{aligned}
	 &\wt{C}_{h}^{1-2\ve-\delta}(A\mcol C)_{\rho} + 3\log(\delta) - 3\log3 - \log(1-2\ve) \\
	 & \hspace{1.8cm} \leq \wt{C}^{\ve}_{F}(A\mcol C)_{\rho} \\
	 & \hspace{2.2cm} \leq \wt{C}^{1-\ve+3\delta}_{h}(A\mcol C)_{\rho} + \log(\nu^{2}/\delta^{4}) \ ,
\end{aligned}
\end{equation}
where $\wt{C}^{\ve}_{F}(A\mcol C)_{\rho},\wt{C}^{\ve}_{h}(A \mcol C)_{\rho}$ are given in Definitions \ref{def:EA-assisted-corr-of-form} and \ref{def:hyp-test-common-info} respectively and $\nu := |\text{spec}(\rho_{AC})|$, which is the number of distinct eigenvalues of $\rho_{AC}$ 
\end{proposition}
We note that, unlike Theorem \ref{thm:one-shot-DSS-main-theorem}, this is not provably tight to terms that scale as $O(\sqrt{n})$ due to the discrepancy of the $\ve$ term in the smoothing in the achievability and converse results. We are however able to show that the rate of these variants are equivalent to first-order to distributed source simulation proper for separable states.

\begin{tcolorbox}[width=\linewidth, sharp corners=all, colback=white!95!black, boxrule=0pt,frame hidden]
\begin{theorem}\label{thm:first-order-asymptotic-equivalence}
	Let $\rho \in \Sep\Density(A\mcol C)$. The rates of distributed source simulation with or without uniform randomness, the entanglement-assisted, and private distributed source simulation all are given by the common information:
\begin{align*}
	\hspace*{-3mm} C(A\mcol C)_{\rho} 
	&= \lim_{\ve \to 0} \lim_{n \to \infty} \left[\frac{1}{n} C^{\ve}_{F}(A^{n}\mcol C^{n})_{\rho^{\otimes n}} \right] \\
	 &= \lim_{\ve \to 0} \lim_{n \to \infty} \left[ \frac{1}{n} C^{\ve}_{U,F}(A^{n}\mcol C^{n})_{\rho^{\otimes n}} \right] \\ 
	 &= \lim_{\ve \to 0} \lim_{n \to \infty} \left[ \frac{1}{n} \wt{C}^{\ve}_{F}(A^{n}\mcol C^{n})_{\rho^{\otimes n}} \right].
\end{align*}
\end{theorem}
\end{tcolorbox}

\section{Operational Quantities for One-Shot Distributed Source Simulation}\label{sec:operational-quantities}
Having introduced the formal background we will use throughout this work, we are in a position to introduce one-shot distributed source simulation (DSS). We first introduce this in the simplest manner and then relate this to quantum Markov chains to simplify how we characterize this task. Although the relation between Markov chains and DSS is well-known, we establish this relation somewhat more directly. This both technically makes certain subsequent sections' ideas more natural and may make the ideas clearer to a reader less familiar with DSS.

In principle, one-shot distributed source simulation is `simply' constructing some target state $\rho_{AC}$ up to some tolerated error $\varepsilon$ from some shared randomness and local operations. We note that we say shared randomness as the original classical randomness, $p_{X} = \sum_{x} p(x) \dyad{x}$, is copied resulting in perfectly correlated randomness which we denote
\begin{align}\label{eq:perf-corr-rand}
\chi^{|p}_{XX'} = \sum_{x} p(x) \dyad{x}_{X} \otimes \dyad{x}_{X'} \ .
\end{align} See Fig. \ref{fig:DSS-fig} for visualization.
\begin{figure}[ht]
    \begin{center}
        \begin{tikzpicture}
            \tikzstyle{porte} = [draw=black!50, fill=black!20]
                \draw
                (0,0) node (qy) {$p_{X}$}
                ++(1.5,0) node[porte] (copy) {Copy}
                ++(2,0.75) node[porte] (Alice-Proc) {$\Phi_{X \to A}$}
                ++(1.75,0) node (A) {$A$}
                ++(0,-1.5) node (C) {$C$}
                ++(-1.75,0) node[porte] (Bob-Proc) {$\Psi_{X' \to C}$}
                ++(-2,-0.25) node (X') {$X'$}
                ++(0,2) node (X) {$X$}
                ++(4.25,-1) node (approx) {$\approx_{\ve}$}
                ++(0.8,0) node (pxz) {$\rho_{AC}$}
                ;
                \draw
                (2.25,-1.5) node (chi) {$\chi_{XX'}^{|p}$}
                ;
                \draw
                (4.75,-1.5) node (rhotilde) {$\wt{\rho}_{AC}$}
                ;
                \draw [draw=blue!50, thick , densely dashed ] (2.25,1.2) -- (2.25,-1.2)
                ;
                \draw [draw=blue!120, thick , densely dashed ] (4.75,1.2) -- (4.75,-1.2)
                ;
                \path[draw = black, ->] (qy) -- (copy); 
                \path[draw=black, ->] (copy) |- (Alice-Proc);
                \path[draw=black, ->] (Alice-Proc) -- (A);
                \path[draw=black, ->] (copy) |- (Bob-Proc);
                \path[draw=black, ->] (Bob-Proc) -- (C);
        \end{tikzpicture}
    \end{center}
    \caption{Diagram of distributed source simulation of a quantum state. After the copying procedure (at the light blue line), the two parties share a perfectly correlated state $\chi^{|p}_{XX'}$. After their local processing (at the dark blue line), the parties share a state $\wt{\rho}_{AC}$ which should be approximately the target state $\rho_{AC}$.}
    \label{fig:DSS-fig}
\end{figure} 

Following Wyner, our interest is in \textit{how much} shared randomness, measured in number of bits, is necessary. We define the one-shot correlation of formation as the amount of randomness necessary for the task. We choose this terminology as it aligns well with terminology from resource theories such as the entanglement of formation \cite{Wootters-1998a,Chitambar-2019a}. Like Wyner, we consider the case of uniform randomness, but we also consider the case where we let the randomness be non-uniform. We now formalize all of this, starting with the notion of approximate simulation.
\begin{definition}\label{def:ve-sim}
Let $\sigma_{B} \in \Density(B)$ and $\Density(X \otimes B) \ni \rho_{XB} = \sum_{x} p(x)\dyad{x} \otimes \rho^{x}_{B}$ be a CQ state. We say $\rho_{XB}$ is a $\ve$-simulation of $\sigma_{B}$ if $\|\sum_{x} p(x)\rho^{x}_{B} - \sigma_{B}\|_{1} \leq \ve$. We denote this $\rho_{XB} \sim_{\ve} \sigma_{B}$.
\end{definition}
\begin{definition}
Let $\sigma_{AC} \in \Density(A \otimes C)$ and $\Density(AXC) \ni \rho_{AXC} = \sum_{x \in \Sigma} p(x) \rho_{A}^{x} \otimes \dyad{x} \otimes \rho_{C}^{x}$ where $p \in \cP(\Sigma)$. We say $\rho_{AXC}$ is a $\varepsilon$-distributed source simulation of $\sigma_{AC}$ if $\|\sum_{x} p(x) \rho_{A}^{x} \otimes \rho_{C}^{x} - \sigma_{AC} \|_{1} \leq \ve$.
\end{definition}
We could now use the definition of $\varepsilon$-distributed source simulation to define one-shot correlation of formation. However, we believe it is clearer to reduce the definition to being in terms of a Markov chain and the definition of $\ve$-simulation.
\begin{proposition}\label{prop:ve-DSS-to-MC-cond}
A $\varepsilon$-distributed source simulation of $\sigma_{AC}$ is a $A-X-C$ Markov chain that is a $\varepsilon$-simulation of $\sigma_{AC}$.
\end{proposition}
To establish this, we need the following theorem which will be relevant for much of this work.
\begin{theorem}(\cite{Hayden-2004a})\label{thm:QMC-equivalences}
The following are equivalent
\begin{enumerate}
	\item $\rho_{ABC}$ is a (short) quantum Markov chain (QMC), denoted $A-B-C$ or $\rho_{A-B-C}$.
	\item There exists a CPTP map $\cR: B \to B \otimes C$ such that $(\id_{A} \otimes \cR)(\rho_{AB}) = \rho_{ABC}$.
	\item There exists a CPTP map $\ol{\cR}: B \to  A \otimes B$ such that $(\ol{\cR} \otimes \id_{C})(\rho_{BC}) = \rho_{ABC}$.
	\item $I(A\mcol C|B) = 0$, where $I(A\mcol C|B)$ is the conditional mutual information.
	\item There exists a finite alphabet $\cJ$ such that there exists a decomposition of $B = \oplus_{j \in \cJ} b^{L}_{j} \otimes b^{R}_{j}$ such that 
	$$ \rho_{ABC} = \bigoplus_{j \in \cJ} \rho_{Ab^{L}_{j}} \otimes \rho_{b^{R}_{j}C} \ . $$
\end{enumerate}
\end{theorem}
All of these results in effect say that the $A$ space and $C$ space are independent so long as one has access to the $B$ space. It is then trivial to prove Proposition \ref{prop:ve-DSS-to-MC-cond}.
\begin{proof}[Proof of Proposition \ref{prop:ve-DSS-to-MC-cond}] There are various ways to prove $\rho_{AXC}$ is a QMC. For our case, note $\cR : \dyad{x}_{B} \mapsto \dyad{x}_{B} \otimes \rho_{C}^{x}$ and same idea for $\ol{\cR}$. Then letting $\sigma_{AC}$ act as $\sigma_{B}$, and $\rho_{AXC}$ as $\rho_{XB}$ in Definition \ref{def:ve-sim} completes the proof.
\end{proof}
With this established, we can define our correlation of formation measures. We note if we write a minimization/infimization over $\wt{\rho}_{A-X-C}$, this means we restrict to optimizing over QMC with a classical register $X$, and we remind the reader that if we write $\pi$ as a register, it means that the marginal on that register is the uniform distribution on a (classical) space $X$. Later, we will write minimization over $A-X-C$ without being a superscript when it is clear with respect to what state the QMC is being considered. Lastly, we always define the minimization/infimization over an empty set to be $+\infty$. 

\begin{definition}\label{def:corr-of-form}
	Let $\ve \in [0,1]$ and $\rho_{AC} \in \Density(A \otimes C)$. The correlation of formation is 
\begin{equation}\label{eq:corr-form-defn}
	\begin{aligned}
    & C^{\ve}_{F}(A\mcol C)_{\rho} \\
    & \hspace{4mm} :=\inf_{\wt{\rho}_{A-X-C}} \left\{H_{0}(X)_{\wt{\rho}} : \|\wt{\rho}_{AC} - \rho_{AC}\|_{1} \leq \ve \right\} \ .
	\end{aligned} 
\end{equation}
Moreover, the one-shot \textit{uniform} correlation of formation is defined as 
\begin{equation}\label{eq:corr-form-unif-defn}
\begin{aligned}
    & C^{\ve}_{U,F}(A\mcol C)_{\rho} \\
    & \hspace{4mm} := \inf_{\wt{\rho}_{A-\pi-C}} \left\{H_{0}(X)_{\wt{\rho}} :  \|\wt{\rho}_{AC} - \rho_{AC}\|_{1} \leq \ve \right\} .
\end{aligned}
\end{equation} 
\end{definition}
\begin{remark}
	 A careful reader may be baffled why we use twice the trace distance rather than the trace distance itself. This is because we will need to use the purified distance smoothing in our correlation measures. If we consider $\rho \approx_{\ve}^{L1} \wt{\rho}$, then $\rho \approx_{\sqrt{\ve}} \wt{\rho}$ whereas if we use trace distance directly there is a factor of 2. This is a problem if we want our results to hold for the full parameter range as in the converse we must convert to purified distance and $\sqrt{x} : [0,1] \to [0,1]$ but $\sqrt{2 x} : [0,1] \to [0,\sqrt{2}]$.
\end{remark}
Note that by construction, the one-shot uniform correlation of formation may be viewed as one-shot distributed source simulation. This can be seen as follows. By definition we are considering QMC $\wt{\rho}_{A-\pi-C}$, so the $X$ register is uniform and there exist local channels $\cR,\ol{\cR}$ that prepare the $A$ and $C$ registers from $X$. This means $X$ is the uniform randomness input, $\cR,\ol{\cR}$ are the local channels in the distributed source simulation, and by definition of the Hartley entropy \eqref{eq:Hartley-entropy} , $H_{0}(X)$ is measuring the minimum number of bits of uniform randomness necessary. Thus we have defined our operational quantity of primary interest.

We now briefly relate the existence of a quantum Markov chain to separability. We formally establish this as we frequently appeal to this idea throughout the paper.
\begin{lemma}\label{lem:sep-marg-of-QMC}
A quantum state $\rho_{AC} \in \Density(A \otimes C)$ has a Markov chain extension $\rho_{A-B-C} : \Tr_{B} \rho_{A-B-C} = \rho_{AC}$ if and only if $\rho_{AC} \in \Sep\Density(A\mcol C)$.
\end{lemma}
\begin{proof}
We prove both directions. \\
($\Rightarrow$) Let $\rho_{AC} \in \Sep\Density(A\mcol C)$. Then by definition, $\rho_{AC} = \sum_{x} p(x) \rho_{A}^{x} \otimes \rho_{C}^{x}$ for some finite alphabet $\cX$, $p \in \cP(\cX)$, and sets of quantum states $\{\rho_{A}^{x}\}_{x \in \cX}$, $\{ \rho_{C}^{x} \}_{x \in \cX}$. It follows $\sum_{x} p(x) \rho^{x}_{A} \otimes \dyad{x} \otimes \rho^{x}_{C}$ is a QMC extension as $\cR(\cdot) := \sum_{x} \Tr[\dyad{x} \cdot] \rho^{x}_{C} \otimes \dyad{x}$ and same idea for $\ol{\cR}$. \\
($\Leftarrow$) If one has a QMC $\rho_{A-B-C}$, then by Theorem \ref{thm:QMC-equivalences}, $\rho_{A-B-C} = \bigoplus_{j \in \cJ} p(j) \rho_{Ab^{L}_{j}} \otimes \rho_{b^{R}_{j}C}$ where the $\rho_{Ab^{L}_{j}}$, $\rho_{b^{R}_{j}C}$ are density matrices on the respective subspaces. It follows
\begin{align*}
 \Tr_{B}(\rho_{A-B-C}) &= \bigoplus_{j \in \cJ} p(j) \Tr_{b^{L}_{j}}(\rho_{Ab^{L}_{j}}) \otimes \Tr_{b^{R}_{j}}(\rho_{b^{R}_{j}C}) \\
 &= \sum_{j \in \cJ} p(j) \rho_{A}^{j} \otimes \rho_{C}^{j} \ \in \Sep\Density(A\mcol C) \ ,
\end{align*}
where we have used that for $B = \oplus_{j \in \cJ} b^{L}_{j} \otimes b^{R}_{j}$, we may decompose $\Tr_{B} = \oplus_{j \in \cJ} \Tr_{b^{L}_{j} \otimes b^{R}_{j}}$ and then $\Tr_{b^{L}_{j}} \otimes \Tr_{b^{R}_{j}}$ will distribute across the tensor product of the states decomposition as is normal. This completes the proof.
\end{proof}

We also will make use of the following straightforward proposition.
\begin{proposition}\label{prop:QMC-unif-close-enough}
For all $\delta \in (0,1)$, if there exists a QMC $\rho_{A-X-C}$, there exists $\widehat{\rho}_{A-\pi-C}$ such that $\rho_{AC} \approx_{\delta}^{\Lone} \widehat{\rho}_{AC}$. 
\end{proposition}
\begin{proof}
Let $\rho_{A-X-C} = \sum_{x \in \cX} p(x) \rho_{A}^{x} \otimes \dyad{x} \otimes \rho^{x}_{C}$. Then there exists finite alphabet $\cX'$ and $\{k_{x}\} \subset \{0,...,|\cX'|\}$ such that $\sum_{x} k_{x} = |\cX'|$ and $\sum_{x \in \cX} |k_{x}/|\cX'| - p(x)| \leq \delta$ since the rationals are dense in $[0,1]$.  Define $\wt{\rho}_{A-\pi-C} = \frac{1}{|\cX'|} \sum_{x' \in \cX'} \rho^{f(x')}_{A} \otimes \dyad{x'} \rho^{f(x')}_{C}$ where $f : \cX' \to \cX$ is such that for each $x \in \cX$, it maps $k_{x}$ of the elements of $\cX'$ to that $x$. Note this means $\wt{\rho}_{A}^{x'} = \rho_{A}^{f(x')}$ for all $x' \in \cX'$ not that we are actually indexing by $f(x')$ which would erase information we need to preserve the Markov chain condition.  It follows by our construction
\begin{align*}
 &\left\|\Tr_{X'}(\wt{\rho}_{A-\pi-C}) - \Tr_{X}(\rho_{A-X-C}) \right\|_{1} \\
 =&  \left\| \sum_{x \in \cX} \frac{k_{x}}{|\cX'|} \rho^{x}_{A} \otimes \rho^{x}_{C} - \sum_{x \in \cX} p(x) \rho^{x}_{A} \otimes \rho^{x}_{C} \right\|_{1} \\ 
 \leq&  \left\| \sum_{x \in \cX} \frac{k_{x}}{|\cX'|} \dyad{x}^{\otimes 2} - \sum_{x} p(x) \dyad{x}^{\otimes 2} \right\|_{1} \\
 =& \sum_{x} \left|\frac{k_{x}}{|\cX'|} - p(x)\right| \leq \delta \ ,
\end{align*}
the first inequality uses the data-processing inequality for the one-norm and the preparation channels $\Phi_{1}(\cdot) = \Tr[\dyad{x} \cdot] \rho^{x}_{A}$ and $\Phi_{2}(\cdot) = \Tr[\dyad{x} \cdot]\rho^{x}_{C}$.
\end{proof}

\section{Max Wyner Common Information Quantities}\label{sec:one-shot-corr-meas}
In this section we introduce the (smooth and non-smooth) one-shot correlation measures we use to characterize one-shot distributed source simulation and study their basic properties with a focus on the properties we will want later. There are three key points to this section, which we divide into three subsections. The first subsection introduces the correlation measures we will use. In principle there are many possible one-shot measures from which we could choose given the multitude of max mutual information quantities (Definition \ref{def:max-mutual-information-quantities}) as well as other divergences we could use to induce the mutual information (some of which are considered in the appendices). We introduce the correlation measures that allow us to get the tightest one-shot rates. However, we also introduce one variation that doesn't seem to get the tightest one-shot rates, but is the one most manageable for our later discussion on trying to derive a strong asymptotic equipartition property for our measures.

In the second subsection, we provide two methods for establishing cardinality bounds on auxiliary (classical) random variables which apply for different correlation measures. This is important when working with correlation measures that involve an auxiliary random variable \edit{for two reasons. Mathematically, it allows} us to \edit{guarantee the optimal is attained, replacing infimizations with minimizations, which is needed if we need to identify the optimizer in a proof. Moreover, physically, it tells us a task bounded by these quantities can be achieved with finite resources.} The two methods we provide are a generalization of the traditional support lemma (Lemma \ref{lem:gen-support-lemma}) and the other may be seen as a more direct application of Carath\'{e}odory's theorem (Lemma \ref{lem:card-bound-for-doubly-max}).

Finally, in the third subsection we introduce the smooth max Wyner common information quantities. In principle there is some freedom also in how we smooth these quantities. In the main text, we simply introduce the `correct' smooth quantities where the target state is smooth prior to evaluating the max Wyner common information. In the appendix, we introduce variants of distributed source simulation where the smoothing is better addressed by smoothing the chosen Markov chain extension instead.

\subsection{Max Wyner Common Informations}
We begin with the definition of our forms of max Wyner common information. As we only consider Wyner common information in this paper, we just call these common information quantities.
\begin{definition}[Max Wyner Common Information Quantitites]\label{def:max-Wyner-CIs}
Let $\rho_{AC} \in \Density(A\edit{\otimes}C)$. We then have the doubly max Wyner common information,
\begin{equation}\label{eq:upuparrow-max-common-information}
	C^{\edit{\upuparrow}}_{\max}(A\mcol C)_{\rho} := \min_{A-X-C} I^{\upuparrow}_{\max}(AC\mcol X)_{\rho} \ , 
\end{equation} 
the max Wyner common information,
\begin{equation}\label{eq:max-common-information}
	C_{\max}(A\mcol C)_{\rho} := \min_{A-X-C} I^{\uparrow}_{\max}(X\mcol AC)_{\rho} \ , 
\end{equation}
and the flipped max Wyner common information,
\begin{equation}\label{eq:flipped-max-common-information}
	C^{F}_{\max}(A\mcol C)_{\rho} := \min_{A-X-C} I^{\uparrow}_{\max}(AC \mcol X)_{\rho} \ ,
\end{equation}
\edit{where without loss of generality $\vert \cX \vert \leq \vert A \vert^{2} \vert C \vert^{2}$ (Lemmas \ref{lem:card-bound-for-doubly-max} \& \ref{lem:cardinality-bound-for-SMCI}), which justifies the minimization.}
\end{definition}
\begin{remark}
	The flipped max Wyner common information is only used in the appendix when considering nuances in attempting to establish a strong converse. However, to justify its validity, it should have all the properties we consider for the other quantities. For this reason, we include it in the main text.
\end{remark}
First, note that we have restricted to a classical register $X$ in Definition \ref{def:max-Wyner-CIs}, but in principle we could have defined it in terms of a quantum Markov chain extension. We could of course argue it is because we are interested in an operational interpretation that will require the classical register. However, \edit{as we prove in the appendix,} this restriction can be made without loss of generality for any mutual information that satisfies data processing.
\begin{lemma}\label{lem:classical-seed-suff}
Let $\rho_{ABC}$ be a $A-B-C$ Markov chain. Then there always exists a $A-X-C$ Markov chain $\rho_{AXC}'$ such that $\rho_{AC}'=\rho_{AC}$ and $\mbb{I}^{x}(AC\mcol B)_{\rho} \geq \mbb{I}^{x}(AC\mcol X)_{\rho'}$, where $\mbb{I}$ is any mutual information measure that satisfies data-processing and $x \in \{\upuparrow,\uparrow,\downarrow\}$ following Eqs.~\eqref{eq:Iupupmax-defn}-\eqref{eq:Idownmax-defn}.
\end{lemma}

\subsection{Cardinality Bounds for Max Wyner Common Information Quantities}\label{subsec:cardinality-bounds}
\edit{As stated in Definition \ref{def:max-Wyner-CIs}, the max Wyner common informations are defined to be minimizations, but this is only justified because we can restrict the cardinality of the auxiliary random variable so that the optimization is over a compact set.} In this subsection, we establish the\edit{se} cardinality bounds. We split this into two \edit{sub}sections. The first provides a `direct proof' for certain measures where we say `direct' as the proof is very similar to how one proves Carath\'{e}odory's theorem. The second \edit{sub}section is longer and introduces a general extension of the support lemma from classical network information theory \cite{el-gamal-2011a} and uses this to establish a cardinality bound. We believe both methods may be useful in other settings.

\subsubsection{A Direct Cardinality Bound Method for Max Wyner Common Information}

\begin{lemma}\label{lem:card-bound-for-doubly-max}
	Let $\rho_{AC} \in \Density(A\otimes C)$. Then 
\begin{align}
	\inf_{\rho_{A-X-C}} I^{\uparrow}_{\max}(X:AC)_{\rho} = \underset{\substack{\rho_{A-X-C}: \\ |\cX| = \edit{|A|^{2}|C|^{2}}}}{\min} I^{\uparrow}_{\max}(X:AC)_{\rho} \label{eq:card-bound-for-max-Wyner} \\
 \inf_{\rho_{A-X-C}} I^{\upuparrow}_{\max}(AC:X)_{\rho} = \underset{\substack{\rho_{A-X-C}: \\ |\cX| = \edit{|A|^{2}|C|^{2}}}}{\min} I^{\upuparrow}_{\max}(AC:X)_{\rho} \label{Eq:card-bound-for-doubly-max}
 \  .
 \end{align}
 That is to say, the max Wyner common information and the doubly max Wyner common information both satisfy the given cardinality bound.
\end{lemma}
\begin{proof}
We focus on proving \eqref{Eq:card-bound-for-doubly-max} and then explain the modifications for proving \eqref{eq:card-bound-for-max-Wyner}.
 
In the case that the set of Markov chain extensions is empty, the result holds trivially as both sides are $+\infty$ by our convention.  So assume that $\rho_{AC} \in \Sep\Density(A\mcol C)$ and let $\rho_{A-X-C}=\sum_{x\in\cX}p(x)\sigma^x_A\otimes\dyad{x}\otimes \sigma_C^{x}$ be a finite-dimensional QMC such that $\rho_{AC}=\sum_{x\in\cX}p(x)\sigma_A^x\otimes \sigma_C^x$. Let $\gamma := D_{\max}(\rho_{A-X-C}\Vert\rho_{AC}\otimes\rho_X)= I^{\upuparrow}_{\max}(AC \mcol X)$. By the definition of max divergence, a direct calculation will verify $\sigma_A^x\otimes \sigma_C^x\leq \exp(\gamma)\rho_{AC}$ for every $x$.  We will show there exists another QMC extension $\hat{\rho}_{A-X-C}$ with $I^{\upuparrow}_{\max}(AC \mcol X)_{\hat{\rho}} \leq \gamma$ and $|\cX|=\edit{|A|^{2}|C|^{2}}$.

Define the set $\mbb{P} := \{ q(x) \geq 0 : \rho_{AC} = \sum_{x \in \cX} q(x)\sigma_{A}^{x} \otimes \sigma_{C}^{x} \}$, which is a convex collection of probability vectors in $\mbb{R}^{|\cX|}$ as the states $\sigma_{A}^{x},\sigma_{C}^{x}$ are trace one and $\rho_{AC}$ is trace one.  We claim that the extreme vectors of $\mbb{P}$ have no more than $\edit{|A|^2|C|^2}$ nonzero elements.  Indeed, suppose that $\edit{\rho_{AC}} = \sum_{x=1}^{r} q(x)\sigma_{A}^{x} \otimes \sigma_{C}^{x}$ with $r\geq|A|^2|C|^2\edit{+1}$ where $q \in \mbb{P}$. \edit{As the space of Hermitian operators on $A \otimes C$ is a real vector space of dimension $|A|^{2}|C|^{2}$, there will be a linear dependence among the $\{\sigma_{A}^{x} \otimes \sigma_{C}^{x}\}_{x=1}^r$ by our assumption on the size of $r$. By definition of linear dependence, we can write $0=\sum_{x=1}^{r} c(x)(\sigma_{A}^{x} \otimes \sigma_{C}^{x})$ for some real numbers $c(x)$ not all equaling zero. By taking the trace of both sides and using that density matrices having unit trace, we may conclude $\sum_{x=1}^{r} c(x) = 0$. As there is a finite number of entries of $c$ and $q(x) > 0$ for all $x \in \{1,...,r\}$, there exists a $\delta > 0$ that is sufficiently small so that $q(x) \pm \delta c(x) > 0$ for all $x \in \{1,...,r\}$, and} we have $\rho_{AC}=\sum_{x=1}^{r} [q(x)\pm \delta c(x)]\sigma_{A}^{x} \otimes \sigma_{C}^{x}$ 
as may be verified by linearity. \edit{Moreover,} $(q(x)+\delta c(x))_{x\in\cX},(q(x)-\delta c(x))_{x\in\cX} \in\mbb{P}$, i.e. each is a valid probability distribution, \edit{as follows from the non-negativity and that $\sum_{x=1}^{r} c(x) = 0$}. This implies that $(q(x))_{x\in\cX}$ cannot be extremal in $\mbb{P}$ since $q(x)=\frac{1}{2}[q(x)+\delta c(x)]+\frac{1}{2}[q(x)-\delta c(x)]$ with $(q(x)\pm \delta c(x))_{x\in\cX}\in\mbb{P}$. This shows that an extremal vector in the set has no more than $\edit{|A|^{2}|C|^{2}}$ elements as promised. Therefore, using an extremal $\hat{q}(x)$ we can write $\rho_{AC}=\sum_{x=1}^{\edit{|A|^{2}|C|^{2}}}\hat{q}(x)\sigma_A^x\otimes\sigma_C^x$, and since $\sigma_A^x\otimes \sigma_C^x\leq \exp(\gamma)\rho_{AC}$ for all $x$, $\widehat{\rho}_{A-\hat{X}-C}=\sum_{x}\hat{q}(x)\sigma_A^x\otimes\dyad{x}\otimes\sigma_C^x$ is a QMC extension of $\rho_{AC}$ that satisfies $I^{\upuparrow}_{\max}(AC\mcol \hat{X})_{\widehat{\rho}} = D_{\max}(\widehat{\rho}_{A-\hat{X}-C}\Vert\widehat{\rho}_{AC}\otimes\widehat{\rho}_{\hat{X}})\leq \gamma$. This shows for any QMC extension, we may find another one with at most the original doubly max mutual information value and the promised cardinality bound. As such, the right hand side of the lemma statement is a lower bound. Since the left hand side of \eqref{Eq:card-bound-for-doubly-max} infimizes over the set on the right hand side, we obtain an equality.

In the case of $I^{\uparrow}_{\max}(X \mcol AC)_{\rho}$, the proof method is nearly identical. The only difference is that for a given $\rho_{AC} \in \Sep\Density(A \mcol C)$ and resulting Markov chain extension $\rho_{A-X-C}=\sum_{x\in\cX}p(x)\sigma^x_A\otimes\dyad{x}\otimes \sigma_C^{x}$, by definition of $I^{\uparrow}_{\max}(X \mcol AC)$, the minimizer is a pair $\gamma$ and $\tau_{AC}$ such that $\sigma^{x}_{A} \otimes \sigma^{x}_{C} \leq \exp(\gamma) \tau_{AC}$ for all $x \in \cX$, where this last relation again follows via direct calculation. Then one may follow the same argument identically to obtain the same cardinality bound for this measure. This completes the proof.
\end{proof}

\subsubsection{The Generalized Support Lemma and Cardinality Bounds for Common Informations}
The previous method for cardinality bounds does not seem to work for $I^{\uparrow}_{\max}(AC:X)_{\rho}$, and thus $C^{F}_{\max}(A\mcol C)_{\rho}$, because a direct calculation will verify that it does not simplify to an entry-wise bound in the same fashion as in the above proof. \edit{For the same reason, it is unclear that we can obtain a cardinality bound for the Wyner common information itself (recall \eqref{eq:Wyner-common-information}, which is defined with mutual information).} It follows we need a different method. Our method divides into two pieces: a generalization of the convex cover method from classical network information theory \cite[Appendix C]{el-gamal-2011a}, which works effectively whenever all constraints on the problem can be written as expectations over the random variable, and second that $I^{\uparrow}_{\max}(AC\mcol X)_{\rho}$ can be expressed in terms of an expectation over the $X$ random variable (Lemma \ref{lem:Imax-by-expectation}). We provide our generalization of the support lemma is as much generality as possible as we believe it may have applications in other settings. For clarity, we provide the proof of the generalized support lemma

\paragraph{Generalized Support Lemma} We first establish our generalization of the convex cover lemma by generalizing the support lemma. We begin by stating Carath\'{e}odory's theorem of which the support lemma may be viewed as a corollary.
\begin{proposition}(Fenchel-Eggleston-Carath\'{e}odory)\label{prop:caratheodory} Any point in a convex closure of a connected compact set $\mathscr{R} \in \mbb{R}^{d}$ can be represented as a convex combination of at most $d$ points in $\mathscr{R}$.
\end{proposition}
Now we present the general lemma.
\begin{lemma}\label{lem:gen-support-lemma}(Generalized Support Lemma)
Let $\cW$ be a topological space. Let the generalized state space $\mathscr{S}(A)$ be a connected, compact subset of $\Pos(A)$ and $\{\rho_{w}\}_{w \in \cW} \subseteq \mathscr{S}(A)$ be a set of \edit{indexed elements of the state space}. \edit{For all $j \in [k]$ let $f_{j}:\mathscr{S}(A) \to \mbb{R}$ be continuous, i.e. we consider a set of real-valued continuous functions where the domain is the state space,} $\mathscr{S}$. Then for any Borel measure $\mu$ of $\cW$, there exists $q \in \cP(\cW')$ where $|\cW'| \leq k$ and  $\{\sigma_{w'}\}_{w' \in \cW'} \subset \mathscr{S}(A)$ such that
$$ \int_{\cW} f_{j}(\rho_{w}) \mu(dW) = \sum_{w' \in \cW'}  f_{j}(\sigma_{w'})q(w')\ . $$
\end{lemma}
\begin{proof}
Our proof is a direct extension of the proof given for the traditional support lemma by Csiszar and K\"{o}rner \cite[Lemma 15.4]{Csiszar-2011a}. By assumption, $\mathscr{S}(A)$ is a compact, connected subset. By assumption each $f_{j}$ is continuous, so the image of $f_{j}(\rho)$ is both connected and compact. Define $F(\rho) := (f_{1}(\rho),...,f_{k}(\rho))$ and the set $\cR := \{F(\rho), \rho \in \mathscr{S}(A)\}$, which is connected and compact as product preserves these properties. Moreover, defining
$$ r_{j} \equiv \int_{W} f_{j}(\rho_{w})\mu(dW) \quad \forall j \in [k] \ , $$
we have $(r_{1},...,r_{k})$ is an element of the convex closure of $\cR$. Therefore, applying Proposition \ref{prop:caratheodory}, there exist $k$ points of $\cR$, which we denote $\{F(\sigma_{j})\}_{j \in [k]}$, along with a distribution $q \in \cP([k])$ such that
$$(r_{1},...,r_{k}) = \sum_{j \in [k]} q(j) F(\sigma_{j}) \ . $$
By definition of $F(\rho)$, we can conclude $r_{j} =\sum_{j} q(i) f(\sigma_{i})$ for all $i \in [k]$. Letting $\cW' = [k]$ completes the proof.
\end{proof}
First, we note the reason we talk in terms of generalized state spaces that are subsets of the positive density matrices is that, for example, this would allow for cardinality bounds on subnormalized states which may be of use given smooth measures. In fact, any closed convex subset of the (possibly subnormalized) density matrices would work, since it would be compact and all convex sets are (path-)connected. Moreover, the generalized state space may be the product space of closed convex subsets of the (possibly subnormalized) density matrices, since the product of connected, compact sets are also connected and compact, which is useful for network settings.\footnote{Formally, your state space is then $\mathscr{S}(A) \times \mathscr{S}(B)$ and functions which are defined on $\Pos(A \otimes B)$ would be extended to being on the state space via composition with the map $(\rho,\sigma) \mapsto \rho \otimes \sigma$.} Note you can also restrict to the support of some state space as needed.

\paragraph{Cardinality Bound for $C_{\max}$} 
We now can use the generalized support lemma to bound the cardinality of the the max common information. This relies on a technical lemma about mutual information quantities which we relegate to an appendix and summarize here.
\begin{lemma}\label{lem:Imax-by-expectation}
Let $\rho_{ACX}$ be classical on $X$. Then, 
$$\exp(I^{\uparrow}_{\max}(AC\mcol X)_{\rho}) = \sum_{x} p_{x} \exp(D_{\max}(\rho_{AC}^{x}\Vert \rho_{AC})) \ , $$
where \edit{recall or note that $D_{\max}(\cdot \Vert \rho_{AC}): \Density(\text{supp}(\rho_{AC})) \to \mbb{R}$ is a continuous function where $\text{supp}(\rho_{AC})$ is the support of $\rho_{AC}$.} Moreover, $I^{\upuparrow}_{\max},I^{\downarrow}_{\max}$ do not seem to satisfy such an averaging statement.
\end{lemma}
\begin{proof}
One uses Corollary \ref{corr:Imax-cond-on-X} in the appendix with the replacements $A \to AC$, $B \to \mbb{C}$. To simplify the RHS term, note, as defined in Corollary \ref{corr:Imax-cond-on-X}, $I^{\uparrow}_{\max}(\rho_{AC\mbb{C}}^{x}\Vert \rho_{AC}) = D_{\max}(\rho_{AC}^{x}\Vert \rho_{AC})$ for each $x$. \edit{The continuity of $D_{\max}$ over the specified domain follows from $D_{\max}(P \Vert Q)$ being continuous whenever $P,Q \geq 0$, $P \ll Q$, and $P \neq 0$ \cite{Tomamichel-2015a}.} Finally taking an exponential gets the form in the lemma. That $I^{\upuparrow}_{\max}(AC\mcol X)$ and $I^{\downarrow}_{\max}(AC\mcol X)$ do not seem to satisfy such an averaging statement may be seen from Propositions \ref{prop:Iupup-max-cq-side} and \ref{prop:Idown-max-cq-side} respectively.
\end{proof}
It is this property of $I_{\max}^{\uparrow}$ we now use to establish cardinality bounds for $\min_{A-X-C} I^{\uparrow,\ve}_{\max}$ which in turn establishes cardinality bounds for $C_{\max}(A\mcol C)$ in the case $\ve = 0$. 

\begin{lemma}\label{lem:cardinality-bound-for-SMCI}
Let $\rho_{AC} \in \Density(A\otimes C)$ and $\ve \geq 0$. Then without loss of generality 
$$ \inf_{A-X-C} I^{\uparrow,\ve}_{\max}(AC\mcol X)_{\rho} = \underset{\substack{\rho_{A-X-C}: \\ |\cX| \leq |A|^{2}|C|^{2}}}{\min} I^{\uparrow,\ve}_{\max}(AC\mcol X)_{\rho} \ . $$
Moreover, in the case $\rho_{AC} = p_{XZ}$ is fully classical, then $|Y| \leq |X||Z|$.
\end{lemma}
\begin{proof}
In the case that the set of Markov chain extensions is empty, the result holds trivially as both sides are $+\infty$ by our convention. Thus, we assume the set of Markov chain extensions is non-empty. This can be split into two cases. The first case is when $\ve = 0$. Let $\rho_{A-X-C}$ be any QMC extension of $\rho_{AC}$. Then $\rho_{A-X-C} = \sum_{x} p(x) \rho^{x}_{A} \otimes \dyad{x}_{X} \otimes \rho^{x}_{C}$. Let $\{\rho_{A|C}^{x}:= \rho_{A}^{x} \otimes \rho_{C}^{x}\}_{x \in \cX} \subset \edit{\Density(\text{supp}(\rho_{AC}))}$ \edit{where the support containment follows as otherwise $\text{supp}(\rho_{AC}) = \text{supp}(\sum_{x} \rho^{x}_{A\vert C}) \not \subset \text{supp}(\rho_{AC})$, which is a contradiction}. Let $\{M_{k}\}_{k \in \cK}$ be the elements of a minimal informationally complete POVM\footnote{\edit{Recall that a POVM is a set of positive operators that sum to identity, $\{M_{x}\}_{x} \subset \Pos(A)$ and $\sum_{x} M_{x} = \mbb{1}_{A}$. If $\{M_{x}\}$ is a set of linearly independent POVM elements that span the space $\Herm(A)$, then it is called `informationally complete' (IC). The naming is justified as, for an IC POVM, the probabilities $\{\Tr[M_{x}\rho]\}_{x}$ specify the state $\rho$ for any $\rho \in \Density(A)$. As $\vert \Herm(A) \vert = \vert A \vert^{2}$, an IC POVM only exists if there are at least $\vert A \vert^{2}$ linearly independent POVM elements. It is thus `minimal' if there are exactly $\vert A \vert^{2}$ linearly independent POVM elements. It is a common exercise and known that such a minimal IC POVM exists in every dimension. We refer the reader to \cite[Section 3.3.1]{Gour-2024a} for further information.}} (IC POVM) on the space, i.e.\ $|\cK| = |A|^{2}|C|^{2}$. Consider the following functions: $f_{k}(\cdot) := \Tr(M_{k} \cdot M_{k}^{\ast})$ for $k \in [|\cK|-1]$ and $f_{obj}(\cdot) := \exp(D_{\max}(\cdot\Vert \rho_{AC}))$. Then,
\begin{gather}
    \hspace{-2mm}\Tr(M_{k}\rho_{AC}M_{k}^{\ast})= \sum_{x} p(x)
    f_{k}(\rho_{A|C}^{x}) = \Pr[\text{Outcome k}] \\
     \sum_{x} p(x) f_{obj}(\rho_{A|C}^{x}) = \exp(I^{\uparrow}_{\max}(AB\mcol X)_{\rho}) \ ,
\end{gather}
where we have used Lemma \ref{lem:Imax-by-expectation}.
\edit{By the continuity claim in Lemma \ref{lem:Imax-by-expectation}, we know that $f_{obj}(\cdot)$ with its domain taken to be $D(\text{supp}(\rho_{AC}))$ is continuous. Therefore, we satisfy the criteria to apply Lemma \ref{lem:gen-support-lemma}. Applying Lemma \ref{lem:gen-support-lemma}, we may conclude there} exists a distribution $q \in \cP(X')$ where $|X'| \leq |A|^{2}|C|^{2}$ and states $\{\sigma_{A|C}^{x}:= \sigma_{A}^{x} \otimes \sigma_{C}^{x}\}_{x \in \cX} \subset \edit{\Density(\text{supp}(\rho_{AC})) \subset D(A \otimes C)}$ \edit{such that we satisfy the equations above.} As $\{f_{k}\}_{k}$ are all but one POVM element of an IC POVM, these constraints guarantee that the output state is indeed $\rho_{AC}$. As the solution is $\sum_{x} q(x) \sigma^{x}_{A} \otimes \dyad{x}_{X} \otimes \sigma^{x}_{C}$, it is a QMC extension of $\rho_{AC}$. The last constraint guarantees the max mutual information is satisfied. Noting that the minimization we are considering is a subset of the infimization, this establishes an equality. This completes the proof for the non-smooth case. 

In the smooth case, we let $\ve \in (0,1)$. Let $\rho_{A-X-C}$ be any QMC extension of $\rho_{AC}$ and let $\wt{\rho}_{AXC} \in \Bve(\rho_{A-X-C})$ denote the optimizer of $I^{\ve}_{\max}(AC:X)_{\rho_{A-X-C}}$. By Proposition \ref{prop:restricting-to-classical-registers}, we know $\wt{\rho}_{AXC}$ is classical on $X$ and by \cite[Lemma 22]{Ciganovic-2013a}, we know that $\wt{\rho}_{AXC}$ is normalized without loss of generality. Thus, without loss of generality, $\wt{\rho}_{AXC} = \sum_{x} \wt{p}(x) \dyad{x}_{X} \otimes \wt{\rho}_{AC}^{x}$ where $\wt{p} \in \cP(\cX)$ and $\{\wt{\rho}_{AC}^{x}\}_{x \in \cX} \subset \Density(A \otimes C)$. The rest of the proof is identical to the smooth case except we consider the state space $\Density(A \otimes C)$, but since we didn't guarantee the optimizer was a QMC to begin with, we don't need to guarantee the restricted version is. This completes the proof.
\end{proof}

\edit{We also note the Wyner common information admits the same cardinality bound via the same method. We will use this bound later. We note that our cardinality bound improves upon the previous bound of $\vert \cX \vert \leq 2\vert A \vert^{2} \vert C \vert^{2}$ that was established by Hayashi \cite[Proof of (8.158)]{Hayashi-2006a}.
\begin{proposition}\label{prop:card-bound-common-info}
	Let $\rho_{AC} \in \Density(A \otimes C)$, then
	$$ \inf_{A-X-C} I(AC:X)_{\rho} = \min_{\substack{\rho_{A-X-C} \\ \vert \cX \vert \leq \vert A \vert^{2} \vert C \vert^{2}}} I(AC:X)_{\rho} \ . 
	$$
\end{proposition}
\begin{proof}
	Note that for any $\rho_{ACX}$, $I(AC:X) = D(\sum_{x} p(x) \rho^{x}_{AC} \otimes \dyad{x} \vert \rho_{AC} \otimes \sum_{x} p(x) \dyad{x}) = \sum_{x} p(x) D(\rho^{x}_{AC} \Vert \rho_{AC})$ as is well-known (see e.g.~\cite[Exercise 11.8.8]{Wilde-2011a}). Similarly $D(\cdot \Vert \rho_{AC}) : D(\text{supp}(\rho_{AC})) \to \mbb{R}$ is a continuous function \cite{Tomamichel-2015a}. This is the equivalent of Lemma \ref{lem:Imax-by-expectation} for mutual information. The proof then is the same as Lemma \ref{lem:cardinality-bound-for-SMCI}.
\end{proof}
}

\subsection{Smooth Max Common Information}\label{subsec:smooth-MCI}
Having established our definitions of max common information, in this subsection we turn to defining smooth versions. This comes with various technical nuances due to the network setting (equivalently, Markov chain extension structure) of the max Wyner common information quantities and this subsection is devoted to establishing the correct definitions and the properties that guarantee this. We end the section with some properties of these quantities that are worth noting.

The first complication in defining smooth max Wyner common information quantities is there seem to be two ways of smoothing $C_{\max}$: one could smooth the state we start with, $\rho_{AC}$, or we could replace the max mutual information quantity itself. The latter is in effect like smoothing the QMC extension, which would suggest it is the wrong approach, and indeed it turns out to only be directly relevant for a variant of one-shot distributed source simulation, which we provide in an appendix. We thus define the following smooth max Wyner common informations:
\begin{definition}\label{def:smooth-max-Wyner-CIs}
Let $\rho_{AC} \in \Density(A\edit{\otimes}C)$ and $\ve \in (0,1)$. Then we define the smooth doubly max common information,
\begin{align}\label{eq:smooth-doubly-MCI}
C^{\upuparrow,\ve}_{\max}(A\mcol C)_{\rho} := \min_{\wt{\rho} \in \Bve(\rho)} C_{\max}^{\edit{\upuparrow}}(A\mcol C)_{\wt{\rho}} \ ,
\end{align}
the smooth max common information (SMCI),
\begin{align}\label{eq:SMCI}
C^{\ve}_{\max}(A\mcol C)_{\rho} := \min_{\wt{\rho} \in \Bve(\rho)} C_{\max}(A\mcol C)_{\wt{\rho}} \ ,
\end{align}
and the smooth flipped max common information,
\begin{align}\label{eq:smooth-flipped-MCI}
C^{F,\ve}_{\max}(A\mcol C)_{\rho} := \min_{\wt{\rho} \in \Bve(\rho)} C_{\max}^{\edit{F}}(A\mcol C)_{\wt{\rho}} \ .
\end{align} 
\end{definition}
There is an implicit issue in the above definitions. Namely, the smoothing ball optimizes over sub-normalized states, but we have not defined a notion of sub-normalized quantum Markov chain extensions. We resolve this in two steps. First, we define such a generalization so that the above definitions can be well-defined.
\begin{definition}\label{def:QMC-ext-for-subnormalized}
Let $\wt{\rho}_{AC} \in \Density_{\leq}(A \otimes C)$. We define the set of QMC extensions as
\begin{equation}\label{eq:QMC-extensions}
	\begin{aligned}
		\hspace{-2mm} \QMC(\wt{\rho}) := \left\{ \begin{array}{c} \wt{\rho}_{ABC} \coloneq (\cR \circ \ol{\cR})(\wt{\rho}_{B}):  \\
		 \Tr_{B}[\wt{\rho}_{ABC}] = \wt{\rho}_{AC} \, , \\
		\cR_{B}(\wt{\rho}_{B}) = \wt{\rho}_{BC} \, , \,
		\ol{\cR}_{B}(\wt{\rho}_{B}) = \wt{\rho}_{AB} \end{array}  \right\} \ ,
	\end{aligned}
\end{equation}
where $\cR \in \Channel(B,BC)$, $\ol{\cR} \in \Channel(B,AB)$.
\end{definition}
It is easy to see the above definition is a generalization of a quantum Markov chain as if $\wt{\rho}_{ABC} \in \QMC(\wt{\rho})$, then $\Tr[\wt{\rho}]^{-1}\wt{\rho}_{ABC}$ is a quantum Markov chain extension of $\wt{\rho}_{AC} := \Tr[\wt{\rho}]^{-1}\wt{\rho}_{AC}$ by Theorem \ref{thm:QMC-equivalences} as $\cR,\ol{\cR}$ are the corresponding recoverability maps. We note any definition that recovers a condition in Theorem \ref{thm:QMC-equivalences} under scaling by $\Tr[\wt{\rho}]^{-1}$ would work as that is all we need for our purposes. We have chosen the above so it works for short quantum Markov chain that are quantum on the $B$ register, but there are simpler definitions if one a priori restricts to $A-X-C$ QMC extensions, which may be used to formally establish results in Section \ref{sec:multi-receiver}.

Having defined the subnormalized generalization of a QMC extension, we show that in fact in all three definitions the optimizer is always a normalized state, so in fact this is ultimately unnecessary. This comes from two facts: we are able to re-normalize the state and keep it in the purified distance ball as was established in \cite{Ciganovic-2013a} (See Lemma \ref{lem:re-norm-in-ball}) and that the max divergence satisfies the normalization property \cite{Tomamichel-2015a}: $D_{\max}(a\rho \Vert b \sigma) = D_{\max}(\rho \Vert \sigma) + \log(a) - \log(b)$. We provide the proofs in full in the appendix (Propositions \ref{prop:SMCI-maxed-by-normal-state} and \ref{prop:SDMCI-maxed-by-normal-state}).

\paragraph{Data-Processing of Common Information Measures}
While we won't need to apply it directly at any point in this work, it is worth noting that (smooth) common information degrades under \textit{local} processing, which a correlation measure should. 
\begin{proposition}\label{prop:local-DPI-of-com-inf}
For $\ve \in [0,1)$, $C^{\ve}_{\max},C^{F,\ve}_{\max},C^{\upuparrow\edit{,\ve}}_{\max}$ are all monotonic under local CPTP maps on both spaces.
\end{proposition}
\edit{The proof is provided in the appendix.} Furthermore, thinking ahead, \edit{this result} tells us that if it takes $\alpha$ amount of randomness to distributed source simulate a target state $\rho_{AC}$ and there are local maps $\Phi,\Psi$ such that $\rho'_{A'C'} = (\Phi \otimes \Psi)(\rho_{AC})$, then it can only require at most the same amount of randomness, i.e.\  $\leq \alpha$, to simulate $\rho'_{A'C'}$.

\paragraph{A Remark on Computability}
One convenient property of one-shot entropic quantities is that for small dimensions they can be solved easily as they form semidefinite programs \cite{Tomamichel-2015a}. However, here we also have the constraint that we are optimizing over Markov chains. This not only makes it hard to solve in general, but actually means that there must be instances where it is NP-hard to solve, because \edit{determining if a state is $\ve$-close to a separable is known to be NP-hard \cite{Gharibian-2008a}.} For completeness, a formal proof is provided in the appendix.
\begin{proposition}\label{prop:NP-hardness}
There exist families of quantum states such that computing $C^{\ve}_{\max}(A\mcol C)_{\rho}$ is NP-hard.
\end{proposition} 

We note \edit{this result does not} say anything about computational complexity when the target distribution is fully classical and thus separable a priori. This also does not imply the value cannot be determined for a fixed state $\rho_{AC}$ as computational complexity statements are asymptotic claims. We also note this is not a problem in terms of establishing our results beyond that it means we cannot compute the answer efficiently in the sense of computational complexity.

\section{Derivation of One-Shot Rate}\label{sec:one-shot-rate}
In this section we derive our one-shot rate, i.e. Theorem \ref{thm:one-shot-DSS-main-theorem}. We split this into a subsection on establishing the achievability and a subsection on establishing the converse.

\subsection{Achievability}
We begin with the derivation of the achievability rate. We highlight that our goal is to find a method that uses soft-covering in such a manner that it is particularly tight and general. We do this through the use of an error exponent result for convex splitting that reduces to soft-covering (See \cite[Remark 1]{Cheng-2023b}) from \cite{Cheng-2023b}. This is a similar method to deriving a one-shot achievability bound for privacy amplification (resp.~multipartite state-splitting) in \cite{Tomamichel-2015a} (resp.~\cite{George-2024a}). We note this means we are not using the recent tight to second-order soft-covering result from \cite{Shen-2022a}. Indeed, it is not clear if one can use that result as it involves a correction term that scales in the spectrum of the state, which becomes an issue when we take the second smoothing.\footnote{In particular, we will smooth the state to which we apply the soft-cover code. This results in the correction term of \cite{Shen-2022a} scaling as $\wt{\nu} := \log(\spec(\wt{\rho}_{n}))$ where $\wt{\rho}_{n} \in \Bve(\rho^{\otimes n})$. It is not clear we can still guarantee $\wt{\nu}$ scales as $O(\log(n))$.} Similarly, we note that we do not use the error exponent result for soft-covering directly from \cite{Cheng-2023a}, which seems hard to apply our method to due to the $L_{p}$-spaces used in that work.

\begin{tcolorbox}[width=\linewidth, sharp corners=all, colback=white!95!black, boxrule=0pt,frame hidden]
\begin{lemma}\label{lem:achievability-bounds-for-one-shot-DSS}
	Let $\ve_{1},\ve_{2},\ve \in (0,1)$ such that $\ve_{1} + \ve_{2} \leq \ve$. Let $\rho_{AC} \in \Density(A\edit{\otimes}C)$ such that $2E_{T}(A \mcol C)_{\rho} \leq \ve_{1}$. Then
	\begin{equation}
	\begin{aligned}
		C^{\ve}_{F}(A \mcol C)_{\rho} &\leq C^{\ve}_{U,F}(A \mcol C)_{\rho} \\
		 &\leq C^{\sqrt{\ve_{1}}}_{\max}(A\mcol C)_{\rho}  +\frac{5}{2}\log\left(\frac{5}{\ve_{2}}\right)\\
		 &\leq C^{\upuparrow,\sqrt{\ve_{1}}}_{\max}(A\mcol C)_{\rho}  +\frac{5}{2}\log\left(\frac{5}{\ve_{2}}\right) \ .
	\end{aligned}
	\end{equation}
\end{lemma}
\end{tcolorbox}
\begin{proof}
The proof follows in three steps. First, we find a way to  re-express the error exponents of \cite{Cheng-2023b} in terms of a smooth measure for soft-covering. This is similar to the direct proof for privacy amplification done in \cite{Tomamichel-2015a}. Second, we convert this error exponent bound into a one-shot bound for soft-covering. Third, we use this code to construct an achievable strategy for one-shot DSS.

We begin with some preliminaries. First define for any Sandwiched R\'{e}yi divergence $\alpha \in [1,\infty)$,
\begin{align}
	I^{\uparrow,\ve}_{\alpha}(A \mcol B)_{\rho} := \min_{\wt{\rho} \in \Bve(\rho)} \min_{\sigma \in \Density(B)} D_{\alpha}(\wt{\rho} \Vert \wt{\rho}_{A} \otimes \sigma_{B}) \ .
\end{align}
Next, we note the optimizer is always normalized. To see this, let $\wt{\rho}_{AB}$ be the optimizer, define $\hat{\rho}_{AB} := \Tr[\wt{\rho}_{AB}]^{-1} \wt{\rho}_{AB}$. Note that this is still contained in the purified distance ball $\Bve(\rho)$ \cite[Lemma 21]{Ciganovic-2013a}. Then $D_{\alpha}(\wt{\rho} \Vert \wt{\rho}_{A} \otimes \sigma_{B}) = D_{\alpha}(\hat{\rho} \Vert \hat{\rho}_{A} \otimes \sigma_{B})$ as $\sigma_{B}$ is normalized and using the normalization condition for R\'{e}nyi divergences \cite{Tomamichel-2015a}. Moreover, note that if $A$ is classical, the optimizer $\wt{\rho}$ is classical on $A$ as this is a minimization and one may apply data-processing with the appropriate pinching channel.

Next, we will apply the error exponents for \textit{convex splitting} from \cite{Cheng-2023b}. As explained prior to the proof, this is sufficient because, as noted in \cite[Remark 1]{Cheng-2023b}, if one considers a CQ state $\rho_{XB}$, the error measure they are bounding reduces to the error rate for soft-covering, which is all we need to bound.

Fix $\rho_{XB}$. Let $\wt{\rho}_{XB},\sigma_{B}$ be the normalized optimizer for the modified smooth R\'{e}nyi mutual information $I^{\ve}_{\alpha}(X \mcol B)_{\rho}$ for some specific but not yet determined $\alpha \in (1,2)$. Then by applying \cite[Theorem 6]{Cheng-2023b} for the choice $\wt{\rho}_{XB}$ and $\tau = \wt{\rho}_{X}$, we have 
\begin{align}
	\mbb{E}_{\cC} \Vert \wt{\rho}^{\cC}_{B} - \wt{\rho}_{B} \Vert_{1} \leq 2^{2/\alpha-1} 2^{-\frac{\alpha-1}{\alpha}(\log M - I^{\uparrow}_{\alpha}(X \mcol B)_{\wt{\rho}})} \quad \forall \alpha \in (1,2) \ ,
\end{align} 
where $\wt{\rho}^{\cC}_{B} = \frac{1}{M} \sum_{x \in \cC} \wt{\rho}^{x}_{B}$ is the random codebook strategy and the expectation is with respect to all random codebook strategies from drawing codewords i.i.d. according to $\wt{p}_{X}$ induced by $\wt{\rho}_{X}$. Now, by choosing the specific but not specified choice of $\alpha \in [1,2]$, $I^{\uparrow}_{\alpha}(X \mcol B)_{\wt{\rho}} = I^{\uparrow,\ve}_{\alpha}(X \mcol B)_{\rho}$ by our assumption. Finally, for our choice of $\alpha$ and using triangle inequality, we have
\begin{align}
	\Vert \rho_{B} - \wt{\rho}_{B}^{\cC} \Vert_{1} \leq&  \Vert \rho_{B} - \wt{\rho}_{B} \Vert_{1} + 2^{2/\alpha-1} 2^{-\frac{\alpha-1}{\alpha}(\log M - I^{\uparrow,\ve}_{\alpha}(X \mcol B)_{\rho})} \\ 
	\leq & 4\ve + 2^{2/\alpha-1} 2^{-\frac{\alpha-1}{\alpha}(\log M - I^{\uparrow,\ve}_{\alpha}(X \mcol B)_{\rho})} \ , \label{eq:smooth-soft-cover-error-exponents}
\end{align}
where we used the relationship between trace distance and purified distance on normalized states, \eqref{eq:Purified-Dist-FvdG}. Note this actually works for any $\alpha \in [1,2]$ by choosing $\wt{\rho},\sigma$ appropriately for each. This completes the first step.

With modified error exponents, we derive the achievable bounds we want. We start by deriving achievable bounds for soft-covering for some error $\ve_{2} \in (0,1)$ in terms of $I^{\uparrow,\ve}_{\max}(X \mcol B)_{\rho}$ for $\ve \in (0,\ve_{2}/4)$. This can be established via the following implications:
\begin{align}
	& \log(M) = I^{\uparrow,\ve}_{\max}(X \mcol B)_{\rho} -\frac{5}{2}\log(2^{-1/5}(\ve_{2}-4\ve)) \label{eq:smooth-soft-cover-achievable} \\
	\Rightarrow  & \log(M) - I^{\uparrow,\ve}_{\alpha}(X \mcol B)_{\rho} \geq -\frac{5}{2}\log(2^{-1/5}(\ve_{2}-4\ve)) \notag \\
	\Leftrightarrow & 4\ve + 2^{1/5}2^{-2/5(\log(M) - I^{\uparrow,\ve}_{\alpha}(X \mcol B)_{\rho})} \edit{\leq \ve_{2}} \label{eq:achievability-reordered}
\end{align}
where the first implication follows from Sandwiched R\'{e}nyi divergences increasing monotonically as $\alpha$ increases. Noting that the LHS of \edit{\eqref{eq:achievability-reordered}} is \edit{the RHS of} \eqref{eq:smooth-soft-cover-error-exponents} for \edit{the choice} $\alpha = 5/3$, we have \eqref{eq:smooth-soft-cover-achievable} is an achievable rate \edit{with approximation error at most $\ve_{2}$}. In particular, as we were working with expectation over a random codebook, this implies such a code exists.

Lastly, we obtain our distributed source simulation bounds. For clarity, let $\ve_{\text{tot}} \in (0,1)$ be the total tolerated error. Let $\ve_{1},\ve_{2} \in (0,1)$ such that $\ve_{1} + \ve_{2} \leq \ve_{\text{tot}}$. As in the lemma statement, we assume $2E_{T}(A \mcol C)_{\rho} \leq \ve_{1}$. Then there exists $\sigma_{AC} \in \Sep \Density(AC)$ within $\ve_{1}$ distance of $\rho_{AC}$. By triangle inequality, for a $\ve_{2}$-approximate soft-covering code of QMC extension of such $\sigma_{AC}$, which we have already shown exists, we have 
\begin{equation}
\begin{aligned}
	\Vert \rho_{AC} - \sigma_{AC}^{\cC} \Vert_{1} &\leq \Vert \rho_{AC} - \sigma_{AC} \Vert_{1} + \Vert \sigma_{AC} - \sigma_{AC}^{\cC} \Vert_{1} \\
	&\leq \ve_{1} + \ve_{2} \leq \ve_{\text{tot}} \ .
\end{aligned}
\end{equation}
This shows any such state is $\ve_{\text{tot}}$-achievable rate for distributed source simulation so long as the $\ve_{1}$ bound holds. Thus, noting that the soft-covering strategy uses a uniform classical register, we have
\begin{align*}
	& C^{\ve_{\text{tot}}}_{U,F}(A \mcol C)_{\rho} \\
	\leq& \min_{\substack{ \sigma \in \Sep\Density(AC) \\ \Vert \rho - \sigma \Vert_{1} \leq \ve_{1} }} \edit{\inf_{\sigma_{A-X-C}}} I_{\max}^{\uparrow,\ve}(X\mcol AC)_{\sigma} \\
	& \hspace{28mm} -\frac{5}{2}\log(2^{-1/5}(\ve_{2}-4\ve)) \\
	\leq&  \min_{\sigma \in \mathscr{B}^{\sqrt{\ve_{1}}}(\rho) \cap \Density(AC)} \edit{\inf_{\sigma_{A-X-C}}} I_{\max}^{\uparrow,\ve}(X\mcol AC)_{\sigma} \\
	& \hspace{28mm} -\frac{5}{2}\log(2^{-1/5}(\ve_{2}-4\ve)) \\
	\leq& \min_{\sigma \in \mathscr{B}^{\sqrt{\ve_{1}}}(\rho)\cap \Density(AC)} \inf_{\sigma_{A-X-C}} I_{\max}^{\uparrow}(X\mcol AC)_{\sigma}  \\
	& \hspace{28mm} -\frac{5}{2}\log(2^{-1/5}(\ve_{2}-4\ve))+ g(\ol{\ve}) \\
	=&  \min_{\sigma \in \mathscr{B}^{\sqrt{\ve_{1}}}(\rho)\cap \Density(AC)}  \min_{\substack{\sigma_{A-X-C}: \\ \vert \cX \vert = \vert A \vert^{2} \vert C \vert^{2}}} I_{\max}^{\uparrow}(X\mcol AC)_{\sigma}  \\
	& \hspace{28mm} -\frac{5}{2}\log(2^{-1/5}(\ve_{2}-4\ve))+ g(\ol{\ve}) \\
	=& C^{\sqrt{\ve_{1}}}_{\max}(A\mcol C)_{\rho}  -\frac{5}{2}\log(2^{-1/5}(\ve_{2}-4\ve)) \ ,
\end{align*}
where the first inequality uses \eqref{eq:smooth-soft-cover-achievable}, the second uses that $P(\rho,\sigma) \leq \sqrt{\Vert \rho -\sigma \Vert_{1}}$, the third uses that smooth mutual information quantities monotonically increase as smoothing decreases, \edit{the first equality uses Lemma \ref{lem:card-bound-for-doubly-max},} and the final equality is by definition of $C^{\ve}_{\max}$ and that it is always obtained by a normalized state (Proposition \ref{prop:SMCI-maxed-by-normal-state}).  Note that because we replace the smoothing parameter on the max mutual information term with no smoothing in the above derivation, we have no constraint on $\ve$ beyond it satisfying $0< \ve < \ve_{2}/4$. To simplify our expression and satisfy these bounds,we let $\ve = \ve_{2}/5$. This results in
\begin{align}
	C^{\ve_{\text{tot}}}_{U,F}(A \mcol C)_{\rho} \leq & C^{\sqrt{\ve_{1}}}_{\max}(A\mcol C)_{\rho}  +\frac{5}{2}\log\left(\frac{5}{\ve_{2}}\right) + \frac{1}{2}   \ .
\end{align} 
This proves the first achievable rate. The second follows from the fact $I^{\upuparrow}_{\max}$ always upper bounds $I^{\uparrow}_{\max}$. This completes the proof.
\end{proof}

\begin{remark}
We remark that while we omit it from this paper, it is also possible to get an achievable bound in terms of the flipped max Wyner common information, $C^{F,\ve}_{\max}(A\mcol C)_{\rho}$ (See Definition \ref{def:smooth-max-Wyner-CIs}) that is still provably tight to terms scaling as $O(\sqrt{n})$ by using the relation between max mutual information quantities from \cite{Ciganovic-2013a}. This conversion will pick up another correction term (that can be made to scale as $O(\log(n))$ as the smoothing parameter scales as $1/n$. This is worth noting as the flipped max Wyner common information seems the easiest to work with for considering stronger converses (See Section \ref{subsec:on-a-strong-converse}).
\end{remark}

\subsection{Converse}
We now derive our one shot converses. The key idea is to relate the Hartley entropy of the classical register $H_{0}(X)_{\rho}$ with the common information of the perfectly correlated state induced by the classical register and then use data processing. We break this into two parts for readability.

\subsubsection{Relation between Hartley Entropy and Common Information of Perfectly Correlated States}
We begin by establishing our relationships between Hartley entropy and max mutual information. We establish two such identities, which we will then use for establishing converses for non-uniform and uniform correlation of formation. We begin with the identity for the non-uniform case.

\begin{proposition}\label{prop:Cmax-of-perfectly-correlated}
For any distribution $p \in \cP(\cX)$, let $\chi^{|p}_{XX'} = \sum_{x} p(x)\dyad{x} \otimes \dyad{x}$. Then $C_{\max}(X\mcol X')_{\chi} = H_{0}(X)_{\chi}$. 
\end{proposition}
\begin{proof}
First note that the optimal Markov chain is $p_{XX'\wt{X}} = \sum_{x} p(x) \dyad{x}^{\otimes 3}$ whose recovery maps are merely copying. While intuitive, one also may make this rigorous in the following manner. The initial state will have to be $\sum_{y} p'(y) \dyad{y}$. By symmetry, the recovery maps for both parties will be conditional distributions $\{q(x|y)\}_{y \in \cY}$. These conditional distributions will in fact have to be deterministic as otherwise the $X$ and $X'$ spaces won't be perfectly correlated. Thus the set $\cY$ can be partitioned into sets $\cY_{x_{i}} := \{y \in \cY: q(x_{i}|y) = 1\}$ and then we can apply a coarse graining map $\Phi$ that takes $\cY_{x_{i}} \to x_{i}$ for all $i$. Note $\Phi(p_{XX'Y}) = p_{XX'\wt{X}}$ and as it is a coarse graining map, by data-processing, $I_{\max}^{\uparrow}(XX'\mcol \wt{X})_{p_{XX'\wt{X}}} = I_{\max}^{\uparrow}(XX'\mcol \wt{X})_{\Phi(p_{XX'Y})} \leq I^{\uparrow}_{\max}(XX'\mcol Y)_{p_{XX'Y}}$. Thus, as we want to minimize $I^{\uparrow}_{\max}$, this establishes we have the optimal Markov chain. 

With this established, by Corollary \ref{corr:Imax-cond-on-X}, we have 
\begin{equation}\label{eq:Cmax-of-max-corr-1}
\begin{aligned}
I^{\uparrow}_{\max}(XX'\mcol \wt{X})_{p_{XX'\wt{X}}}  = \log(\sum_{x} p(x) \exp(I^{\uparrow}_{\max}(p^{x}_{XX'}\Vert p_{X}))) \ . 
\end{aligned}
\end{equation}
Noting $p^{x}_{XX'} = \dyad{x}^{\otimes 2}$, 
\begin{align*}
    & I^{\uparrow}_{\max}(p^{x}_{XX'}\Vert p_{X}) \\ 
    =& \min_{q \in \cP(\cX')} \Bigg\{\lambda: \dyad{x}^{\otimes 2} \leq 2^{\lambda} p_{X} \otimes \sum_{ x'}q(x') \dyad{x'} \Bigg\} \\
    =& \min\left\{ \lambda:  \dyad{x}^{\otimes 2} \leq 2^{\lambda} p(x') \dyad{x'} \otimes \dyad{x} \, \forall x' \right\} \\
    =&  \min \left\{\lambda: 1 \leq 2^{\lambda} p(x) \right\}  \Rightarrow \lambda = p(x)^{-1} \ ,
\end{align*}
where in the first line we have just used the definition and Proposition \ref{prop:restricting-to-classical-registers}, in the second we have used that it is clear that the choice $q_{X} = \dyad{x}$ will allow $\lambda$ to be minimized, and then we are dealing with diagonal operators so the bound must hold entry-wise, the third is because the L.H.S. only has support on $\dyad{x}^{\otimes 2}$ and this completes the argument. Therefore, plugging this into \eqref{eq:Cmax-of-max-corr-1}, we 
\begin{align*}
I^{\uparrow}_{\max}(XX'\mcol \wt{X})_{p_{XX'\wt{X}}} 
=& \log(\sum_{x} p(x) \exp(\log(1/p(x)))) \\
=& \log(\sum_{x} 1 ) = \log(|\cX|) = H_{0}(X)_{\chi} \ .
\end{align*}
This completes the proof.
\end{proof}

We now establish the relation we will need for the non-uniform correlation of formation.
\begin{proposition}\label{prop:Imax-double-Hartley-relation}
$I_{\max}^{\upuparrow}(X:X')_{\chi^{|\pi}} = H_{0}(X)_{\chi^{|\pi}}$ where we remind the reader $\chi^{|\pi}_{XX'}$ is defined via the uniform distribution.
\end{proposition}
\begin{proof}
By definition of $I_{\max}^{\upuparrow}$ and $D_{\max}$, 
\begin{align*}
	I_{\max}^{\upuparrow}(X:X')_{\chi^{|\pi}} =& \min \{\gamma: |X|^{-1}\delta_{XX'} \leq 2^{\gamma} |X|^{-2} \mbb{1}_{XX'}\} \ ,
\end{align*}
where $\delta_{XX'} = \sum_{x,x' \in \cX} \delta_{x,x'} \dyad{x}_{X} \otimes \dyad{x'}_{X'}$ and $\delta_{x,x'}$ is the Kronecker delta. By simplifying, this means we want $0 \leq (2^{\gamma}|X|^{-1} - 1)$, which means $\gamma \geq \log(|X|) = H_{0}(X)_{\chi^{|\pi}}$. As it's a minimization, we get the equality. This completes the proof.
\end{proof}

\subsubsection{One-Shot Converses for Correlation of Formation} 

We now present the one-shot converses. We begin with our primary tasks of interest, the uniform and non-uniform one-shot distributed source simulation. We note the square root in the correlation measure is due to purified distance smoothing rather than something fundamental per se. We again stress that the results hold for the entire range of smoothing parameters so long as the state is separable.
\begin{lemma}\label{lem:one-shot-formation-converse}
Let $\ve \in (0,1)$. Let $\rho_{AC} \in \Density(A \otimes C)$ such that $2E_{T}(\rho) < \ve$. 
\begin{align*}
 \max\{C^{\sqrt{\ve}}_{\max}(A\mcol C)_{\rho} \, , \, C^{F,\sqrt{\ve}}_{\max}(A\mcol C)_{\rho} \} &\leq C^{\ve}_{F}(A\mcol C)_{\rho} \\
 C^{\upuparrow,\sqrt{\ve}}_{\max}(A\mcol C)_{\rho} &\leq C^{\ve}_{U,F}(A\mcol C)_{\rho} \ .
\end{align*}
In particular, the bound holds for any $\rho_{AC} \in \Sep\Density(A\mcol C)$.
\end{lemma}
\begin{proof}
We will prove the bound for $C^{\ve}_{F}(A\mcol C)_{\rho}$ and then explain how the proof is nearly identical in the other cases. Let $C^{\ve}_{F}(A\mcol C)_{\rho}=n < \infty$ where $n$ is finite because $E_{T}(\rho) < \ve$ so there is $\widehat{\rho}_{AC} \in \Sep\Density(A\mcol C)$ such that $\widehat{\rho}_{AC} \approx_{\ve}^{\Lone} \rho_{AC}$. This means, $P(\widehat{\rho}_{AC},\rho) \leq \sqrt{\ve}$. Moreover, by Lemma \ref{lem:sep-marg-of-QMC} and Proposition \ref{prop:QMC-unif-close-enough}, this means there exists $\widehat{\rho}_{A-\wt{X}-C}$ such that $\|\widehat{\rho}_{AC} - \rho_{AC} \|_{1} \leq \ve$ and $|\wt{X}| = n$ is finite. We can use the distribution from $\widehat{\rho}_{\wt{X}}$ to define $\chi^{|p}$, i.e. $\chi^{|p}_{\wt{X}'\wt{X}\wt{X}''} := \sum_{x \in \wt{X}} \widehat{p}(x) \dyad{x}^{\otimes 3}$. The recovery maps \edit{for $\hat{\rho}_{A-\wt{X}-C}$} are preparation channels \edit{that may be expressed as} $\cR_{\wt{X} \to \wt{X}A}$ \edit{and $\ol{\cR}_{\wt{X} \to \wt{X}C}$.} Define $\cR_{\wt{X} \to A} \equiv \Tr_{\wt{X}} \circ \cR$ and likewise for $\ol{\cR}$. It follows $\widehat{\rho}_{A-\wt{X}-C} = (\edit{\cR_{\wt{X}' \to A} \otimes \ol{\cR}_{\wt{X}'' \to C}})(\chi^{|p}_{\wt{X}'\wt{X}\wt{X}''})$. Then we have
\begin{align*}
    C^{\sqrt{\ve}}_{\max}(A\mcol C)_{\rho}
    =& \min_{\wt{\rho} \in \Bvec{\sqrt{\ve}}(\rho)} \min_{\substack{\wt{\rho}_{A-\wt{X}-C}\edit{:} \\ \edit{\vert \cX \vert = \vert A \vert^{2} \vert C \vert^{2}} }} I^{\uparrow}_{\max}(\wt{X}\mcol AC)_{\wt{\rho}} \\
    \leq & I^{\uparrow}_{\max}(AC\mcol \wt{X})_{\widehat{\rho}} \\
    = & I^{\uparrow}_{\max}(\wt{X} \mcol AC)_{(\edit{\cR \otimes \ol{\cR}})(\chi)}\\
    \leq &I^{\uparrow}_{\max}(\wt{X} \mcol XX')_{\chi} \\
    =& I^{\uparrow}_{\max}(\wt{X} \mcol X)_{\chi}  \\
    =& H_{0}(X)_{\chi} = n = C^{\ve}_{F}(A\mcol C)_{\rho} \ ,
\end{align*}
where the first equality is by definition, the first inequality is by our choice of $\widehat{\rho}$ being feasible, the second equality is by the equivalence established previously, the second inequality is by data processing, the third equality is by the invariance of $I^{\uparrow}_{\max}(A \mcol B)$ under an isometry\footnote{A short proof: Let $\sigma_{B}$ be such that $\rho_{AB} \leq 2^{I^{\uparrow}_{\max}(A\mcol B)_{\rho}} \rho_{A} \otimes \sigma_{B}$. 
Then 
$V\rho_{AB}V^{\ast} = \rho_{ABC} \leq 2^{I^{\uparrow}_{\max}(A\mcol B)_{\rho}} \rho_{A} \otimes V\sigma_{B}V^{\ast}$ where $V$ is the isometry.
As max mutual information uses a minimization, this completes the proof.} and that $X \to XX'$ is just the isometry that copies in the basis, and the last steps are by Proposition \ref{prop:Cmax-of-perfectly-correlated} and our assumption respectively. In the case that one considers $C^{F,\sqrt{\ve}}_{\max}(A \mcol C)_{\rho}$, note that the isometric invariance step guarantees we can still apply Proposition \ref{prop:Cmax-of-perfectly-correlated} identically. In the case that one considers $C^{\upuparrow,\sqrt{\ve}}_{\max}(A \mcol C)_{\rho}$ and $C^{\ve}_{U,F}(A \mcol C)_{\rho}$, the proof structure is again identical but one uses Proposition \ref{prop:Imax-double-Hartley-relation}. This completes the proof.
\end{proof}

\paragraph{On Data Processing and Reversibility in Distributed Source Simulation} We end with some remarks on the role of data-processing in this problem as this is the key idea in the one-shot converse. In particular, it is easy to see that for perfectly correlated states $\chi^{|p}$, we do not need to apply data processing and thus our results would be tight if we measure error and smooth our measures using the same distance measure. This should not be surprising as the correlated states are the states where the Wyner common information and G\'{a}cs-K\"{o}rner common information \cite{Gacs-1973a} are the same \cite[Exercise 16.30]{Csiszar-2011a}, which shows these are the states where correlation can be asymptotically simulated and re-extracted reversibly. In this sense, our method for establishing a converse is sensitive to the information lost in the distributed-source simulation in the operationally appropriate manner.

\section{On Asymptotics of One-Shot Distributed Source Simulation}\label{sec:asymptotics}
We now turn to considering asymptotics of one-shot distributed source simulation. Theorem \ref{thm:one-shot-DSS-main-theorem} shows that the study of second-order asymptotics, and thus a strong converse, is reduced to establishing these properties for our correlation measures. Establishing this turns out to be a very difficult problem both due to the auxiliary random variable and due to our lack of understanding of the smooth entropy calculus, which were motivations for this work. Due to the difficulty, we are only able to establish a weak asymptotic equipartition property (AEP) for the smooth max Wyner common information, \eqref{eq:SMCI}. Both the achievability and converse require different methods than standard due to the auxiliary random variable. Afterwards, we provide an in-depth discussion on strategies for establishing a strong AEP for smooth max Wyner common information and, as a technical contribution and evidence such results are possible, we establish the first strong AEP for $I^{\uparrow,\ve}_{\max}(A\mcol B)_{\rho}$ (Theorem \ref{thm:strong-AEP-for-Imax}).

\subsection{Weak Asymptotic Equipartition Property}
In this subsection we establish the following theorem.
\begin{theorem}\label{thm:weak-AEP}
	Let $\rho_{AB} \in \Sep\Density(A\edit{\otimes}B)$. Then,
	\begin{align}
		\lim_{\ve \to 0} \lim_{n \to \infty} \frac{1}{n} C^{\ve}_{\max}(A^{n} \mcol C^{n})_{\rho^{\otimes n}} = C(A \mcol C)_{\rho} \ .
	\end{align}
\end{theorem}
\begin{remark}
	Note that the regularized relative entropy is known to be faithful \cite{Piani-2009a}. This means that if we require the error to tend towards zero, this is the most general result we could establish. 
\end{remark}

As noted previously, the auxiliary random variable in the correlation measure causes issues in establishing this using pre-existing tools. Below we explain the barrier to pre-existing tools for achievability and then provide our proof.

\paragraph{Achievability} 
At least for $I^{\uparrow,\ve}_{\max}(A\mcol B)_{\rho}$, the achievability of the weak AEP is established via a chain rule to smooth min and max-entropies \cite{Berta-2011a}.\footnote{One can define other max mutual information quantities that establish their achievability via smooth max-divergence, see \cite{Anshu-2020a}.} However, the proof of the smooth chain rule in \cite{Berta-2011a} at least seems to be at odds with the fact that we do not smooth over the auxiliary random variable in our definition of smooth max Wyner common information (See Definition \ref{def:max-Wyner-CIs}). This is the major barrier for lifting previous methods. Instead, we appeal directly to the definition of max divergence and use results on quantum typicality and conditional typicality. To avoid stating many \edit{definitions and} properties of quantum typical sets, we refer to the relevant results in \cite[\edit{Chapters 14 and 15}]{Wilde-2011a} when needed\edit{, but for clarity we do recall the definitions of strongly typical set, strong conditionally typical subspace, and strong conditionally typical projector from that reference.
\begin{definition}\label{def:strongly-typical-set}
	Given a sequence $x^{n} \in \cX^{n}$, for any $x \in \cX$, let $N(x \vert x^{n}) = \vert \{i: x_{i} = x\} \vert$, i.e.~the number of appearances of $x$ in $x^{n}$. Then for a fixed distribution $p_{X} \in \cP(\cX)$, the $\delta$-strongly typical set $T^{X^n}_{\delta}$ is the set of all sequences $x^{n}$ such that their empirical distribution defined by $\{\frac{1}{n}N(x \vert x^{n})\}_{x \in \cX}$ is entry-wise $\delta$-far from the distribution $p_{X}$ and otherwise has no support:
	\begin{equation}
	\begin{aligned}
		\hspace{-2mm} T^{X^{n}}_{\delta} \equiv \left\{ \begin{array}{ll} x^{n} \in \cX^{n} : \forall x \in \cX, &  \\
		\left\vert \frac{1}{n}N(x\vert x^{n}) - p_{X}(x) \right\vert \leq \delta & \hspace{-2mm} \text{if } p_{X}(x) > 0 \\
		\frac{1}{n}N(x \vert x^{n}) = 0 & \hspace{-2mm}\text{otherwise} \end{array} \right\}. 
	\end{aligned}
	\end{equation}	 
\end{definition}
\begin{definition}\cite[Definition 15.2.3]{Wilde-2011a} \label{def:cond-typical-subspace}
Fix a classical-quantum state $\rho_{XB} = \sum_{x \in \cX} p(x)\dyad{x}_{X} \otimes \rho_{B}^{x}$. For each $x \in \cX$, denote the spectral decomposition $\rho^{x}_{B} \equiv \sum_{y \in \cY} p_{Y \vert X}(y \vert x) \dyad{y_{x}}$ where that $p_{Y \vert X}$ is a conditional distribution follows from $\rho^{x}_{B}$ being a quantum state. For any sequence $x^{n} \in \cX^{n}$, the strong conditionally typical subspace $T^{\delta}_{B^{n} \vert x^{n}}$ is
\begin{align}
	\hspace{-2mm} T^{\delta}_{B^{n}\vert x^{n}} \equiv \text{span}\left\{ \otimes_{x \in \cX} \ket{y_{x}^{I_{x}}}_{B^{I_{x}}} : \forall x , \, y^{I_{x}} \in T_{\delta}^{(Y \vert x)^{\vert I_{x}\vert}} \right\} ,
\end{align}	
where $I_{x} \coloneq \{i: x_{i} = x\}$ is an indicator set that selects the indices $i$ in the sequence $x^{n}$ for which the $i^{th}$ symbol $x_{i}$ is equal to $x \in \cX$, $B^{I_{x}}$ selects the systems from $B^{n}$ where the classical sequence $x^{n}$ is equal to the symbol $x$, $\ket{y^{I_{x}}_{x}}$ is some string of states from the set $\{\ket{y_{x}}\}$, $y^{I_{x}}$ is a classical string corresponding to this string of states,  $Y\vert x$ is a random variable with distribution $p_{Y \vert X}(y\vert x)$, and $\vert I_{x} \vert$ is the cardinality of the indicator set $I_{x}$.
\end{definition}
\begin{definition}\label{def:strong-conditionally-typical-projector}
	Fix a classical-quantum state $\rho_{XB} = \sum_{x \in \cX} p(x)\dyad{x}_{X} \otimes \rho_{B}^{x}$. Then a strong conditionally typical projector for sequence $x^{n}$ is a tensor product of typical projectors \cite[Definition 15.1.3]{Wilde-2011a} of the $\rho^{x}_{B}$,
	\begin{align}
		\Pi^{\delta}_{B^{n} \vert x^{n}} \equiv \bigotimes_{x \in \cX} \Pi^{\rho_{x},\delta}_{B^{I_{x}}} \ ,
	\end{align}	
	where $I_{x}$ is as defined in Definition \ref{def:cond-typical-subspace} and $B^{I_{x}}$ indicates the systems for which a particular  typical projector for $\rho^{x}_{B}$ projects. 
\end{definition}

We refer the reader to \cite{Wilde-2011a} for further details.}
\begin{lemma}\label{lem:max-wyner-common-info-achievability}
	For $\rho_{AC} \in \Sep \Density(A \mcol C)$ and $\ve \in (0,1)$, 
	$$ \lim_{n \to \infty} \frac{1}{n} C^{\ve}_{\max}(A^{n} \mcol C^{n})_{\rho^{\otimes n}} \leq C(A \mcol C)_{\rho} \ . $$
\end{lemma}
\begin{proof}
If $\rho_{AC} \not \in \Sep\Density(A\mcol C)$, this is trivial as $C(A\mcol C) = +\infty$. We therefore assume $\rho_{AC} \in \Sep\Density(A\mcol C)$ for the rest of the proof. 

First we construct a Markov chain distribution whose marginal is contained in $\Bvec{\ve}(\rho_{AC}^{\otimes n})$ for sufficiently large $n$. Let $\rho_{A-X-C}$ be the optimizer of $C(A:C)_{\rho}$\edit{, which is obtained due to Proposition \ref{prop:card-bound-common-info}.} It follows it is of the form $\sum_{x \in \cX} p(x) \rho_{A}^{x} \edit{\otimes \dyad{x}} \otimes \rho_{C}^{x}$. Then we consider 
\begin{align*} 
\tau^{n}_{A^{n}X^{n}C^{n}} 
:=& \Pr[x^{n} \in T\edit{_{\delta}^{X^{n}}}]^{-1} \sum_{x^{n} \in T\edit{_{\delta}^{X^{n}}}} p(x^{n}) \tau_{A^{n}}^{x^{n}} \edit{\otimes \dyad{x^{n}}} \otimes \tau_{C^{n}}^{x^{n}}  \ ,
\end{align*}
where $T\edit{_{\delta}^{X^{n}}}$ is the strongly typical set for $\rho_{X}$ \edit{(Definition \ref{def:cond-typical-subspace})} , $\tau_{A^{n}}^{x^{n}} := \Tr[\rho_{A^{n}}^{x^{n}} \Pi^{\delta}_{A^{n}|x^{n}}]^{-1}\Pi^{\delta}_{A^{n}|x^{n}}\rho_{A^{n}}^{x^{n}}\Pi^{\delta}_{A^{n}|x^{n}} $ where $\Pi^{\delta}_{A^{n}|x^{n}}$ is the \edit{strong conditionally typical projector (Definition \ref{def:strong-conditionally-typical-projector})} \edit{and for any $x^{n} \in \cX^{n}$, $\rho^{x_{n}}_{A^{n}} \equiv \otimes_{i \in [n]} \rho^{x_{i}}$. We define $\tau_{C^{n}}^{x^{n}}$ similarly.}
This is a Markov chain by its algebraic structure, so we just need to verify its purified distance can be made arbitrarily small as $n$ grows. We will instead just use trace norm as purified distance goes to zero as trace norm does.
\begin{align*}
    & \left\| \tau^{n}_{\widehat{A}\widehat{X}\widehat{C}} - \rho_{A-X-C}^{\otimes n} \right\|_{1} \\
    = & \Bigg\| \Pr[x^{n} \in T\edit{_{\delta}^{X^{n}}}]^{-1} \sum_{x^{n} \in T\edit{_{\delta}^{X^{n}}}} p(x^{n}) \tau_{A^{n}}^{x^{n}} \edit{\otimes \dyad{x^{n}}} \otimes \tau_{C^{n}}^{x^{n}} \\
    & \hspace{5mm} - \sum_{x^{n} \in \cX^{n}} p(x^{n})  \rho_{A^{n}}^{x^{n}} \edit{\otimes \dyad{x^{n}}} \otimes \rho_{C^{n}}^{x^{n}} \Bigg\|_{1} \\
    = & \Bigg\| \sum_{x^{n} \in T\edit{_{\delta}^{X^{n}}}} \left(\Pr[x^{n} \in T\edit{_{\delta}^{X^{n}}}]^{-1} p(x^{n}) - p(x^{n})\right)\dyad{x^{n}} \\
    & \hspace{15mm} \otimes \left(\tau_{A^{n}}^{x^{n}} \otimes \tau_{C^{n}}^{x^{n}} - \rho_{A^{n}}^{x^{n}} \otimes \rho_{C^{n}}^{x^{n}}\right) \Bigg\|_{1} + \Pr[x^{n} \not \in T\edit{_{\delta}^{X^{n}}}] \\
    =& \sum_{x^{n} \in T\edit{_{\delta}^{X^{n}}}} \left(\Pr[x^{n} \in T\edit{_{\delta}^{X^{n}}}]^{-1} p(x^{n}) - p(x^{n})\right) \\
    & \hspace{15mm} \cdot \left\| \tau_{A^{n}}^{x^{n}} \otimes \tau_{C^{n}}^{x^{n}} - \rho_{A^{n}}^{x^{n}} \otimes \rho_{C^{n}}^{x^{n}} \right\|_{1} + \Pr[x^{n} \not \in T^{\delta}_{X^{n}}] \ ,
\end{align*}
where everything so far is just expanding definitions. We now bound the trace norm for a single $x^{n}$. To simplify notation, let $\Pi:= \Pi^{\delta}_{A^{n}|x^{n}} \otimes \Pi^{\delta}_{C^{n}|x^{n}}$ and $\Pi^{\perp} := (\mbb{I}^{A^{n}C^{n}} - \Pi)$. Then we may establish the inequality
\begin{align*}
   & \left\| \tau_{A^{n}}^{x^{n}} \otimes \tau_{C^{n}}^{x^{n}} - \rho_{A^{n}}^{x^{n}} \otimes \rho_{C^{n}}^{x^{n}} \right\|_{1} \\
   = & \Big\| 
   \tau_{A^{n}}^{x^{n}} \otimes \tau_{C^{n}}^{x^{n}} - \Pi \rho_{A^{n}}^{x^{n}} \otimes \rho_{C^{n}}^{x^{n}}\Pi  \\
   &\hspace{5mm} -\left(\rho_{A^{n}}^{x^{n}} \otimes \rho_{C^{n}}^{x^{n}} - \Pi\rho_{A^{n}}^{x^{n}} \otimes \rho_{C^{n}}^{x^{n}}\Pi\right) \Big\|_{1}  \\
   \leq & \left\| 
   \tau_{A^{n}}^{x^{n}} \otimes \tau_{C^{n}}^{x^{n}} - \Pi \rho_{A^{n}}^{x^{n}} \otimes \rho_{C^{n}}^{x^{n}}\Pi \right\|_{1} \\
   &\hspace{5mm} + \left\|\rho_{A^{n}}^{x^{n}} \otimes \rho_{C^{n}}^{x^{n}} - \Pi\rho_{A^{n}}^{x^{n}} \otimes \rho_{C^{n}}^{x^{n}}\Pi \right\|_{1} \\
   =& \left(\Tr[\rho_{A^{n}}^{x^{n}} \Pi^{\delta}_{A^{n}|x^{n}}]^{-1} \cdot \Tr[\rho_{C^{n}}^{x^{n}} \Pi^{\delta}_{C^{n}|x^{n}}]^{-1} - 1\right) \\
   &\hspace{5mm} + \left(1-\Tr[\rho_{A^{n}}^{x^{n}} \Pi^{\delta}_{A^{n}|x^{n}}] \cdot  \Tr[\rho_{C^{n}}^{x^{n}} \Pi^{\delta}_{C^{n}|x^{n}}]\right) \\
    \leq & \left(\frac{2\ve' - {\ve'}^{2}}{(1-\ve')^{2}}\right) + 2\ve' - {\ve'}^{2}  \\
    \leq & 2  \frac{\ve'}{(1-\ve')^{2}} + 2\ve' \\ 
    \leq& 4\ve' + 2\ve' = 6 \ve' \ ,
\end{align*}
where the first inequality is the triangle inequality, the following equality is using the definition of $\tau^{x^{n}}_{A^{n}} \otimes \tau^{x^{n}}_{C^{n}}$, the second inequality is for \edit{arbitrary $\ve' \in (0,1)$ and} sufficiently large $n$ \edit{by} using the `unit probability' property of \edit{the strongly conditional typical projector \cite[Property 15.2.4]{Wilde-2011a}}, and \edit{in the final inequality} we have \edit{further assumed} that $\ve' \leq 1/2$ so that $\ve'/(1-\ve')^{2} \leq 2\ve'$. Now we plug this back into our equation to get
\begin{align*}
    & \left\| \tau^{n}_{\widehat{A}\widehat{X}\widehat{C}} - \rho_{A-X-C}^{\otimes n} \right\|_{1} \\
    =& 6\ve' \cdot \sum_{x^{n} \in T^{\delta}_{X^{n}}} (\Pr[x^{n} \in T^{\delta}_{X^{n}}]^{-1} p(x^{n}) - p(x^{n})) \\
   &\hspace{5mm} + \Pr[x^{n} \not \in T^{\delta}_{X^{n}}] \\
    \leq & 6\ve' \cdot \left(\Pr[x^{n} \in T^{\delta}_{X^{n}}]^{-1} \Pr[x^{n} \in T^{\delta}_{X^{n}}] - \Pr[x^{n} \in T^{\delta}_{X^{n}}]\right) \\
   &\hspace{5mm} + \Pr[x^{n} \not \in T^{\delta}_{X^{n}}] \\
    \leq& 6\ve' \cdot(1 - (1-\ve')) + \ve' \\
    =& 6{\ve'}^{2} + \ve' \leq  4\ve' \ ,
\end{align*}
where we again use that $n$ must be sufficiently large and in the final inequality we have used that we already assumed $\ve' \leq 1/2$, so $6{\ve'}^{2} \leq 3 \ve'$.

\sloppy With this construction done, we bound the value of $I^{\uparrow}_{\max}(X^{n} \mcol A^{n}C^{n})_{\tau^{n}}$. First note that $\left(\tau^{n}_{X^{n}}\right)^{-1/2} = \Pr[x^{n} \in T^{\delta}_{X^{n}}]^{1/2} \sum_{x^{n} \in T^{\delta}_{X^{n}}} \frac{1}{p(x^{n})} \dyad{x^{n}}$. Starting from the definition:
\begin{align*}
    & \exp\left(I^{\uparrow}_{\max}(X^{n} \mcol A^{n}C^{n})_{\tau^{n}} \right) \\
    =& \min_{\sigma_{A^{n}C^{n}} \in \Density(A^{n}C^{n})} \Vert (\tau_{X^{n}}^{n} \otimes \sigma)^{-1/2} \tau^{n}_{X^{n}A^{n}C^{n}} (\tau_{X^{n}}^{n} \otimes \sigma)^{-1/2} \Vert_{\infty} \\
    =& \min_{\sigma_{A^{n}C^{n}} \in \Density(A^{n}C^{n})} \Big \Vert \sum_{x^{n} \in T^{\delta}_{X^{n}}} \dyad{x^{n}} \\
    & \hspace{3.5cm} \otimes \sigma^{-1/2} \edit{(}\tau_{A^{n}}^{x^{n}}\otimes \tau_{C^{n}}^{x^{n}}\edit{)} \sigma^{-1/2} \Vert_{\infty} \\
    =& \min_{\sigma_{A^{n}C^{n}} \in \Density(A^{n}C^{n})} \max_{x^{n} \in T^{\delta}_{X^{n}}} \Big \Vert \sigma^{-1/2} \edit{(}\tau_{A^{n}}^{x^{n}}\otimes \tau_{C^{n}}^{x^{n}}\edit{)} \sigma^{-1/2} \Vert_{\infty}
\end{align*}
where the second equality is using the orthonormality of the classical basis to collapse the three independent sums of $x^{n}$ and to cancel the $\Pr[x^{n} \in T^{\delta}_{X^{n}}]$ terms.

We now need to bound the operators within the operator norm. First, using properties of strong conditional quantum typicality \cite[Section 15.2.4]{Wilde-2011a},
\begin{equation}\label{eq:tau-dominance}
\begin{aligned}
\tau_{A^{n}}^{x^{n}} =&   \Tr[\rho_{A^{n}}^{x^{n}} \Pi^{\delta}_{A^{n}|x^{n}}]^{-1}\Pi^{\delta}_{A^{n}|x^{n}}\rho_{A^{n}}^{x^{n}}\Pi^{\delta}_{A^{n}|x^{n}} \\
\leq&  (1-\ve)^{-1}2^{-n(H(A|X)-\delta'')}\Pi^{\delta}_{A^{n}|x^{n}} \ .
\end{aligned}
\end{equation}
Second, we will make the \textit{choice} that $\sigma_{A^{n}C^{n}}$ is the normalized projector onto the typical subspace of $\rho_{AC}^{\otimes n}$ embedded into $A^{n} \otimes C^{n}$, $\sigma_{A^{n}C^{n}} := \Tr[\Pi^{\delta}_{A^{n}C^{n}}]^{-1}\Pi^{\delta}_{A^{n}C^{n}}$. Then, using we were considering a minimization over $\sigma_{A^{n}C^{n}}$, we have
\begin{align*}
    & \exp\left(I^{\uparrow}_{\max}(X^{n} \mcol A^{n}C^{n})_{\tau^{n}} \right) \\
    \leq & \max_{x^{n} \in T^{\delta}_{X^{n}}} \Big \Vert \left(\Tr[\Pi^{\delta}_{A^{n}C^{n}}]^{-1}\Pi^{\delta}_{A^{n}C^{n}}\right)^{-1/2} \tau_{A^{n}}^{x^{n}}\otimes \tau_{C^{n}}^{x^{n}} \\
    & \hspace{35mm} \cdot \left(\Tr[\Pi^{\delta}_{A^{n}C^{n}}]^{-1}\Pi^{\delta}_{A^{n}C^{n}}\right)^{-1/2} \Big \Vert_{\infty} \\
    \leq & \Tr[\Pi^{\delta}_{A^{n}C^{n}}](1-\ve)^{-2}2^{-n(H(A|X)-\delta'')} 2^{-n(H(C|X)-\delta'')} \\
    &\hspace{5mm} \cdot  \max_{x^{n} \in T^{\delta}_{X^{n}}} \Big \Vert \Pi^{\delta}_{A^{n}C^{n}} \Pi^{\delta}_{A^{n}|x^{n}}\otimes \Pi^{\delta}_{C^{n}|x^{n}} \Pi^{\delta}_{A^{n}C^{n}}\Big\Vert_{\infty} \\
    \leq & \Tr[\Pi^{\delta}_{A^{n}C^{n}}](1-\ve)^{-2}2^{-n(H(A|X)-\delta'')} 2^{-n(H(C|X)-\delta'')} \\
    \leq& (1-\ve)^{-2}2^{n(H(AC)+c\delta)} 2^{-n(H(A|X)-\delta'')} 2^{-n(H(C|X)-\delta'')} \ , 
\end{align*}
where the first inequality is due to the minimization, the second inequality is \eqref{eq:tau-dominance}, the third inequality is that an orthonormal projector is a contraction so the operator norm term is bounded above by one, and the final inequality is the dimension bound of the typical projector. 
Finally, we take the logarithm, to obtain
\begin{align*}
    & I^{\uparrow}_{\max}(X^{n} \mcol A^{n}C^{n})_{\tau^{n}} \\
     \leq& nH(AC)_{\rho} -nH(A|X)_{\rho} - nH(A|X)_{\rho} \\
     &\hspace{5mm} -2\log(1-\ve) +c\delta +2\delta'' \\
    =& n[H(AC)_{\rho} -H(AC|X)_{\rho}] -2\log(1-\ve) +c\delta +2\delta'' \\
    =& nI(AC:X)_{\rho} -2\log(1-\ve) +c\delta +2\delta'' \\
    =& nC(A:C)_{\rho} -2\log(1-\ve) +c\delta +2\delta''
\end{align*}
where the first equality uses the Markov chain structure of $\rho$ and the second equality uses $H(AC)-H(AC|X) = I(AC:X)$. Dividing by $n$ and taking the limit completes the proof.
\end{proof}

\paragraph{Converse} We now establish the converse. Again, the main barrier is working with the non i.i.d. structure of the auxiliary random variable. We get around this by reducing the smooth max Wyner common information to the smooth Wyner common information, using a common strategy for single-letterizing a measure in network information theory, and using continuity arguments for the entropies.
\begin{definition}
Let $\rho \in \Density(A\otimes C)$. The smooth common information is
$$ C^{\ve}(A\mcol C)_{\rho} := \min_{\wt{\rho} \in \Bve(\rho)} \min_{A-X-C} I(AC\mcol X)_{\wt{\rho}} \ . $$
\end{definition}
\begin{lemma}\label{lem:general-conditional-chain-rule}
$H(A^{n}_{1}|B) \leq \sum_{i=1} H(A_{i}|B)$, with saturation if and only if there exists a labeling of $[n]$ such that $\rho_{A_{i}^{n}B}$ is a $A_{i} - B - A_{i+1}^{n}$ Markov chain for all $i \in [n-1]$. In other words, with saturation only if $A_{i}$ can be generated from $B$ for all $i \in [n]$.
\end{lemma}
\begin{lemma}\label{lem:SMCI-converse}
Let $\ve \in (0,1)$. For $\rho \in \Density(A \otimes C)$,
$$ \lim_{\ve \to 0} \lim_{n \to \infty} \left[\frac{1}{n} C^{\ve}_{\max}(A^{n}\mcol C^{n})_{\rho^{\otimes n}}\right] \geq C(A\mcol C)_{\rho} \ . $$ 
\end{lemma}
\begin{proof}[\hypertarget{proof:SMCI-weak-converse-direct}{\proofname}]
Let $\sigma_{A^{n}C^{n}} \in \Bve(\rho_{AC}^{\otimes n})$ be the minimizer of $C^{\ve}_{\max}(A^{n}\mcol C^{n})_{\rho_{AC}^{\otimes n}}$\edit{, which exists as $\sigma_{A^{n}C^{n}} \in D(A^{n} \otimes C^{n})$ as discussed in Section \ref{subsec:smooth-MCI} and the cardinality bound on the auxiliary random variable from Lemma \ref{lem:card-bound-for-doubly-max}.} Then
\begin{align*} 
C^{\ve}_{\max}(A^{n}\mcol C^{n})_{\rho^{\otimes n}} 
=& C_{\max}(A^{n}\mcol C^{n})_{\sigma} \geq C(A^{n}\mcol C^{n})_{\sigma} \\
=& I(A^{n}C^{n}\mcol X)_{\sigma} \ ,
\end{align*}
where we used that $I^{\uparrow}_{\max}(A\mcol B) \geq I(A\mcol B)$ and that $\sigma_{A^{n}XC^{n}}$ is a $A^{n} - X - C^{n}$ Markov chain by definition of common information. Now we decompose the final right hand side of this chain of inequalities.
\begin{align*}
    &I(A^{n}C^{n}\mcol X)_{\sigma} \\
    =& \left[ H(A^{n}C^{n})_{\sigma} - H(A^{n}C^{n}|X)_{\sigma} \right] \\
    =& \left[ H(A^{n}C^{n})_{\sigma} - \sum_{i=1}^{n} H(A_{i}C_{i}|X)_{\sigma} \right]  \\
    =&\Bigg[ H(A^{n}C^{n})_{\sigma} 
    - \sum_{i=1}^{n} H(A_{i}C_{i})_{\sigma} 
    + \sum_{i=1}^{n} H(A_{i}C_{i})_{\sigma} \\
    & \hspace{5mm} - \sum_{i=1}^{n} H(A_{i}C_{i}|X)_{\sigma} \Bigg]  \\
    =&  H(A^{n}C^{n})_{\sigma} 
    - \sum_{i=1}^{n} H(A_{i}C_{i})_{\sigma} + \sum_{i=1}^{n} I(A_{i}C_{i}:X)_{\sigma}  \ ,
\end{align*}
where the first equality is a well-known chain rule. The second equality is because we consider a $A^{n}-X-C^{n}$ Markov chain, which implies the marginal $A_{i}XC_{i}$ is a $A_{i}-X-C_{i}$ Markov chain for all $i \in [n]$ and  Lemma \ref{lem:general-conditional-chain-rule} tells us the chain rule is saturated for Markov chains. The final identity is again using the same chain rule as the first equality. Therefore, dividing this by $n$ we have
\begin{equation}\label{eq:converse-bound1}
\begin{aligned}
		\hspace{-1mm} \frac{1}{n}C^{\ve}_{\max}(A^{n}\mcol C^{n})_{\rho^{\otimes n}}
		\geq & \frac{1}{n} H(A^{n}C^{n})_{\sigma} - \frac{1}{n} \sum_{i=1}^{n} H(A_{i}C_{i})_{\sigma} \\
    & \hspace{5mm}  + \frac{1}{n} \sum_{i=1}^{n} I(A_{i}C_{i}:X)_{\sigma} \ .
\end{aligned}
\end{equation}

We now aim to introduce $C^{\ve}(A\mcol C)_{\rho}$ into the above bound. Consider $\tau_{A-X-C} \equiv \sigma_{A_{k}-X-C_{k}}$ where $k := \mrm{argmin}_{k \in [n]} I(A_{i}C_{i}:X)_{\sigma}$. This is a Markov chain for the reason explained above and $\sigma_{A_{k}C_{k}} \in \Bve(\rho_{AC})$ as purified distance only decreases under partial trace. It follows that $\tau$ is a feasible point for $C^{\ve}(A:C)$, so we have
$$ C^{\ve}(A:C) \leq I(A_{k}C_{k}:X)_{\sigma} \leq \frac{1}{n} \sum_{i=1}^{n} I(A_{i}C_{i}:X)_{\sigma} \ , $$
where the second inequality is the minimizer lower bounds the average. Combining this with \eqref{eq:converse-bound1}, 
\begin{equation*}
\begin{aligned}
\frac{1}{n}C^{\ve}_{\max}(A^{n}\mcol C^{n})_{\rho^{\otimes n}} 
		\geq& C^{\ve}(A:C)_{\rho} + \frac{1}{n} H(A^{n}C^{n})_{\sigma} \\
		& \hspace{18mm}
    - \frac{1}{n} \sum_{i=1}^{n} H(A_{i}C_{i})_{\sigma} \ .
\end{aligned}
\end{equation*}

Note by assumption $\sigma_{A^{n}C^{n}} \in \Bve(\rho^{\otimes n}_{AC})$, and $\sigma_{A_{i}C_{i}} \in \Bve(\rho_{AC})$ for all $i \in [n]$ as explained earlier. As purified distance upper bounds trace distance, we may use the Fannes-Audenaert inequality to conclude
\begin{align*}
    H(A^{n}C^{n})_{\sigma} &\geq  nH(AC)_{\rho} + \ve n \log(|AC|) + h_{2}(\ve) \\
    H(A_{i}C_{i})_{\sigma} & \leq H(AC)_{\rho} - \ve \log(|AC|) - h_{2}(\ve) \ ,
\end{align*}
where we have used that the von Neumann entropy is additive over tensor products in the first inequality.
Plugging these in and cancelling the von Neumann entropy terms, we obtain 
\begin{align*}
\frac{1}{n}C^{\ve}_{\max}(A^{n}\mcol C^{n})_{\rho^{\otimes n}} 
 \geq& C^{\ve}(A\mcol C)_{\rho} + \left(1+\frac{1}{n}\right)h_{2}(\ve) \\
 & \hspace{5mm} + 2\ve \log(|AC|) \ .
\end{align*}
Then if one lets $n \to \infty$ and then $\ve \to 0$ on both sides, one obtains the result. Note that this held for all density matrices because if $\Bve(\rho)$ contains a separable state, a Markov chain exists and so both $C^{\ve}_{\max},C^{\ve}$ are finite and otherwise, both are infinite.
\end{proof}

\subsection{On a Strong Asymptotic Equipartition Property}\label{subsec:on-a-strong-converse}

\sloppy Given that Theorem \ref{thm:one-shot-DSS-main-theorem} shows that our measure is tight enough for second-order asymptotics, it is reasonable to discuss barriers to establishing this. In this section we focus on barriers to establishing strong AEP, i.e. establishing $\lim_{n \to \infty} \frac{1}{n} C^{\ve}_{\max}(A^{n} \mcol C^{n})_{\rho^{\otimes n}} = C(A\mcol C)_{\rho}$ for all $\ve \in (0,1)$. This is a weaker property than a second-order expansion and thus a natural first step. Moreover, it is reasonable to believe as there is a known exponential strong converse in the classical setting \cite{Yu-2018a}.

We first note that establishing such a strong AEP is an issue of establishing a \textit{strong converse} statement, i.e. \edit{for all $\ve \in (0,1)$ ,} $\lim_{n \to \infty} \frac{1}{n} C^{\ve}_{\max}(A^{n} \mcol C^{n})_{\rho^{\otimes n}} \edit{\geq C(A\mcol C)_{\rho}}$. This is because the achievability statement (Lemma \ref{lem:max-wyner-common-info-achievability}) already holds for all $\ve \in (0,1)$. Thus we are looking for new tools for establishing this inequality, which we state as a conjecture.
\begin{conjecture}\label{conj:strong-converse-max-Wyner-common-info}
For at least $\rho_{AC} \in \Sep\Density(A\mcol C)$, for all $\ve \in (0,1)$,
\begin{align}
	\lim_{n \to \infty} \frac{1}{n} C^{\ve}_{\max}(A^{n} \mcol C^{n})_{\rho^{\otimes n}} \edit{\geq C(A\mcol C)_{\rho}} \ .
\end{align}
\end{conjecture}

As a first step towards constructing new tools for establishing this converse claim, we consider the seemingly easier problem of establishing the same sort of strong converse claim for $I^{\uparrow,\ve}_{\max}(A\mcol C)_{\rho}$. If such a claim was not possible, it seems less likely that establishing such a result with a non i.i.d. auxiliary random variable would be feasible. However, we are able to resolve this open problem.
\begin{theorem}\label{thm:strong-AEP-for-Imax}
Let $\ve \in (0,1)$ and $\rho \in \Density(A \otimes B)$. Then, $$ \lim_{n \to \infty} \left[ \frac{1}{n} I^{\ve}_{\max}(A^{n}\mcol B^{n})_{\rho^{\otimes n}} \right] = I(A\mcol \edit{B})_{\rho} \ . $$
\end{theorem}
The basic proof idea is to refine certain properties of smooth entropies from \cite{Berta-2011a,Ciganovic-2013a} so that they are tight enough to establish this. Due to the need to largely re-produce previous proofs with slightly tighter bounds to establish this theorem, we provide it in the appendix. We note that Theorem \ref{thm:strong-AEP-for-Imax} also provides another method for establishing strong converses, which may be of value in other settings as well.

The hope would then be to lift the sort of chain rules used to establish Theorem \ref{thm:strong-AEP-for-Imax}, which is really about establishing the strong converse claim, to establish Conjecture \ref{conj:strong-converse-max-Wyner-common-info}. We do not succeed in doing this, but we do reduce the problem to an unknown property of the min-entropy, which we now provide.
\begin{proposition}\label{prop:open-problem}
Let $\rho_{AC} \in \Density(A \otimes C)$. A strong AEP for flipped smooth max common information (See Definition \ref{def:smooth-max-Wyner-CIs}) holds if for all $\ve \in (0,1)$, there exists $\ol{\ve} \in (\ve,1)$ such that $0 < \ve + \delta(\ol{\ve}) < 1$ where $\delta(\ol{\ve}):= 2\sqrt{\ol{\ve}(1-\ol{\ve})}$ so that
\begin{equation}
\begin{aligned}\label{eq:new-min-entropy-property}
& \lim_{n \to \infty} \left[\frac{1}{n} \max_{\wt{\rho}\in\Bvec{\ve+\delta{(\ol{\ve})}}(\rho^{\otimes n})} \, \max_{A^{n}-X-C^{n}} H_{\min}(A^{n}C^{n}|X) \right ] \\
\leq& \max_{A-X-C} H(AC|X)_{\rho} \ .
\end{aligned}
\end{equation}
\end{proposition}
Naturally, it is our conjecture that \eqref{eq:new-min-entropy-property} holds. We make some brief remarks upon the difficulties in establishing this conjecture.

First, we can immediately see \eqref{eq:new-min-entropy-property} being sufficient is itself a converse statement in the sense that we are looking for the regularized smooth quantity to be upper bounded by the conditional von Neumman entropy term. However, \eqref{eq:new-min-entropy-property} is distinct from the smooth min-entropy strong converse \cite{Tomamichel-2015a} as we take the extension on the purified state and have a non-i.i.d.\ random auxiliary variable to deal with still. Moreover, it is unclear how to switch the smoothing from before taking the extension to after, even if we restrict to separable states where both resulting sets are non-empty. Moreover, this switching of optimizations, one would still have the non-i.i.d.\ auxiliary random variable to handle, so this would not be a sufficient solution. 

Second, note that we consider $A^{n}-X-C^{n}$, which technically is a parallel process where maps only act on $X$ once to generate $A^{n}$,$C^{n}$. However, this structure implies $A_{i}-A^{i-1}_{1}XC^{i-1}_{1}-C_{i}$ for all $i \in [n]$ \cite{Hayashi-2006a}. This means our process can be forced to look sequential. It would then be natural to wonder if one may use the recent entropy accumulation theorem (EAT) results \cite{Dupuis-2020a,Metger-2022a} in this setting. We have not found this to be the case, but provide a bit more detail as to why in the appendix.

To summarize, it appears that a strong converse for our measure would require an i.i.d.\ reduction that takes into account the Markov chain structure. This does not align with the EAT nor a traditional de Finetti theorem where a purification is involved. To the best of our knowledge, none of the problems for establishing a strong converse that we have noted are addressed in the smooth entropy framework. We do however note there are certainly other ways of establishing strong converses in the face of auxiliary variables even in the quantum setting, e.g. \cite{Cheng-2021a}.

\section{Extending to Many Receivers}\label{sec:multi-receiver}
We have now established the general framework of one-shot distributed source simulation and its relation to smooth max common information. We now explain that it is straightforward to generalize beyond simulating a bipartite distribution. In the classical case this was addressed by Liu et al. \cite{Liu-2010a}. In that work if one goes to the appendix where they prove it, they just point out it is in effect the same as proof as before. Indeed, this observation lifts to our setting with one nuance: it is not clear how to argue the minimizer for multipartite systems should be classical. This is because the proof of Proposition \ref{prop:restricting-to-classical-registers} uses the decomposition of a quantum Markov chain and it is not clear how to generalize this to more systems. Regardless, this is in effect irrelevant with regards to the operational task at hand, so we focus on the initial resource being classical randomness that is copied.

We begin with notation. $A - X - C$ is very natural as it shows the two registers splaying out from $X$. We can't do this for more than two registers. As such, if we imagine we want $\rho_{A_{1}},...,\rho_{A_{m}}$ each generated from initial classical register $X$ independently, then we will write this as $X \seedto{} \edit{A^{m}}$. \edit{In a straightforward extension of separability, such an $X$ always exists when $\rho_{A^{m}}$ is a fully separable state (see \cite{Horodecki-2009a} for further discussion on multipartite entanglement and separability).
\begin{definition} 
	For integer $m \geq 2$ and Hilbert spaces $A_{1},A_{2},...,A_{m}$, a state is fully separable if there exists a finite alphabet $\cX$, probability distribution $p \in \cP(\cX)$, and sets of quantum states $\{\sigma^{x}_{A_{i}}\}_{x \in \cX}$ for each $i \in [m]$ such that $\rho_{A^{m}} = \sum_{x \in \cX} p(x) \otimes_{i \in [m]} \sigma^{x}_{A_{i}}$. For fixed $m$ and Hilbert spaces $A_{1},...,A_{m}$, we denote the space of fully separable states as $\Sep\Density(\mcol A^{m} \mcol)$.
\end{definition}
We may also define the multipartite trace distance of entanglement for a state $\rho_{A^{m}}$:
\begin{align}
	E_{T}(\mcol A^{m} \mcol)_{\rho} \coloneq \inf_{\sigma_{A^{m}} \in \Sep\Density(\mcol A^{m} \mcol)} \text{TD}(\rho, \sigma) \ . 
\end{align}
}
 
With this notation established we can define the following operational tasks.
\begin{definition}
Let $\rho_{\edit{A^{m}}} \in \Density(\edit{A^{m}})$. Let $\ve \in (0,1)$. The one-shot correlation of formation is defined as
\begin{equation}
\begin{aligned}
   \hspace{-1mm} C^{\ve}_{F}(\mcol \edit{A^{m}}\mcol )_{\rho}
    = \min\Big \{H_{0}(X)_{\wt{\rho}} : \wt{\rho}_{X\seedto{}\edit{A^{m}}} \, \& \, \wt{\rho} \sim_{\ve} \rho_{\edit{A^{m}}} \Big\} .
\end{aligned}
\end{equation} 
Moreover, the one-shot \textit{uniform} correlation of formation, which is the one-shot common information, is defined as 
\begin{equation}
\begin{aligned}
   \hspace{-1mm} C^{\ve}_{U,F}(\edit{\mcol A^{m}\mcol })
    =  \min\Big\{H_{0}(X)_{\wt{\rho}} : \wt{\rho}_{\pi\edit{\seedto{}A^{m}}} \, \& \,  \wt{\rho} \sim_{\ve} \rho_{\edit{A^{m}}} \Big\}  ,
\end{aligned}
\end{equation} 
where we remind the reader $\pi$ means the register is uniform.
\end{definition}
We can \edit{then} define the extended smooth correlation measures\edit{:}
\begin{equation*}
    \begin{aligned}
    C_{\max}^{\edit{\ve}}(\mcol \edit{A^{m}}\mcol )_{\rho_{\edit{A^{m}}}} 
    :=& \min_{\wt{\rho} \in \Bve(\rho_{\edit{A^{m}}})} \min_{\wt{\rho}_{X \seedto{scale=0.5} \edit{A^{m}}}} I^{\uparrow}_{\max}(\edit{X \mcol A^{m}})_{\wt{\rho}_{X \seedto{} \edit{A^{m}}}} \edit{ , } 
    \end{aligned}
\end{equation*}
\edit{as well as the multipartite Wyner common information:
\begin{align}
	C(\mcol A^{m} \mcol)_{\rho} = \min_{\rho_{X \seedto{scale=0.5} A^{m}}} I(A^{m}\mcol X)_{\rho} \ ,
\end{align}
where \edit{it} may be established to be a minimization by following the cardinality arguments of Section \ref{sec:one-shot-corr-meas}.} We can establish the one-shot converse the same way as before using data-processing by extending the $\chi^{|p}$ to more parties. We can establish achievability using the one-shot random covering as before. 

Similarly, the weak AEP will follow the same way. The achievability of $C^{\edit{\ve}}_{\max}(\mcol \edit{A^{m}} \mcol )$ can be proven \edit{for a fully separable state} using strong typicality in the same fashion as before since we can still decompose the optimizer of $C(\mcol \edit{A^{m}}\mcol )$ as $\sum_{x \in \cX} p(x) \dyad{x} \bigotimes_{i \in [n]} \rho^{x}_{A_{i}}$ and construct a state using strong conditional typicality from that. The converse \edit{is} established the same as before with the replacement $\edit{A^{n}C^{n} \mapsto A^{1,n}A^{2,n}...A^{m,n}}$ where the \edit{first superscript} here denotes the party label. \edit{In summary, this achieves the following straightforward multipartite extensions of Theorems \ref{thm:one-shot-DSS-main-theorem} and \ref{thm:weak-AEP}.
\begin{theorem}
	Let $\ve_{1},\ve_{2},\ve \in (0,1)$ such that $\ve_{1}+\ve_{2} \leq \ve$. Let $\rho_{A^{m}} \in \Density(A^{m})$ such that $2E_{T}(\mcol A^{m} \mcol)_{\rho} \leq \ve_{1}$. Then,
	\begin{equation}
	\begin{aligned}
		\hspace{-3mm} C_{\max}^{\sqrt{\ve}}(\mcol A^{m} \mcol )_{\rho_{A^{m}}} \leq& C^{\ve}_{F}(\mcol A^{m} \mcol )_{\rho} \\ 
		\leq& C_{\max}^{\sqrt{\ve_{1}}}(\mcol A^{m} \mcol )_{\rho_{A^{m}}} + \text{corr}(\ve_{2}), 
	\end{aligned}
	\end{equation}
	where $\text{corr}(\ve_{2}) =  \frac{5}{2}\log(\frac{5}{\ve_{2}}) + \frac{1}{2}$. Furthermore, if $\rho_{A^{m}} \in \Sep\Density(\mcol A^{m} \mcol)$, then $\lim_{\ve \to 0} \lim_{n \to \infty} \frac{1}{n} C^{\ve}_{\max}(\mcol A^{m} \mcol)_{\rho^{\otimes n}} = C(\mcol A^{m} \mcol)_{\rho}$.
\end{theorem}
} Thus we have extended all our results to multiple parties.

\paragraph{Monotonicity in Number of Parties}
In \cite{Liu-2010a} the authors note, albeit in the classical setting, that given $\rho_{\edit{A^{m}}}$ and \edit{$\rho_{A^{k}} \equiv \Tr_{A^{m}_{k-1}}(\rho_{A^{m}}) = (\Tr_{A_{m}} \circ \Tr_{A_{m-1}} \circ \cdots \circ \Tr_{A_{k-1}})(\rho_{A^{m}})$ for $k < m$}, then $C(\mcol \edit{A^{m}} \mcol ) \geq C(\mcol \edit{A^{k}} \mcol )$, i.e. that common information can only decrease as you decrease the number of parties. They suggest (1) this is surprising and (2) no such property is known to hold for mutual information. We wish to briefly address these points in case they provide conceptual clarity for the reader.

First, the authors suggest this monotonicity is surprising because ``if the information is \textit{common} it ought to be non-increasing when more random variables are included." This suggests the authors view common information as measuring the intersection of the randomness of the states (random variables). However, the common information is measuring the randomness needed to produce each random variable independently, i.e. to generate a common variable, and this is like measuring the union of the randomness of the states in some sense. Whether or not one agrees this is what ``common'' should denote, if one views it in this fashion, it is clear that it must increase when you add more random variables.

Second, that the common information has such a monotonic property is an immediate corollary of the mutual information having the same property, which in the quantum information community is called the data-processing inequality (not to be confused with the data-processing inequality for a Markov chain as is common in classical information theory). We show this in the following generic manner. 
\begin{proposition}
Given $\rho_{\edit{A^{m}}}$ and $\rho_{\edit{A^{k}}} \equiv \Tr_{A^{m}_{\edit{k-1}}}(\rho_{\edit{A^{m}}})$, $\mbb{C}(\mcol \edit{A^{m}} \mcol ) \geq \mbb{C}(\mcol \edit{A^{k}}\mcol )$ where $\mbb{C}$ is the Wyner common information defined using any mutual information $\mbb{I}$ satisfying data-processing \edit{and admits a cardinality bound}.
\end{proposition}
\begin{proof}
Let $\rho^{\star}_{\edit{A^{m}}X}$ be the minimizer of $\mbb{C}(\mcol \edit{A^{m}}\mcol )_{\rho}$ \edit{as exists by our assumption of admitting a cardinality bound}. Then,
\begin{align*}
    \mbb{C}(\mcol \edit{A^{m}}\mcol )_{\rho} =& \mbb{I}(\edit{A^{m}}:X)_{\rho^{\star}} \\
    \geq& \mbb{I}(\edit{A^{k}}:X)_{\Tr_{A^{m}_{\edit{k-1}}}(\rho^{\star})} \\ 
     \geq& \min_{\rho_{X \seedto{scale=0.5} \edit{A^{k}}}} \mbb{I}(\edit{A^{k}}\mcol X)_{\rho} 
      = \mbb{C}(\mcol \edit{A^{k}}\mcol )_{\rho} \ ,
\end{align*}
where the first inequality is the data-processing inequality and the second is using  minimizing $\Tr_{A^{m}_{k}}(\rho^{\star})$ is one classical register for generating $\rho_{\edit{A^{k}}}$ and so we can further minimize over other extensions.
\end{proof}

\section{Entangled State Source Simulation}\label{sec:ent-source-sim}
We have now established the limits of distributed source simulation in the one-shot and asymptotic setting in terms of smooth entropic quantities. These results however have only applied to separable states, and so it is worthwhile to ask what can be said about entangled states. It is known that one cannot convert entangled states with zero communication and no auxilliary resource \cite{Hayden-2003a,pure-state-revisited}. However, it is also known that there exists a sufficiently large (pure) entangled state that can be used to generate any (pure) entangled state up to small error with zero communication \cite{van-2003a}. Specifically, what the authors show is the following.
\begin{proposition}\label{prop:embezzling}(\cite{van-2003a}) For any $\ve > 0$ and a target bipartite pure state $\ket{\varphi}_{AB}$ with Schmidt rank $m$, the state $\ket{\mu(n)}_{A'B'} = \frac{1}{\sqrt{H_{n}}} \sum_{j=1}^{n} \frac{1}{\sqrt{j}}\ket{j}_{A'}\ket{j}_{B'}$ is such that for $n > m^{1/\ve}$ there exist unitaries $U_{AA'}, W_{BB'}$ so that
\begin{align*}
& F(U \otimes W (\ket{\mu(n)}_{A'B'}\ket{0}_{A}\ket{0}_{B}),\ket{\mu(n)}_{A'B'} \otimes \ket{\varphi}_{AB}) \\
& \hspace{70mm} \geq 1 - \ve \ ,
\end{align*}
where $F(\cdot,\cdot)$ is the fidelity and $H_{n} = \sum_{k=1}^{n} \frac{1}{k}$ is the Harmonic number.
\end{proposition}
This means that Alice and Bob, when given the initial state $\ket{\mu(n)}$, can prepare the target state $\ket{\varphi}_{AB}$ using local operations and zero communication. In this sense it seems the natural extension of distributed source simulation to the quantum setting. However, there are technical distinctions. In standard distributed source simulation, the classical register remains correlated with the output registers $A,C$, but when one embezzles, the $A,C$ registers are decoupled from the initial registers as is captured by the tensor product. This decoupling has the further advantage of allowing the initial state to be used in further protocols in exchange for further degrading the total approximation, which we note none of the correlation of formations could guarantee. We remark that when such a decoupling property holds for the input and output, such an input is known as a `catalyst.' That is, the state $\ket{\mu(n)}$ in Proposition \ref{prop:embezzling} is a catalyst (See \cite{Datta-2023a} for a recent review on catalytic transformations). These similarities and distinctions are summarized in Fig \ref{fig:DSS-vs-EDSS}. Given these distinctions, we separately analyze the cases where the input resource is required to stay decoupled and the case where it is not. We call these `Catalytic Entangled Source Simulation' and `Entangled Source Simulation' to distinguish between them. The primary result of this section is that entangled source simulation with the seed restricted to an embezzling state (Proposition \ref{prop:embezzling}) and free classical correlation is a natural analogue of distributed source simulation to the fully quantum setting and obtain a near-tight characterization of this setting (Theorem \ref{thm:Emb-EoS}). The identification of this being the appropriate analogue is obtained through analysis of different variants of catalytic and non-catalytic entangled state source simulation, which makes up the rest of the section.

\begin{figure}
\centering
\begin{subfigure}[b]{0.95\columnwidth}
   \begin{center}
        \begin{tikzpicture}
            \tikzstyle{porte} = [draw=black!50, fill=black!20]
                \draw
                (0,0) node (qy) {$p_{X}$}
                ++(1.25,0) node[porte] (copy) {Copy}
                ++(1.5,0.9) node[porte] (Alice-Proc) {$\Phi_{X \to A}$}
                ++(1.5,0) node (A) {$A$}
                ++(0,-1.7) node (C) {$C$}
                ++(-1.5,0) node[porte] (Bob-Proc) {$\Psi_{X' \to C}$}
                ++(2.2,0.75) node (approx) {$\approx_{\ve}$}
                ++(0.75,0) node (pxz) {$\rho_{AC}$}
                ;
                \draw
                ++ (3,0) node (X)[style = {porte, dashed}, fill=white] {$X$};
                \path[draw = black, ->] (qy) -- (copy); 
                \path[draw=black, ->] (copy) |- (Alice-Proc);
                \path[draw=black, ->] (Alice-Proc) -- (A);
                \path[draw=black, ->] (copy) |- (Bob-Proc);
                \path[draw=black, ->] (Bob-Proc) -- (C);
                \path[draw=black, ->] (copy) -- (X);
        \end{tikzpicture}
    \end{center}
   \caption{Distributed Source Simulation}
   \label{subfig:DSS-ent-state-comp} 
\end{subfigure}
\par\bigskip	
\begin{subfigure}[b]{0.95\columnwidth}
   \begin{center}
        \begin{tikzpicture}
            \tikzstyle{porte} = [draw=black!50, fill=black!20]
            		\draw ++(-1.5,0) node (sigma) {$\sigma_{A'C'}$};
                \draw ++(-1,0.8) node (A') {$A'$};
                \draw ++(-1,-0.8) node (C') {$C'$};
                \draw ++(0.25,1) node[porte] (A-prep) {$\Phi_{A' \to AA'}$};
                \draw ++(0.25,-1) node[porte] (C-prep) {$\Phi_{C' \to CC'}$};
                \draw ++(3.5,1) node (A) {$A$};
                \draw ++(3.5,0.25) node (C) {$C$};
                \draw ++(2,0.25) node (Apout)[style = {porte, dashed}, fill=white]  {$A'$};
                \draw ++(2,-1) node (Cpout)[style = {porte, dashed}, fill=white]  {$C'$};
                \draw ++(4.75,0.75) node (rhoout) {$\rho_{AC}$};
                \draw ++(4,0.75) node (approx) {$\approx_{\ve}$};
                \draw[->] (A-prep) -- (A);
                \draw[->] (A-prep) to[out=+0,in=+180] (Apout);
                \draw[->] (C-prep) to[out=+0,in=+180] (C);
                \draw[->] (C-prep) -- (Cpout);
                \draw[->] (sigma) to[out=+0,in=+180] (A-prep);
                \draw[->] (sigma) to[out=+0,in=+180] (C-prep);
        \end{tikzpicture}
    \end{center}
   \caption{\centering Entangled Source Simulation}
   \label{subfig:entangeld-source-sim}
\end{subfigure}
\par \bigskip
\begin{subfigure}[b]{0.95\columnwidth}
\begin{center}
	     \begin{tikzpicture}
            \tikzstyle{porte} = [draw=black!50, fill=black!20]
            		\draw ++(0,0) node (sigma) {$\sigma_{\wt{A}\wt{C}}$};
                \draw ++(0.5,0.8) node (A') {$A'$};
                \draw ++(0.5,-0.8) node (C') {$C'$};
                \draw ++(1.75,1) node[porte] (A-prep) {$\Phi_{\wt{A} \to AA'}$};
                \draw ++(1.75,-1) node[porte] (C-prep) {$\Phi_{\wt{C} \to CC'}$};
                \draw ++(3.75,1) node (A) {$A$};
                \draw ++(3.75,0.25) node (C) {$C$};
                \draw ++(3.75,-0.25) node (Apout) {$A'$};
                \draw ++(3.75,-1) node (Cpout) {$C'$};
                \draw ++(5.25,0.75) node (rhoout) {$\rho_{AC}$};
                \draw ++(4.5,0) node (approx) {$\approx_{\ve}$};
                \draw ++(5.25,0) node (otimes) {$\otimes$};
                \draw ++(5.25,-0.75) node (sigmaout) {$\sigma_{A'C'}$};
                \draw[->] (A-prep) -- (A);
                \draw[->] (A-prep) to[out=+0,in=+180] (Apout);
                \draw[->] (C-prep) to[out=+0,in=+180] (C);
                \draw[->] (C-prep) -- (Cpout);
                \draw[->] (sigma) to[out=+0,in=+180] (A-prep);
                \draw[->] (sigma) to[out=+0,in=+180] (C-prep);
        \end{tikzpicture}
    \end{center}
   \caption{Catalytic Entangled Source Simulation}
   \label{subfig:embez-source-sim}
\end{subfigure}
\caption{Comparison between distributed source simulation and the entangled state versions. Dashed boxes denote that those registers are not considered in the approximation criterion. (a) Distributed source simulation where the $X$ register is possibly strongly correlated to $A$ and $C$. (b) Entangled source simulation where the input is an arbitrary quantum state and an appropriate marginal of the output must achieve the target state to tolerable error $\ve$. For arbitrary $\sigma_{A'C'}$, this depicts Definition \ref{def:ent-of-simulation}, but if one restricts $\sigma_{A'C'}$ to an embezzling state then this depicts Definition \ref{def:ve-ent-of-sim}. (c) Entangled source simulation where the auxiliary state is required to be output approximately decoupled from the simulated state. This depicts Definition \ref{def:emb-ent-of-sim}.}
\label{fig:DSS-vs-EDSS}
\end{figure}

\subsection{Catalytic Entangled Source Simulation}
We begin with the case where we require the input resource to stay decoupled, i.e. be catalytic. We start with this more restrictive case as it allows us to introduce ideas we make use of in the more general setting. We define the catalytic entanglement of simulation, which measures the amount of entanglement (with respect to a specific choice of measure) necessary to source simulate a state in a distributed manner and is named such so as to avoid any confusion with entanglement of formation. To define this, we will need the definition of entanglement rank \cite{Watrous-Book}.
\begin{definition}
For all $r \geq 1$, the set of all operators $R \in \Pos(A \otimes B)$ for which there exists a finite alphabet $\cX$ and collection of linear operators $\{M_{x}\}_{x \in \cX} \subset \Lin(A,B)$ such that $\mrm{rank}(M_{x}) \leq r$ for all $x \in \cX$ and 
$$R = \sum_{x \in \cX} \opvec(M_{x})\opvec(M_{x})^{\ast} \  $$
are called the operators of entanglement rank $r$. Here $\opvec: \Lin(A,B) \to A \otimes B$ is defined via $\opvec(\ket{i}\bra{j}) = \ket{j}\ket{i}$. We denote the set of operators $R_{AB}$ with entanglement rank less than or equal to $r$ as $\mrm{Ent}_{r}(A\mcol B)$.

 We say a density matrix $\rho_{AB} \in \Density(A \otimes B)$ has entanglement rank $r'$ if it is contained $\mrm{Ent}_{r'}(A\mcol B)$ but not $\mrm{Ent}_{r'-1}(A \mcol B)$. For notational simplicity, we define $\mrm{Ent}_{A\mcol B}: \Density(A \otimes B) \to \mbb{N}_{\geq 1}$ as the function that takes a density matrix and returns its entanglement rank. The subscript is because the partitioning is relevant as will be shown in a following proposition.
\end{definition}
The set of separable operators is equivalent to the set $\mrm{Ent}_{1}(A\mcol B)$ and if $A \cong B \cong \mbb{C}^{d}$, then all positive semidefinite operators are contained within $\mrm{Ent}_{d}(A\mcol B)$. More generally, the entanglement rank cone $r$ contains all non-negative scalings of convex combinations of vectors who all have at most Schmidt number $r$ (this is alluded to in \cite{Watrous-Book}; see \cite{George-2022a} for further explanation). This in particular implies the entanglement rank of a pure state is its Schmidt rank as we would want and possibly expect.

\begin{definition}\label{def:emb-ent-of-sim}
Let $\rho_{AC} \in \Density(A \otimes C)$ and $\ve \in (0,1)$. The $\ve$-catalytic entanglement of simulation is the logarithm of the minimal entanglement rank of a bipartite state such that under local operations it may be converted to $\rho$ up to $\ve$-error as measured by fidelity. Formally,
\begin{equation*}
    \begin{aligned}
& C^{\ve}_{\mrm{Cat-EoS}}(A\mcol C)_{\rho} := \log\min\{r \geq 1: \exists \sigma \in \mrm{Ent}_{r}(A' \mcol C') :\\
& \hspace{52mm}
 (\Phi \otimes \Psi)(\sigma) \approx_{\ve}^{F} \sigma \otimes \rho \} \ ,
    \end{aligned}
\end{equation*}
where $\Phi \in \Channel(A',A' \otimes A)$ and $\Psi \in \Channel(C',C' \otimes C)$ and $\sigma \approx_{\ve}^{F} \rho$ means $F(\sigma,\rho) \geq 1 - \ve$. (See Fig.\ \ref{subfig:embez-source-sim} for diagramatic depiction.)
\end{definition}
It is worthwhile to relate this back to the correlation of formation measure. Much like $C^{\ve}_{F}$, catalytic entanglement of simulation measures the logarithm of the `dimension' of the resource, in this case entanglement rank, but does not care about the uniformity of the resource. Second, recall that distributed source simulation used the recovery maps $\cR:X \to X \otimes C, \ol{\cR}: X \to A \otimes X$. Then in catalytic entanglement of simulation, the $\Phi,\Psi$ act similarly.

Note that while it only measures the entanglement rank, its demand on the ancillary state being (approximately) unchanged and uncorrelated means that it is not clear how one would make use of a classical ancillary state, at least when $\ve$ is sufficiently small. This is because if $\rho$ is built conditionally on $\sigma$, it will not be uncorrelated. For this reason it seems to properly capture the notion of embezzling as being the strategy.

With this definition introduced, we establish achievability bounds and then argue that these bounds should be approximately tight.

\begin{lemma}\label{lem:Schmidt-rank-inv-under-LU}
Let $\rho_{AB} \in \Density(A\otimes B)$. Then $\mrm{SR}(AR:B),\mrm{SR}(A:BR)$ is the same for all purifications $\ket{\psi}_{ABR}$, where $\text{SR}(AR:B)$ is the Schmidt rank of a pure state $\psi_{ABR}$ over the bipartitioning $AR$,$B$.
\end{lemma}
\begin{proof}
By isometric equivalence of purifications, $\ket{\psi}_{R'AB} = (V_{R \to R'} \otimes \mbb{1})\ket{\psi}_{RAB}$. Let $\ket{\psi}_{RAB} = \sum_{i \in [m]} \sqrt{p_{i}} \ket{\phi_{i}}_{RA} \otimes \ket{\varphi_{i}}_{B}$ be its Schmidt decomposition. Then $\ket{\psi}_{R'AB} = \sum_{i \in [m]} \sqrt{p_{i}} V_{R \to R'}\ket{\phi_{i}}_{RA} \otimes \ket{\varphi_{i}}_{B}$ is its Schmidt decomposition as an isometry maps pure states to pure states. An identical argument holds for the other partitioning.
\end{proof}
The following lemma shows it is necessary to take the minimization in the previous lemma.
\begin{proposition}\label{prop:SR-changed-under-partitioning}
For pure state $\ket{\psi}_{RAB}$, in general $\mrm{SR}(AR:B) \neq \mrm{SR}(A:BR)$. Moreover, the exists $\ket{\psi}_{ABR}$ such that $\mrm{SR}(A:BR) - \mrm{SR}(AR:B) = d_{B} - 1$, the maximum possible difference.
\end{proposition}
\begin{proof}
We present an example. Let $\rho_{AB} = \dyad{\phi}_{A} \otimes \pi_{B}$ where we remind the reader $\pi_{B}$ denotes the maximally mixed state. Then $\ket{\psi}_{ABR} = \ket{\phi}_{A} \otimes \ket{\Phi^{+}}_{BB'} \otimes \ket{\phi}_{A'}$ where $R \equiv A'B'$ and $\ket{\Phi^{+}}_{BB'} = \sum_{i \in [|B|]} |B|^{-1/2} \ket{i}_{B}\ket{i}_{B'}$ is the maximally entangled state. It follows $\mrm{SR}(A:BR) = 1$ as it is product across this partitioning. On the other hand,
\begin{align*}
    \ket{\psi}_{ABR} =& \ket{\phi}_{A} \otimes \left( d_{B}^{-1/2} \sum_{i \in [d_{B}]} \ket{i}_{B}\ket{i}_{B'}  \right) \otimes \ket{\phi}_{A'} \\
    =& d_{B}^{-1/2} \sum_{i \in [d_{B}]} \ket{i}_{B} \otimes \ket{\phi}_{A} \otimes \left(\ket{\phi} \otimes \ket{i}\right)_{R}\ .
\end{align*}
as $\{\ket{\phi}^{\otimes 2} \otimes \ket{i}\}_{i}$ form the Schmidt vectors for the $AR$ space, $\mrm{SR}(A:BR) = d_{B}$. This completes the proof.
\end{proof}

\begin{proposition}\label{prop:incompressibility}
For $\ve \in (0,1)$, one can construct $\rho_{AB}$ to accuracy $\ve$ via embezzling using $\ket{\mu(n)}$ for $n > m^{1/\ve}$ where 
$$ m = \min\{\mrm{SR}(AR:B),\mrm{SR}(A:BR)\} \ , $$
where $\psi_{ABR}$ is an arbitrary purification of $\rho_{AB}$.
\end{proposition}
\begin{proof}
As embezzling is for a pure state (Proposition \ref{prop:embezzling}), the strategy must embezzle in a purification of the target state. That is, if the target state is $\rho_{AB}$, the strategy is to consider a purification $\ket{\psi}_{ABR}$ such that $\Tr_{R}[\psi_{ABR}] = \rho_{AB}$. Note by the locality constraints on the unitaries in Proposition \ref{prop:embezzling}, either Alice or Bob aims to embezzle in the purifying space, i.e.\ the $R$ register. Noting that embezzling is a function of the Schmidt rank (the parameter $m$ in Proposition \ref{prop:embezzling}), the goal is to minimize the Schmidt rank of the purification. By Lemma \ref{lem:Schmidt-rank-inv-under-LU}, we know the Schmidt rank is the same for every choice of purification once one fixes the partitioning of the spaces, thus we may consider an arbitrary purification. By Proposition \ref{prop:SR-changed-under-partitioning}, there is a difference in Schmidt rank depending on how the state is partitioned, which physically is a function of who embezzles in the purification, so we take the minimum. This completes the proof.
\end{proof}
\begin{corollary}
Let $\rho_{AC} \in \Density(A \edit{\otimes} C)$ and $\ve \in (0,1)$, then
\begin{align*}
 C^{\ve}_{\mrm{Cat-EoS}}(A\mcol C)_{\rho}  \leq \frac{1}{\ve}\log(\min\{\mrm{Ent}_{AR:C}(\psi),\mrm{Ent}_{A:CR}(\psi)\}) \ ,
\end{align*}
where $\psi_{ACR}$ is an arbitrary purification of $\rho_{AC}$.
\end{corollary}
\begin{proof}
Because the entanglement rank is equal to the Schmidt rank for a pure state, this follows the definition of $C^{\ve}_{EE,S}$ and the previous proposition where we have taken the logarithm.
\end{proof}
There are two questions: the first would be if this strategy, when no classical side-information is allowed, is optimal. Roughly speaking, it is in the sense that \cite{van-2003a} showed that if one allowed LOCC and a state dependent catalyst that the error is bounded below by $\Omega(1/\log(n))$, but the universal embezzling strategy scales as $O(1/\log(n))$ where $n$ is the Schmidt rank of the embezzling state being used. Of course this scaling requires the error demanded be small, i.e. if the error is sufficiently large, it may be feasible to use less entanglement; this question is developed further by authors of this work in a separate paper \cite{pure-state-revisited}. We also stress that $C^{\ve}_{\mrm{EoS,Cat}}$ seems to be captured effectively by embezzling as we already argued why, in general, a classical auxiliary state could not be useful.

The second question would be if this strategy has any notion of compressibility in the sense that it requires less resources for many copies of an i.i.d.\ state. This is not so: we show this strategy scales in the number of copies.
\begin{proposition}
For $\ve \in (0,1)$, one can construct $\rho_{AB}^{\otimes k}$ to accuracy $\ve$ via embezzling using $\ket{\mu(n)}$ for $n > m^{1/\ve}$ where 
$$ m = k \cdot \min\{\mrm{SR}(AR:B),\mrm{SR}(A:BR)\} \ . $$
\end{proposition}
\begin{proof}
Consider a purification of $\rho_{AB}^{\otimes k}$, $\ket{\psi}_{A^{k}B^{k}R'}$. Consider a purification of $\rho_{AB}$, $\ket{\phi}_{ABR}$. It follows $\ket{\phi}^{\otimes k}$ is a purification of $\rho_{AB}^{\otimes k}$. By the isometric equivalence of purifications, there is an isometry or reversed isometry taking $R'$ to $R^{k}$. As this is a local map, it can't change the Schmidt rank. Thus, $\mrm{SR}(A^{k}R':B^{k}) = \mrm{SR}(A^{k}R^{k}:B^{k})$ and likewise for the other partitioning. Finally, $\mrm{SR}(A^{k}R^{k}:B^{k}) = k \cdot \mrm{SR}(AR:B)$, and likewise for the other partitioning. This completes the proof.
\end{proof}

\subsection{Entangled Source Simulation}
We now turn to the more general setting of entanglement of simulation without the restriction to using a catalysis. As highlighted previously, the inclusion of a classical register is not possible in the catalytic setting for small tolerated error. However, distributed source simulation does not require this decoupling condition as the initial classical register $X$ will be correlated with the $A$ and $C$ registers (See Fig. \ref{fig:DSS-vs-EDSS}). It follows catalytic entanglement of simulation is too restrictive to be a proper analogue of distributed source simulation. The question then becomes what is the natural setup to correspond with distributed source simulation. The most general setting would be local operations and shared entanglement (LOSE) with no constraints, that is, the input can be any state and a marginal of the output is the target state up to some tolerated error. This aligns with the notation of Fig.\ \ref{fig:DSS-vs-EDSS}.

However, this unconstrained LOSE setting, if we measure the needed entanglement in terms of entanglement rank, has a simple characterization as we briefly prove following a few definitions.
\begin{definition}\label{def:ent-of-simulation}
Let $\ve \in (0,1)$ and $\rho_{AC} \in \Density(A \otimes C)$. The $\ve$-entanglement of simulation is given by
\begin{align*}
C^{\ve}_{\mrm{EoS}}(A\mcol C)_{\rho} := \log \min & \; \; \mrm{Ent}_{\wt{A}:\wt{C}}(\sigma_{\wt{A}\wt{C}}) 
\\
\text{s.t.} & \; \;  \Tr_{A'C'}(\Phi \otimes \Psi)(\sigma) \approx_{\ve}^{F} \rho \\
& \; \; \sigma \in \Density(\wt{A} \otimes \wt{C}) \ , 
\end{align*}
where $\Phi \in \Channel(\wt{A},A' \otimes A)$, $\Psi \in \Channel(\wt{C},C' \otimes C)$ and $\rho \approx_{\ve}^{F} \sigma$ means $F(\rho,\sigma) \geq 1 - \ve$. (See Fig.\ \ref{subfig:entangeld-source-sim} for diagramatic depiction.)
\end{definition}
\begin{definition}
Let $\rho_{AB} \in \Density(A \otimes B)$. We define the $\ve$-approximate entanglement rank as
$$ \min_{\wt{\rho} \in \mathscr{B}^{\ve}_{F}(\rho)}  \mrm{Ent}_{A \mcol B}(\rho) \ , \ $$
where $\mathscr{B}^{\ve}_{F}(\rho_{AB}) := \{\wt{\rho} \in \Density(A \otimes B): F(\rho,\wt{\rho}) \geq 1 - \ve \}$. This measures the smallest entanglement rank for a state $\sigma$ to be $\ve$-close to $\rho$ under fidelity.
\end{definition}
\begin{proposition}
For $\ve \in [0,1]$, $\rho_{AC} \in \Density(A \otimes C)$,
$$ C^{\ve}_{\mrm{EoS}}(A\mcol C)_{\rho} = \log(\mrm{Ent}^{\ve}_{A:C}(\rho)) \ . $$
\end{proposition}
\begin{proof}
We prove this is a lower bound and then note it can be achieved. It is known that local operations can only decrease entanglement rank \cite{Watrous-Book}. Note that the throwing out of the $A',C'$ spaces are also local operations. Therefore, the input state $\sigma$ must have an entanglement rank that is lower bounded by $\mrm{Ent}^{\ve}_{A:C}(\rho)$ or else it will violate the error restriction. Moreover, the state $\tau_{AC}$ that minimizes $\mrm{Ent}^{\ve}_{A:C}(\rho)$ can be forwarded and the local operations be trivial, so this is also achievable. Taking the logarithm completes the proof.
\end{proof}
We note part of the simplicity of this setting is our choice of measure. \cite{Jain-2012a} analyzed the quantum correlation complexity, which is the same setting but measuring the logarithm of the rank of the initial state used in the protocol rather than its entanglement rank and the analysis in this case is more arduous. 

While entanglement of simulation allows for correlation between the seed, it still does not seem to properly capture distributed source simulation. Namely, we have shown the embezzlement of simulation is achieved by finding a seed that is most similar to the initial while reducing its entanglement rank. This imposes no structure on the input seed whereas distributed source simulation starts from a uniform or non-uniform perfect correlation. In other words, an appropriate analogue should maintain a notion of `universal' seed. As the embezzling states $\ket{\mu(n)}$ in Proposition \ref{prop:embezzling} are a universal family, restricting to seed states that are embezzling states with free classical correlation gives rise to a natural analogue of distributed source simulation for fully quantum target states.\footnote{One might note that maximally entangled states would be the natural equivalent of uniform shared randomness, but \cite{Hayden-2003a} proves this won't work for LOSE. Moreover, we note that there exist other universal families of embezzling states--- see \cite{Leung-2014a} and references therein.}
\begin{definition}\label{def:ve-ent-of-sim}
Let $\ve \in (0,1)$ and $\rho_{AC} \in \Density(A \otimes C)$. The $\ve$-embezzling entanglement of simulation  is
\begin{align*}
 &C^{\ve}_{\mrm{Emb-EoS}}(A\mcol C)_{\rho} \\
:=&\log\min\{r \geq 1: \Tr_{A'C'} \circ (\Phi \otimes \Psi)(\sigma(r)) \approx_{\ve}^{F} \rho \} \ ,
\end{align*} 
where $\sigma(r)_{\wt{A}\wt{C}} := \dyad{\mu(r)} \otimes \sigma_{X_{A}X_{C}}$, $\wt{A} \cong A' \otimes X_{A}$, $\wt{C} \cong C' \otimes X_{C}$, $\Phi \in \Channel(\wt{A},A' \otimes A)$, $\Psi \in \Channel(\wt{C},C' \otimes C)$, and $\sigma \approx_{\ve}^{F} \rho$ means $F(\rho,\sigma) \geq 1 - \ve$. (See Fig.\ \ref{subfig:entangeld-source-sim} for diagramatic depiction where note in this case $\sigma_{A'C'}$ should be the embezzling state.)
\end{definition}

The difference between embezzleable entanglement of simulation, $C^{\ve}_{\mrm{EE,S}}$, and the entanglement of simulation restricted to embezzling states, $C^{\ve}_{\mrm{E|E,S}}$ is of course that we allow for arbitrary classical assistance in the latter which as addressed we cannot in general do with the former. This in particular allows us to distribute a flag state in the latter setting, which results in the following theorem.

\begin{theorem}\label{thm:Emb-EoS}
Let $\rho \in \Density(A \otimes C)$ and $\ve \in [0,1)$. Then 
\begin{align*}
 \log(\mrm{Ent}^{\ve}_{A:C}(\rho)) \leq C^{\ve}_{\mrm{Emb-EoS}}(A\mcol C)_{\rho} \leq \frac{1}{\ve} \log(\mrm{Ent}_{A:C}(\rho)) \ .
\end{align*}
Moreover, for sufficiently small $\ve \in (0,1)$ the upper bound is nearly optimal for pure state $\ket{\psi}$ that is not local unitarily equivalent to $\ket{\mu(r')}$ for any $r'$, and the lower bound is tight when $\ve = 0$ and $\ket{\psi}$ is local unitarily equivalent to some $\ket{\mu(r')}$. These two points imply these bounds are approximately tight although they don't match.
\end{theorem}
\begin{proof}
We first prove the achievability and converse. Then we provide examples where these are (nearly) tight bounds. \\
(\textit{Achievability}) The upper bound is trivial when $\ve = 0$, so we assume $\ve \in (0,1)$. Consider the decomposition of $\rho_{AC} = \sum_{x \in \cX} p(x) \dyad{\psi_{x}}$ such that the max Schmidt rank of $\{\ket{\psi_{x}}\}_{x \in \cX}$ is minimized over all possible decompositions of $\rho_{AC}$. Let $m$ be this max Schmidt rank. That is, 
$$m := \min_{\{p_{X},\{\ket{\psi_{x}}\}_{x} \} : \sum_{x} p(x) \dyad{\psi_{x}} = \rho_{AC}} \max_{x \in \cX} \mrm{SR}(A \mcol C)_{\psi} \ . $$
 Let $\ket{\mu(r)}$ such that $r > m^{1/\ve}$. Then let $\sigma(r) = \dyad{\mu(r)}_{A'C'} \otimes \sum_{x} p(x) \dyad{x}_{X_{A}} \otimes \dyad{x}_{X_{C}}$. Let Alice receive $A'X_{A}$ and Charlie receive $C'X_{C}$. Conditioned on $x$, Alice and Charlie embezzle in $\ket{\psi_{x}}$ using $\mu(r)$. For notational simplicity, we describe this in terms of conditional isometries. Call the set of local isometries conditioned on $x \in \cX$ that do this $\{\cU_{x \in \cX}\}$ and $\{\cV_{x}\}_{x \in \cX}$. These isometries include the local unitaries $U_{x}^{\ast}$,$W_{x}^{\ast}$ from Proposition \ref{prop:embezzling} which we have pushed into the isometries using the isometric invariance of fidelity. Call the total maps that implement this $\Phi$ and $\Psi$. Already tracing out the classical registers and the purification registers,
\begin{align*}
	 & F(\dyad{\mu(r)} \otimes \rho_{AC},\Tr_{A'C'}(\Phi \otimes \Psi)(\sigma)) \\
	\geq & \sum_{x \in \cX}p(x) F(\ket{\mu(r)} \otimes \ket{\psi_{x}}, (\cU_{x} \otimes \cV_{x})\ket{\mu(r)}\ket{0}\ket{0}) \\
	\geq & \sum_{x \in \cX} p(x) (1-\ve) \\
	= & 1-\ve \ ,
\end{align*}  
where the first inequality is going to the isometric picture and using the joint concavity of fidelity and we've used the fidelity guarantee of Proposition \ref{prop:embezzling} in the second inequality. Taking a logarithm of $r$ due to Definition \ref{def:ve-ent-of-sim} completes the achievability. \\
(\textit{Converse}) Let $r' := \mrm{Ent}^{\ve}_{A:C}(\rho)$. By definition, the strategy must be a product map $\Phi \otimes \Psi$. It follows it is a separable map \cite[Definition 6.17]{Watrous-Book}, and thus it cannot increase the entanglement rank \cite[Theorem 6.23]{Watrous-Book}. Thus, the entanglement rank of the initial state $\sigma$ is at least this $r$ value. \\
(\textit{Near Tightness of Upper Bound}) Consider the target state $\rho_{AB}$ is a pure $\phi_{AB}$ such that it is not local unitarily equivalent to any member of $\{\ket{\mu(r)}\}$. There is no advantage to the shared randomness setting as there is an optimal pair of local maps from $\ket{\mu(r')}$ to $\phi$ for any $r'$. Thus, we have reduced the problem to embezzling, which is near optimal as shown in \cite{van-2003a}. \\
(\textit{Tightness of Lower Bound}) Consider the target state $\rho_{AB}$ is local unitarily equivalent to $\ket{\mu(r')}$. Then as $F(\rho,\sigma) = 1$ if and only if $\rho = \sigma$. Thus if $\ve = 0$, one must forward $\ket{\mu(r')}$ instead of any smaller embezzling state.
\end{proof}
While the above result is approximately tight in certain settings, we remark the lower bound is clearly loose in general as the inability to convert one pure state to another with no communication is not only a function of the Schmidt rank, but the Schmidt coefficients \cite{Hayden-2003a,pure-state-revisited}. For example, in \cite{Hayden-2003a}, the authors show that the amount of communication necessary for pure state transmission depends on the `entanglement spread,' which is effectively a function of the maximum Schmidt coefficient and the number of Schmidt coefficients.

\section{Conclusion and Open Problems}\label{sec:conclusion}
In this paper we have considered the task of distributed source simulation in the one-shot setting for fully quantum systems. We introduced various one-shot operational quantities and correlation measures related to distributed source simulation and established near-tight one-shot bounds. In particular, this established a one-shot version of Wyner's common information result \cite{Wyner-1975a} in the smooth entropy framework. In doing so, we generalized the support lemma to preparation channels and showed nuances about smoothing measures when an auxiliary random variable is involved--- ideas which likely will have further use in one-shot quantum network theory. One particular technical point of interest is that we found it is important to \textit{not} be smoothing the auxiliary variable and this led to inducing the measure via $D_{\max}$. We also considered variants of distribute source simulation for comparison with distributed source simulation proper in the appendix to highlight how in a more restricted setting one may instead use a hypothesis testing divergence-induced correlation measure.

We then proceeded to recover asymptotic results from the one-shot results by establishing asymptotic equipartition properties for our one-shot correlation measures. In doing so, we perhaps intuitively extended Wyner's original result \cite{Wyner-1975a}, and Hayashi's extension to separable states \cite{Hayashi-2006a}, by showing that to first-order asymptotically there is no advantage in non-uniform shared randomness in the task (Theorem \ref{thm:first-order-asymptotic-equivalence}). An open question is whether these tasks differ in second-order asymptotics. This is unclear as both achievability results in Theorem \ref{thm:one-shot-DSS-main-theorem} use uniform soft-covering, but the converses are in terms of different measures.

Motivated by Yu and Tan's recent result strong converse in the classical setting \cite{Yu-2018a,Yu-corrections-2018a}, it would be our expectation that at least uniform correlation of formation would admit a strong converse. To remain in the smooth entropy framework, this would require establishing a strong converse for our asymptotic equipartition property. We showed one option would be to establish a new property for conditional min entropy (Proposition \ref{prop:open-problem}) that likely would require new tools. As we explained in that section, current methods do not seem satisfactory to establish this property.  Alternatively, one could establish the strong converse in some other fashion outside of the smooth entropy calculus.

As already alluded to, an open problem related to the establishment of a strong converse would be the establishment of second-order asymptotics, though this may be difficult with the rate being a function of the auxiliary random variable. Indeed, to the best of our knowledge, the second-order asymptotics of distributed source simulation are not known even in the fully classical setting. It has been highlighted previously it is difficult to obtain second-order asymptotics for network coding problems that involve an auxiliary random variable \cite{Tan-2014a}, which was the consistent issue throughout this work. To the best of our knowledge, the only network coding problem with an auxiliary random variable that has had second-order asymptotics established is the Gray-Wyner problem \cite{Watanabe-2016a,Zhou-2016a}.

After establishing the general framework of one-shot distributed source simulation for bipartite states we considered two variations: more parties and an entangled state variation. In the multipartite setting we explained while all the proofs extend in a straightforward manner and made some clarifications with regards to comments in previous work \cite{Liu-2010a}. In the quantum setting, we discussed the ability to use embezzling as a strategy for the equivalent of distributed source simulation as conversion of entangled states to arbitrary error with zero communication is impossible. We considered both the setting where the initial state cannot include classical correlation beyond through entanglement and when it can and gave tight upper and lower bounds that in general do not match.

\section{Acknowledgments}
I.G. would like to thank Vincent Y. F. Tan for providing a draft of Yu's and his monograph on common information \cite{Yu-2022a} prior to its publication. I.G. would also like to thank Felix Leditzky, Marco Tomamichel, and Michael X.\ Cao for helpful discussions on strong converses. We thank an anonymous reviewer for bringing Lemma \ref{lem:characterization-of-variant-of-DSS} to our attention. \edit{We also thank anonymous reviewers for IEEE Transactions in Information Theory for further helpful feedback on the presentation of the work.} The majority of this work was performed while I.G. was at University of Illinois at Urbana-Champaign. This work was supported in part by the National Science Foundation under Grant
2112890. I.G. acknowledges the support of an Illinois Distinguished Fellowship during this work.

\bibliography{References.bib}

\begin{IEEEbiographynophoto}{Ian George}
received his B.A. degrees in physics and philosophy from Kenyon College, Ohio, USA in 2018, his M.Sc. degree in physics (quantum information) from the University of Waterloo, Ontario, Canada in 2020, and his Ph.D. in Electrical and Computer Engineering from University of Illinois at Urbana-Champaign, Illinois, USA in 2024. He is currently a postdoctoral researcher at the Centre for Quantum Technologies, Singapore. His research interests include quantum cryptography, network information theory, and computation. He works broadly on quantum information processing with a focus on developing information-theoretic methods for analyzing quantum cryptographic protocols, communication networks, and noisy quantum computation.
\end{IEEEbiographynophoto}

\begin{IEEEbiographynophoto}{Min-Hsiu Hsieh}
has been the director of the Hon Hai Quantum Computing Research Center in Taiwan since January 2021. Prior to this role, he was an Associate Professor at the University of Technology Sydney in Australia and a member of the UTS Centre for Quantum Software and Information from 2014 to 2020. From 2010 to 2012, he was a postdoctoral research fellow at the University of Cambridge in the UK, where he was affiliated with the Centre for Quantum Information and Foundations. Before that, he worked as a research scientist with the ERATO-SORST Quantum Computation and Information Project in Japan, which was funded by the Japan Science and Technology Agency. During this time, he was also affiliated with the Department of Computer Science at the University of Tokyo. From 2014 to 2018, he held an Australian Research Council Future Fellowship. His research interests include general topics in quantum information and computation. 
\end{IEEEbiographynophoto}

\begin{IEEEbiographynophoto}{Eric Chitambar}
received his B.S. degree in physics from the University of Notre Dame, Indiana, USA in 2005 and his Ph.D. degree in physics from the University of Michigan in Ann Arbor, Michigan, USA, in 2010 under the direction of Prof. Yaoyun Shi.  He served as a post-doctoral researcher at the University of Toronto in Canada and later at the Perimeter Institute for Theoretical Physics in Waterloo, Canada.  Dr. Chitambar joined the physics department at Southern Illinois University in 2012 before moving to the University of Illinois Urbana-Champaign in 2018 where he is currently a Professor of Electrical and Computer.  His research interests include quantum communication and cryptography, entanglement theory, and general quantum resource theories.  His work tackles a broad range of problems, from quantum foundations to applied information theory and experimental protocols.
\end{IEEEbiographynophoto}

\appendix

\section*{Mutual Information Lemmas}
In this section of the appendix, we establish various properties of mutual information measures. Remember there are three versions of (smooth) max mutual information (Definition \ref{def:max-mutual-information-quantities}). In each result we specify which mutual information we mean. Many of these results are relatively straightforward variations of proofs from \cite{Muller-2013a} and/or \cite[Chapter 6]{Tomamichel-2015a}.

\edit{
We begin by establishing Lemma \ref{lem:classical-seed-suff}, which shows mutual information of Markov chains is minimized by a quantum-classical-quantum Markov chain.
\begin{proof}[Proof of Lemma \ref{lem:classical-seed-suff}]
We prove it for $\mbb{I}^{\downarrow}(AC:X)$ case as it is then straightforward to see the same proof method will hold for the other cases.
Let $\rho_{ABC}$ be a $A-B-C$ Markov chain. Then $\rho_{B} = \oplus_{x} p(x) \rho_{b^{L}_{x}} \otimes \rho_{b^{R}_{x}}$. Define the map $\cE: B \to BX$ as $\cE(\cdot) = \sum_{j} \Pi_{x}^{B} \cdot \Pi_{x}^{B} \otimes \dyad{x}_{X}$ where $\{\Pi_{x}\}$ are the mutually orthogonal projectors onto the subspaces $b^{L}_{x} \otimes b^{R}_{x}$. Define $\rho_{ABXC}' := \cE(\rho_{ABC})$.  Then it follows
\begin{align*}
   \mbb{I}^{\downarrow}(AC:B)_{\rho} = & \mbb{D}(\rho_{ABC}\Vert \tau_{AC} \otimes \sigma_{B}) \\
    \geq & \mbb{D}(\rho_{ABXC}'\Vert \tau_{AC} \otimes \sigma_{BX}) \\ \geq & \mbb{D}(\rho_{AXC}'\Vert \tau_{AC} \otimes \sigma_{X}) \\
    \geq & \min_{\substack{\tau' \in \Density(A \otimes C) \\ \sigma' \in \Density(X)}} \mbb{D}(\rho_{AXC}'\Vert \tau_{AC}' \otimes \sigma_{X}') \\
    =& \mbb{I}^{\downarrow}(AC:X)_{\rho'} \ , 
\end{align*}
where the first inequality is DPI using $\cE$, the second inequality is DPI using the partial trace on the $B$ space, and in both cases the map only acts on one side of conditioned tensor product, and the final is just re-minimizing. We can also guarantee $\sigma_{X}'$ is classical by DPI and pinching on the computational basis. Moreover $\rho_{AXC}'$ is $A-X-C$ Markov chain extension of $\rho_{AC}$ trivially as $\cE$ never acted on the $AC$ spaces and its recovery maps are just state preparations maps, e.g.\ $x \mapsto \Tr_{b^{L}_{x}}(\rho_{Ab^{L}_{x}})$. This completes the proof.
\end{proof}
\begin{remark}
It is worth noting why the above isn't proven to be an equality. When one converts the $B$ register to a classical $X$ register, they destroy any entanglement between $A$ (resp.\ $C$) and $b^{L}_{x}$ (resp.\ $b^{R}_{x}$). To recover this entanglement, one needs to apply the recovery map, e.g.\ 
$$\rho_{X} \to \rho_{B} \xrightarrow[]{\cR} \rho_{BC} \xrightarrow[]{\ol{\cR}} \rho_{ABC} \ . $$
However, to preserve the form of mutual information, you can only act on the $B$ space, so it is not possible to evaluate this directly.
\end{remark}
}

\edit{We now establish} that when smoothing a quantum state with classical registers, you can restrict to optimizers that are classical.
\begin{proposition}\label{prop:restricting-to-classical-registers}
Let $\rho_{AXBY} \in \Pos(A\otimes X\otimes B\otimes Y)$ be classical on $X$ and $Y$. For $\ve \in [0,\sqrt{\Tr(\rho)})$, the smoothing ball $\Bve(\rho)$ may be restricted to QCQC states and the optimal $\ol{\tau},\tau$ will be classical on the same registers when optimizing over $\mbb{I}^{\ve,x}(AX\mcol BY)_{\rho}$ for all $x \in \{\downarrow,\uparrow,\upuparrow\}$ where $\mbb{I}$ is any mutual information defined on any R\'{e}nyi divergence $\mbb{D}$.
\end{proposition}
\begin{proof}
By definition, 
$$\mbb{I}^{\ve,x}(AX\mcol BY)_{\rho} = \min_{\wt{\rho} \in \Bve(\rho)} \min_{\ol{\tau} \in S_{|\wt{\rho}}^{AX} , \tau \in S_{|\wt{\rho}}^{BY}}\mbb{D}(\wt{\rho}\Vert \tau \otimes \ol{\tau}) \ , $$
where the sets $S$ will also depend which $x \in \{\downarrow,\uparrow,\upuparrow\}$ we are using. Next, by data-processing, $\mbb{D}((P_{X} \otimes P_{Y})(\wt{\rho})\Vert P_{X}(\tau) \otimes P_{Y}(\ol{\tau})) \leq \mbb{D}(\wt{\rho}\Vert \tau \otimes \ol{\tau})$, where $P_{X},P_{Y}$ are the pinching maps onto the computational bases for the registers $X$  and $Y$. As $\rho_{AXBY}$ is classical on $X$ and $Y$, by data-processing of purified distance, we may restrict minimizing $\Bve(\rho)$ to states that are classical on $X$ as $\rho$ is invariant under pinching on $X$ and $Y$. Furthermore then $S_{|\wt{\rho}}^{AX}$ may be restricted to being classical on $X$ and same idea for the other set with respect to $Y$ as the optimizing choice of $\wt{\rho},\ol{\tau},\tau$ will be classical on those spaces as we showed via DPI. This completes the proof.
\end{proof}

Next, we will need to establish that we can write terms proportional to max mutual information in terms of an expectation on the classical register. This will be broken up into multiple steps. Each step has a conditional entropic equivalent which can be found in \cite{Tomamichel-2015a}.
\begin{proposition}
Let $\rho_{ABX} \in \Density_{\leq}(A\otimes B\otimes X)$ such that it is classical on $X$. That is, $\rho_{ABX} = \sum_{x} p(x) \dyad{x} \otimes \rho_{AB}^{x}$. Let $\alpha' := \alpha - 1$. Then for $\alpha \in (0,1) \cup (1,\infty)$ we have the following
\begin{align*}
   \mbb{I}^{\upuparrow}_{\alpha}(A\mcol BX) 
   &= \frac{1}{\alpha'} \log \left(\sum_{x} p_{x} \exp((\alpha')D_{\alpha}(\rho_{AB}^{x}\Vert \rho_{A} \otimes \rho_{B}^{x}))\right) \\
    \mbb{I}^{\uparrow}_{\alpha}(A\mcol BX)_{\rho}
    &= \frac{\alpha}{-\alpha'} \log \left(\sum_{x} p_{x}\exp(\frac{\alpha'}{\alpha}\mbb{I}_{\alpha}^{\uparrow}(\rho_{AB}^{x}\Vert \rho_{A}))\right) \\
    \mbb{I}^{\downarrow}_{\alpha}(A\mcol BX)_{\rho}&=\min_{\substack{q \in \cP(\cX) \\ \ol{\tau}_{A},\{\tau_{B}^{x}\}_{x}}} \frac{1}{-\alpha'} \log \Bigg(\sum_{x} p_{x}^{\alpha}q(x)^{-\alpha'} \\ 
    & \hspace{25mm} \cdot \exp((\alpha') D_{\alpha}(\rho_{AB}^{x}\Vert \ol{\tau}_{A} \otimes \rho_{B}^{x}))\Bigg) \ ,
\end{align*}
where $\mbb{I}_{\alpha}$ is any Petz or Sandwiched R\'{e}nyi mutual information over the specified ranges and for the middle quantity we define
$$ \mbb{I}_{\alpha}^{\uparrow}(\rho_{AB}\Vert \sigma_{A}) := \min_{\tau_{B}} \mbb{D}_{\alpha}(\rho_{AB}\Vert \sigma_{A} \otimes \tau_{B}) \ . $$
\end{proposition}
\begin{proof}
The proof is largely identical to that of \cite[Proposition 5.1]{Tomamichel-2015a}, but we provide it for completeness.
We begin from the fact that for any CQ states $\rho_{XA},\sigma_{XA}$, it holds
\begin{align*}
&\mbb{D}_{\alpha}(\rho_{XA}\Vert \sigma_{XA}) \\
 =& \frac{1}{\alpha-1} \log \left(\sum_{x} p_{x}^{\alpha}q_{y}^{1-\alpha} \exp((\alpha-1)\mbb{D}_{\alpha}(\rho_{A}^{x}\Vert \sigma_{A}^{x} ))\right) \ . 
\end{align*}
As $I^{\upuparrow}_{\alpha}(A\mcol BX) = D_{\max}(\rho_{ABX}\Vert \rho_{A} \otimes \rho_{BX})$, we have its simplification by direct calculation.
For $I^{\downarrow}_{\alpha}(A\mcol BX)$ we have established the minimizers $\tau_{BX}$ can be restricted to being classical in Proposition \ref{prop:restricting-to-classical-registers}, thus $\mbb{I}_{\alpha}(A\mcol BX)_{\rho} = \min_{\ol{\tau}_{A}, \tau_{BX}} D_{\max}(\rho\Vert \ol{\tau} \otimes \tau_{BX})$ so again a direct calculation can be established. 

Lastly we establish the remaining case. First recall $\mbb{I}^{\uparrow}_{\alpha}(A\mcol BX) = \min_{\tau_{BX}} \mbb{D}_{\alpha}(\rho\Vert \rho_{A} \otimes \tau_{BX})$. It follows from above then that we have
\begin{align*}
    &\mbb{I}^{\uparrow}_{\alpha}(A\mcol BX)_{\rho} \\
     = & \frac{1}{\alpha'} \min_{\tau_{BX}} \log(\sum_{x} p_{x}^{\alpha}q_{x}^{-\alpha'} \exp((\alpha')\mbb{D}_{\alpha}(\rho_{AB}^{x}\Vert \rho_{A} \otimes \tau_{B}^{x}))) \\
    = &  \frac{1}{-\alpha'} \min_{q \in \cP(\cX)} \log(\sum_{x} p_{x}^{\alpha}q_{x}^{-\alpha'} \exp((\alpha-1)\mbb{I}_{\alpha}(\rho_{AB}^{x}\Vert \rho_{A}))) \ ,
\end{align*}
where the first is by definition and the second is by defining $\mbb{I}_{\alpha}^{\uparrow}(\rho_{AB}\Vert \sigma_{A}) := \min_{\tau_{B}} \mbb{D}_{\alpha}(\rho_{AB}\Vert \sigma_{A} \otimes \tau_{B})$. The reason this definition suffices in the second step is because minimizing over $\tau_{BX} = \sum_{x} q_{x} \dyad{x} \otimes \tau_{B}^{x}$ is equivalent to minimizing the set $\{\tau_{B}^{x}\}_{x}$ and distribution $q_{x}$ independently, so for each $x$ we can move the choice of minimizing $\tau_{B}^{x}$ in front of the relative entropy. Next, we define $r_{x} := p_{x}\exp(\frac{\alpha-1}{\alpha}\mbb{I}_{\alpha}^{\uparrow}(\rho_{AB}^{x}\Vert \rho_{A}))$ for every $x \in \cX$. This means $r_{x}^{\alpha}p_{x}^{-\alpha} = \exp((\alpha-1)\mbb{I}^{\uparrow}_{\alpha}(A\mcol B)_{\rho_{AB}^{x}})$ for each $x \in \cX$. Thus we can plug this substitution back in to obtain:
\begin{align*}
 \mbb{I}^{\uparrow}_{\alpha}(A\mcol BX)_{\rho}
  = & \frac{1}{\alpha-1} \min_{q \in \cP(X)} \log(\sum_{x} p_{x}^{\alpha}q_{x}^{1-\alpha}r_{x}^{\alpha}p_{x}^{-\alpha}) \\
   = & \frac{1}{\alpha-1} \min_{q \in \cP(X)} \log(\sum_{x} q_{x}^{1-\alpha}r_{x}^{\alpha}) \ . 
\end{align*}
The last optimization problem is a straightforward minimization over a simplex and thus can be solved using KKT conditions. One could skip over this, but just to be complete, it is provided at the end of the proof. It finds the optimizer is $\ol{q}_{x} = r_{x}/(\sum_{x} r_{x})$ for all $x$, so we will plug this in to get the answer:
\begin{align*}
     \mbb{I}^{\uparrow}_{\alpha}(A\mcol BX)_{\rho} 
    =& \frac{1}{\alpha-1} \log\left( \sum_{x} \ol{q}_{x}^{1-\alpha}r_{x}^{\alpha} \right) \\
    =& \frac{1}{\alpha-1} \log\left( \sum_{x} r_{x}^{\alpha} \left[ \sum_{x'} r_{x'}/r_{x}\right]^{\alpha-1} \right) \\
    =& \frac{1}{\alpha-1} \log\left( \sum_{x} r_{x} \left[\sum_{x'} r_{x'}\right]^{\alpha-1} \right) \ , 
\end{align*}
where the second equality is by definition of $\ol{q}_{x}$ and the third is because $r_{x}^{\alpha} r_{x}^{1-\alpha} = r_{x}$. Continuing onwards,
\begin{align*}
    =&\frac{\alpha}{\alpha-1} \log\left( \left(\sum_{x} r_{x}\right)^{1/\alpha} \left[ (\sum_{x'} r_{x'})\right]^{1-1/\alpha} \right) \\
    =& \frac{\alpha}{\alpha-1} \log \left(\sum_{x} r_{x}\right) \ ,
\end{align*}
where the first line we have multiplied and divided by $\alpha$ and then pulled the $1/\alpha$ factor into the logarithm and distributed. Plugging in the definition of $r_{x}$ completes the proof.

(\textit{Solving the KKT Criteria}) We can move the $\log$ out so we are optimizing 
$$\min \left \{\sum_{x} q_{x}^{1-\alpha}r^{\alpha}_{x} : \sum_{x} q_{x} - 1 = 0 \, , \, -q_{x} \leq 0 \, \forall x \in \cX \right\} \ . $$
Denote the objective function $f(q)$, the equality constraint $h(q)$, and the inequality constraints $g_{i}(q)$. Note $\pdv{}{q_{x}}f(q) = (1-\alpha)(r_{x}/q_{x})^{\alpha}$, $\pdv{}{q_{x}}h(q) = 1$, and $\pdv{}{q_{x}}g_{i}(q) = -\delta_{x,i}$. Thus the Lagrangian constraint is 
$(1-\alpha)\sum_{x} (r_{x}/q_{x})^{\alpha} e_{x} + \lambda \sum_{x} e_{x} - \sum_{x} \mu_{x} e_{x} = 0$. Since this is effectively entry-wise, this means $(\alpha-1)(r_{x}/q_{x})^{\alpha} + \mu_{x} = \lambda$ for all $x$. Note if there is any $x$ such that $q_{x} = 0$, then this would make $\lambda = \infty$, but to be primal feasible there must exists $q_{x} \in (0,1]$ which would make $\lambda$ also finite. This is a contradiction, therefore we can conclude $q_{x} > 0$ for all $x$. If $q_{x} > 0$ for all $x$, then complementary slackness requires $\mu_{x} = 0$ for all $x$. This simplifies the Lagrange constraint so that by moving things around we conclude
$$ \lambda = (\alpha-1)(r_{x}/q_{x})^{\alpha} \Rightarrow q_{x} = r_{x}(\alpha-1)^{1/\alpha}/\lambda  \ . $$
Then by primal feasibility condition
$$ \sum_{x} q_{x} = 1 \Rightarrow \lambda = (\alpha-1)^{1/\alpha}\sum_{x} r_{x} \ .$$
Finally, plugging this value of $\lambda$ into $q_{x}$, we have $q_{x} = r_{x}/(\sum_{x} r_{x})$. Note that the objective function is linear as $\alpha$ is fixed as are the constraints so we have linear constraint qualifications which tells us this is indeed a minimizer.
\end{proof}
Note what the above shows us is that only $\mbb{I}^{\uparrow}_{\alpha}$ can be expressed as a mixture of the same information measure over the conditional states. This gives us the following nice corollary.
\begin{corollary}\label{corr:Imax-cond-on-X}
Let $\rho_{ABX} \in \Density_{\leq}(A\otimes B\otimes X)$ such that it is classical on $X$. Then  
$$I^{\uparrow}_{\max}(A\mcol BX)_{\rho} = \log(\sum_{x} p_{x} \exp(I^{\uparrow}_{\max}(\rho_{AB}^{x}\Vert \rho_{A}))) \ , $$
where 
$$ I^{\uparrow}_{\max}(\rho^{x}_{AB}\Vert \rho_{A}) := \min_{\tau_{B} \in \Density(B)} \mbb{D}_{\max}(\rho^{x}_{AB}\Vert \rho_{A} \otimes \tau_{B}) \ . $$
\end{corollary}
\begin{proof}
By the previous proposition, we know this to hold for $\alpha \in (1,\infty)$. Since $I^{\uparrow}_{\max} := \lim_{\alpha \to \infty} I^{\uparrow}_{\alpha}$, we just need to take the limit on the right hand side. This just means we need to use the product law of limits. By L'Hopital's rule, $\alpha/(\alpha-1)$ and $(\alpha-1)/\alpha$ both go to one and as $\lim_{\alpha \to \infty} D_{\alpha} = D_{\max}$, $I^{\uparrow}_{\alpha}$ is $I^{\uparrow}_{\max}$. This completes the proof.
\end{proof}

We also show explicitly that $I^{\upuparrow}_{\max}$ and $I^{\downarrow}_{\max}$ won't satisfy the property we need for applying the generalized support lemma.
\begin{proposition}\label{prop:Iupup-max-cq-side}
Let $\rho_{AX} \in \Pos(A\otimes X)$ be classical on $X$. Then we have 
$$ I^{\upuparrow}_{\max}(A\mcol X) = \max_{x} D_{\max}(\rho^{x}_{A}\Vert \rho_{A}) \ . $$
\end{proposition}
\begin{proof}
We can just focus on $\rho_{AX} \in \Density(AX)$ by normalization property of $D_{\max}$. So we can write $\rho_{AX} = \sum_{x} p(x) \dyad{x} \otimes \rho^{x}_{A}$. Then we have
\begin{align*}
    &I^{\upuparrow}_{\max}(A\mcol X)_{\rho} \\
    =& D_{\max}(\rho_{AX}\Vert \rho_{A} \otimes \rho_{X}) \\ 
    =& \min \{\lambda : \sum_{x} p(x) \dyad{x} \otimes \rho_{A}^{x}  \leq \exp(\lambda) \sum_{x} (\dyad{x} \otimes \rho_{A}) \} \\
    =& \min \{\lambda : \rho_{A}^{x} \leq \exp(\lambda) \rho_{A} \} \\
    =& \max_{x} D_{\max}(\rho^{x}_{A}\Vert \rho_{A}) \ ,
\end{align*}
where the second equality is the definition of $D_{\max}$ and expanding the states,
\end{proof}
\begin{proposition}\label{prop:Idown-max-cq-side}
Let $\rho_{AX} \in \Pos(A\otimes X)$ be classical on $X$. Then 
$$ I^{\downarrow}_{\max}(A\mcol X)_{\rho_{AX}} = \min_{\substack{q \in \cP(\cX) \\ \tau_{A} \in \Density(A)}} \log \left( \frac{p(x)}{q(x)} \|\tau_{A}^{-1/2} \rho^{x}_{A} \tau_{A}^{-1/2}\|_{\infty} \right) \ . $$
\end{proposition}
\begin{proof}
By definition,
\begin{align*}
	I^{\downarrow}_{\max}(A\mcol X)_{\rho_{AX}}
	= & \min_{\substack{\tau_{A} \in \Density(A)\\ \sigma_{X} \in \Density(X)}} D_{\max}(\rho_{AX}\Vert \tau_{A} \otimes \sigma_{X}) \\
	= &\log  \min_{\substack{q \in \cP(\cX)\\ \tau_{A} \in \Density(A) }} \max_{x} \frac{p(x)}{q(x)} \|\tau_{A}^{-1/2}\rho_{A}^{x}\tau_{A}^{-1/2}\|_{\infty} \ ,
\end{align*}
where we have used that we can restrict $\sigma_{X}$ to classical states by Proposition \ref{prop:restricting-to-classical-registers} and that $D_{\max}(\rho\Vert \sigma) = \log \|\sigma^{-1/2}\rho\sigma^{-1/2}\|$ to simplify in the second line.
\end{proof}

\subsection*{Smooth Max Common Information Lemmas}
Here we establish that the smooth max common informations are always optimized by a normalized state. The proofs are effectively streamlined versions of \cite[Lemma 22]{Ciganovic-2013a}, which make use of the normalization property of the max divergence \cite{Tomamichel-2015a}. We also make use of the following lemma.
\begin{lemma}\label{lem:renorm-in-ball}(\cite[Lemma 21]{Ciganovic-2013a}) Let $\rho_{AB} \in \Density(A \otimes B)$ and $\ve \geq 0$. If $\wt{\rho} \in \Bve(\rho)$, then $\wt{\rho}/\Tr(\wt{\rho}) \in \Bve(\rho)$.
\end{lemma}
\begin{proposition}\label{prop:SMCI-maxed-by-normal-state}
Let $\rho_{AC} \in \Density(A\otimes C)$ and $\ve \geq 0$ such that there exists a subnormalized Markov chain in the smoothing ball. Then for $C^{\ve}_{\max}(A\mcol C)_{\rho}$ and $C^{F,\ve}_{\max}(A \mcol C)_{\rho}$, there exists an optimizer that is a normalized quantum Markov chain state.
\end{proposition}
\begin{proof}
We provide the proof for $C^{\ve}_{\max}(A \mcol C)_{\rho}$, the other case has an identical proof. Let $\wt{\rho}_{A-X-C} \in D_{\leq}(AXC)$ and $\sigma_{AC} \in \Density(AC)$ be the optimizer of $C^{\ve}_{\max}(A \mcol C)_{\rho}$. Let $\tau_{A-X-C} := \Tr[\wt{\rho}]^{-1} \wt{\rho}_{A-X-C}$ which note is a Markov chain extension of $\Tr[\wt{\rho}]^{-1} \wt{\rho}_{AC}$ by Definition \ref{def:QMC-ext-for-subnormalized}. Then,
\begin{align*}
	C^{\ve}_{\max}(A \mcol C)_{\rho} =& D_{\max}(\wt{\rho}_{A-X-C} \Vert \sigma_{AC} \otimes \wt{\rho}_{X})\\
	=& D_{\max}(\Tr[\wt{\rho}] \tau_{A-X-C} \Vert \sigma_{AC} \otimes \Tr[\wt{\rho}] \tau_{X}) \\
	=& D_{\max}(\tau_{A-X-C} \Vert \sigma_{AC} \otimes \tau_{X}) \ ,
\end{align*}
where the last equality uses the normalization property of the max divergence. Then noting that $\cB^{\ve}(\rho) \ni \tau_{A-X-C} = \Tr[\wt{\rho}]^{-1} \wt{\rho}_{A-X-C}$ by Lemma \ref{lem:renorm-in-ball}, , we know $\tau_{A-X-C}$ is feasible, and thus this completes the proof.
\end{proof}

\begin{proposition}\label{prop:SDMCI-maxed-by-normal-state}
Let $\wt{\rho}_{AC} \in \Bve(\rho)$ with QMC extension $\wt{\rho}_{A-X-C}$ optimize $C^{\upuparrow,\ve}_{\max}(A:C)_{\rho}$. Then, $\Tr[\wt{\rho}_{AC}] = 1$, i.e. is normalized.
\end{proposition}
\begin{proof} 
Let $\wt{\rho}_{AC} \in \Bve(\rho)$ with quantum Markov chain extension $\wt{\rho}_{A-X-C}$ optimize $C^{\upuparrow,\ve}_{\max}(A:C)_{\rho}$. Let $\tau_{A-X-C} = \wt{\rho}_{A-X-C}/\Tr[\wt{\rho}]$, which has a classical register $X$ and is a QMC as established in the previous proof. Then we may write
\begin{align*}
	C^{\upuparrow,\ve}_{\max}(A:C)_{\rho} &= D_{\max}(\wt{\rho}_{A-X-C} \Vert \wt{\rho}_{AC} \otimes \wt{\rho}_{X}) \\
	&= D_{\max}(\Tr[\wt{\rho}] \tau_{A-X-C} \Vert \Tr[\wt{\rho}]^{2} \tau_{AC} \otimes \tau_{X}) \\
	&= D_{\max}(\tau_{A-X-C} \Vert \tau_{AC} \otimes \tau_{X}) - \log(\Tr[\wt{\rho}]) \ ,
\end{align*}
where the third equality is using the normalization property of max divergence \cite{Tomamichel-2015a}. Now as $\Tr[\wt{\rho}] \leq 1$, we may conclude $-\log(\Tr[\wt{\rho}]) \geq 0$, which means that $D_{\max}(\tau_{A-X-C} \Vert \tau_{AC} \otimes \tau_{X})$ is at least as good of a minimizer as $\wt{\rho}$. As $\tau_{A-X-C} \in \Bve(\rho)$ and is a normalized density matrix, this completes the proof.
\end{proof}

\edit{Next we prove the data processing inequality.

\begin{proof}[Proof of Proposition \ref{prop:local-DPI-of-com-inf}]
We focus on $C^{\ve}_{\max}(A \mcol C)$ as the other cases are nearly identical. We focus on a map being applied on the $A$ space, by symmetry of the argument this will also establish the other case. Let $\wt{\rho}_{A-X-C}$ be the minimizer of $C^{\ve}_{\max}(A\mcol C)_{\rho}$. Consider $\Phi(\wt{\rho})$ where $\Phi_{A \to A'} \in \Channel(A,A')$. Note the resulting state is still a QMC with recovery map $\Phi \circ \cR_{X \to AX}$ and $(\Tr_{X} \circ \Phi)(\wt{\rho}) \in \Bve(\Phi(\rho))$ by the DPI for purified distance. Then we have,
\begin{align*}
    C_{\max}^{\ve}(A'\mcol C)_{\Phi(\rho)} \leq  I^{\uparrow}_{\max}(X \mcol A'C)_{\Phi(\wt{\rho})} 
     \leq& I^{\uparrow}_{\max}(X \mcol AC)_{\wt{\rho}} \\
      =& C_{\max}^{\ve}(A\mcol C)_{\rho} \ ,
\end{align*}
where the first inequality is our choice of element in the smoothing ball and QMC extension and the second is the DPI. Note we needed local maps because we need to preserve the QMC structure.
\end{proof}}

\edit{ We end by formally showing the smooth max common information can be NP-hard to solve.
\begin{proof}[Proof of Proposition \ref{prop:NP-hardness}]
First, note that if $E_{P}(A\mcol C)_{\rho} \leq \ve$, then there exists a separable state $\wt{\rho}$ such that $\wt{\rho} \in \Bve(\rho)$, so, by Lemma \ref{lem:sep-marg-of-QMC}, it admits a QMC extension, and thus $C^{\ve}_{F}(A:C)_{\rho}$ is finite. In contrast, if not such separable state exists, it does not admit a QMC extension and by our convention the value is infinite (or may be represented as a special symbol from a computing perspective). Therefore, if $C^{\ve}_{F}(A:C)_{\rho}$ is efficient to compute, then we have a method for the ability to determine if the state is within some distance from the space of separable states. It is known determining membership is strongly NP-hard \cite{Gharibian-2008a}, which means even with some tolerated distance from the set of separable states $\beta$ determining whether it is in or out of the set is NP-hard. Therefore this would be in contradiction with being able to compute $C^{\ve}_{\max}$ efficiently.
\end{proof}
}

\section*{Technical Lemmas for Strong Converse Claims} In this section we establish lemmas relevant to Section \ref{subsec:on-a-strong-converse}, which is on the possibility of establishing a strong converse of some smooth max Wyner common information quantity. As noted in the main text, our major contribution is the first strong converse for up-arrow max mutual information, $I^{\uparrow,\ve}_{\max}(A \mcol B)_{\rho}$, which is Theorem \ref{thm:strong-AEP-for-Imax}. As our reduction of the existence of a strong converse for smooth flipped max Wyner common information to a new property of smooth min-entropy follows nearly the same steps as proving Theorem \ref{thm:strong-AEP-for-Imax}, we present each proposition relevant for establishing Theorem \ref{thm:strong-AEP-for-Imax} and then immediately provide the corresponding proposition for establishing Proposition \ref{prop:open-problem} and explain any needed modification for establishing that proof. Finally, at the end of the section, we provide the discussion on trying to use entropy accumulation to establish our strong converse mentioned in Section \ref{subsec:on-a-strong-converse}.

 for how to reduce the asymptotic behaviour of SMCI to properties of the conditional min-entropy. To do so, we first show how to generalize results of \cite{Berta-2011a,Ciganovic-2013a} to get a strong AEP for $I^{\uparrow,\ve}_{\max}$, and then just note how to modify these proofs to establish chain rules for SMCI. We begin by proving the lower bounds in detail as this requires the most alteration from the previous proof \cite{Ciganovic-2013a}. We first refine the basic trick for a lower bound.
\begin{proposition}\label{prop:min-entropy-measure-bound}
Let $\ve \in (0,1)$ and $\rho \in \Density(A)$. Then there exists $0 \leq \Pi \leq \mbb{1}$ such that $[\Pi,\rho] = 0$, $P(\rho,\Pi\rho\Pi) \leq 2\sqrt{\ve(1-\ve)}$ and 
$$ H^{\ve}_{\min}(A)_{\rho} \leq H_{\min}(A)_{\Pi\rho\Pi} \ . $$
\end{proposition}
\begin{proof}
\sloppy By \cite[Lemma 18]{Berta-2011a}, there exists a $\Pi$ as specified such that $\Tr((\mbb{1}-\Pi^{2})\rho) \leq 2\ve$ satisfying the min-entropy bound given. Note $2\ve \geq \Tr((\mbb{1}-\Pi^{2})\rho)=1 - \Tr(\Pi^{2}\rho)$ which gives us $\Tr(\Pi^{2}\rho) \geq 1 - 2\ve$, so $\Tr(\Pi^{2}\rho)^{2} \geq (1-2\ve)^{2}$. Finally this means $\sqrt{1-\Tr(\Pi^{2}\rho)^{2}} \leq \sqrt{1-(1-2\ve)^{2}}$. By \cite[Lemma A.7]{Berta-2011a}, we have $P(\rho,\Pi\rho\Pi) \leq \Tr(\rho)^{-1/2} \sqrt{\Tr(\rho)^{2}-\Tr(\Pi^{2}\rho)^{2}} = \sqrt{1-\Tr(\Pi^{2}\rho)^{2}} \leq \sqrt{1 - (1-2\ve)^{2}} =  2\sqrt{\ve(1-\ve)}$.
\end{proof}

Now we can use this refinement to extend the result of \cite{Ciganovic-2013a} to the full parameter range.

\begin{proposition}\label{prop:Imax-lb-chain-rule}
(\cite[Lemma 6, Extended to Full Parameter Range]{Ciganovic-2013a}) Let $\rho_{AB} \in \Density(A \otimes B)$. Let $0 < \ve < \ol{\ve} < 1$ such that such that $\ve + \delta(\ol{\ve}) \in (0,1)$ where $\delta(\ol{\ve}) := 2\sqrt{\ol{\ve}(1-\ol{\ve})}$. Then it holds $$I^{\uparrow,\ve}_{\max}(A\mcol B)_{\rho} \geq H^{\ol{\ve}-\ve}_{\min}(A)_{\rho} - H^{\ve+\delta(\ol{\ve})}_{\min}(A|B)_{\rho} \ . $$
\end{proposition}
\begin{proof}
The proof is largely identical to the original except that we have refined certain steps. By re-arranging \cite[Lemma B.13]{Berta-2011a} and maximizing over the smoothing ball, we have
\begin{align*}
     H^{\ve+\delta(\ol{\ve})}_{\min}(A|B)_{\rho}
    \geq  \max_{\wt{\rho} \in \Bvec{\ve + \delta(\ol{\ve})}(\rho)} \left[ H_{\min}(A)_{\wt{\rho}} - I^{\uparrow}_{\max}(A\mcol B)_{\wt{\rho}} \right] \ .
\end{align*}
Then,
\begin{align*}
   & H^{\ve+\delta(\ol{\ve})}_{\min}(A|B)_{\rho} \\
   \geq& \max_{\wt{\rho} \in \Bvec{\ve + \delta(\ol{\ve})}(\rho)} \left[ H_{\min}(A)_{\wt{\rho}} - I^{\uparrow}_{\max}(A\mcol B)_{\wt{\rho}} \right] \\
    \geq& \max_{\omega \in \Bvec{\ve}(\rho)} \left[ \max_{\Pi} \left[ H_{\min}(A)_{\Pi\omega \Pi} - I^{\uparrow}_{\max}(A\mcol B)_{\Pi \omega \Pi} \right] \right] \ ,
\end{align*}
where the new maximization is over $0 \leq \Pi_{A} \leq \mbb{1}_{A}$ such that $\Pi\omega\Pi \approx_{\delta(\ol{\ve})} \omega$. This is a lower bound because we restricted smoothing to $\Bve$, so $\omega \approx_{\ve} \rho$ which using purified distance is a metric implies $\Pi \omega \Pi \approx_{\ve + \delta(\ol{\ve})} \rho$. Let $\omega^{\star} \in \Bve(\rho) \cap \Density(A \otimes B)$ be the optimizer of $I^{\ve}_{\max}(A\mcol B)_{\rho}$ which is normalized without loss of generality \cite[Lemma 22]{Ciganovic-2013a}. Then,
\begin{align*}
    H^{\ve+\delta(\ol{\ve})}_{\min}(A|B)_{\rho}
    \geq &  \max_{\Pi} \left[ H_{\min}(A)_{\Pi\omega^{\star} \Pi} - I^{\uparrow}_{\max}(A\mcol B)_{\Pi \omega^{\star} \Pi} \right] \\
    \geq &  \max_{\Pi} \left[ H_{\min}(A)_{\Pi\omega^{\star} \Pi}\right] - I^{\uparrow}_{\max}(A\mcol B)_{\omega^{\star}} \ , 
\end{align*}
where the first we have chosen $\omega^{\star}$ rather than maximizing and the second we have used that $D_{\max}$ actually satisfies data-processing for CPTNI maps. Then as $\omega^{\star}$ is normalized and we range over $\Pi$ such that $\Pi \omega^{\star} \Pi \approx_{\delta(\ol{\ve})} \omega^{\star}$, we know by Proposition \ref{prop:min-entropy-measure-bound} that we can bound the min-entropic term
\begin{align*}
    H^{\ve+\delta(\ol{\ve})}_{\min}(A|B)_{\rho} &  \geq H^{\ol{\ve}}_{\min}(A)_{\omega^{\star}} - I^{\uparrow}_{\max}(A\mcol B)_{\omega^{\star}} \\
    & \geq H^{\ol{\ve}-\ve}_{\min}(A)_{\rho} - I^{\uparrow,\ve}_{\max}(A\mcol B)_{\rho}\ , 
\end{align*}
where the first line is using $\omega^{\star}$ was the optimizer for $I^{\uparrow,\ve}_{\max}(A\mcol B)_{\rho}$. The second line is because if $\wt{\rho} \in \Bvec{\ol{\ve}-\ve}(\rho)$, as $\omega^{\star} \approx_{\ve} \rho$, we can conclude $\wt{\rho} \approx_{\ol{\ve} } \omega^{\star}$ and thus is included in the previous line's optimization. As smooth min-entropy is maximized, this suffices. Re-ordering the terms completes the proof.
\end{proof}
As a corollary, we have the following lower bound on SMCI.
\begin{corollary}
Let $\rho_{AC} \in \Density(A \otimes C)$. Let $0 < \ve < \ol{\ve} < 1$ such that $\ve + \delta(\ol{\ve})$ where $\delta(\ol{\ve}) := 2\sqrt{\ol{\ve}(1-\ol{\ve})}$. Then it holds,
\begin{align*}
C^{\ve}_{\max}(A\mcol C)_{\rho}
\geq & H^{\ol{\ve}-\ve}_{\min}(AC)_{\rho} \\
& \hspace{5mm} - \max_{\wt{\rho} \in \Bvec{\ve + \delta(\ve)}(\rho)} \max_{A-X-C} H_{\min}(AC|X) \ , 
\end{align*}
\end{corollary}
\begin{proof}
The proof is effectively the same as the previous proposition except one must keep track of the minimiziation over Markov chain extensions along with the projection. One will need to minimize over the Markov chain extension under the projection $\Pi$ and maintain the Markov chain property, i.e. consider a $\min_{(\Pi\omega\Pi)_{A-B-C}}  \, I^{\uparrow}_{\max}(AC\mcol B)_{\Pi \omega \Pi}$ term where we still demand $\Pi\omega\Pi \approx_{\delta(\ol{\ve})} \omega$. Note if the feasible set is empty, then this term is infinite and so, as we subtract it, in this setting the lower bound trivially holds, so the proof works when such a $\Pi$ exists or does not.
\end{proof}
The upper bound chain rule for $I^{\ve}_{\max}$ is already sufficient, so we just state it.
\begin{proposition}\label{prop:Imax-ub-chain-rule}(\cite[Lemma B.12]{Berta-2011a}) Let $\ve \in (0,1)$. Then,
\begin{align*}
 I^{\uparrow,\ve}_{\max}(A\mcol B)_{\rho} \leq &   H^{\ve^{2}/48}_{\max}(A)_{\rho} \\
 & \hspace{5mm} - H^{\ve^{2}/48}_{\min}(A|B)_{\rho} - 2\log(\ve^{2}/24) \ .
\end{align*}
\end{proposition}
\begin{corollary}
Let $\ve \in (0,1)$ and $\rho_{AC} \in \Density(A\otimes C)$, then
\begin{align*}
	C^{\ve}_{\max}(A\mcol C)_{\rho} & \leq H^{\ve^{2}/48}_{\max}(AC)_{\rho} - \max_{A-X-C} H^{\ve^{2}/48}_{\min}(AC|X)_{\rho} \\
	&\hspace{40mm} - 2\log(\ve^{2}/24)\ . 
\end{align*}
\end{corollary}
\begin{proof}
The proof is the same as the previous proposition except you push the minimization over Markov extensions through and note the minus sign flips the minimization into a maximization.
\end{proof}
It is clear that the chain rules for $I^{\uparrow,\ve}_{\max}$ can then be used to establish a strong AEP for $I^{\uparrow,\ve}_{\max}$ as we quickly show.
\begin{proof}[Proof of Theorem \ref{thm:strong-AEP-for-Imax}]
Fix $\ol{\ve},\delta(\ol{\ve})$ that satisfy Proposition \ref{prop:Imax-lb-chain-rule} to get the inequalities
\begin{align*}
& H^{\ol{\ve}-\ve}_{\min}(A^{\otimes n})_{\rho^{\otimes n}} - H^{\ve + \delta(\ol{\ve})}_{\min}(A^{n}|B^{n})_{\rho^{\otimes n}} \\
 \leq& I^{\uparrow,\ve}_{\max}(A^{n}\mcol B^{n})_{\rho^{\otimes n}} \\
 \leq& H^{\ve^{2}/48}_{\max}(A^{n})_{\rho^{\otimes n}} - H^{\ve^{2}/48}_{\min}(A^{n}|B^{n})_{\rho^{\otimes n}} - 2\log(\ve^{2}/24) \ . 
\end{align*}
Dividing by $n$ and taking the limit as $n \to \infty$, using the AEP for smooth min and max-entropies \cite{Tomamichel-2015a},
\begin{align*}
H(A)_{\rho} - H(A|B)_{\rho}
\leq&  \lim_{n \to \infty} \left[ I^{\uparrow,\ve}_{\max}(A^{n}\mcol B^{n})_{\rho^{\otimes n}} \right] \\ 
\leq& H(A)_{\rho} - H(A|B)_{\rho} \ . 
\end{align*}
Noting that $H(A) - H(A|B) = I(A\mcol B)$ by standard chain rules completes the proof.
\end{proof}

This general proof method then gets us the following.
\begin{proposition}
Let $\rho_{AC} \in \Density(A \otimes C)$. Let $\ve \in (0,1)$, $\delta \in (0,\ve)$, $\ol{\ve} \in (\ve,1)$ such that $0 < \ve + \delta(\ol(\ve)) < 1$ where $\delta(\ol{\ve}):= 2\sqrt{\ol{\ve}(1-\ol{\ve})}$. Then
\begin{align*}
& H(AC)-  \lim_{n \to \infty} \left[\frac{1}{n} \max_{\wt{\rho}\in\Bvec{\ve+\delta(\ol{\ve})}(\rho^{\otimes n})} \, \max_{A^{n}-X-C^{n}}  H_{\min}(A^{n}C^{n}|X) \right ] \\
& \leq \lim_{n \to \infty}\left[\frac{1}{n} C^{F,\ve}_{\max}(A^{n}\mcol C^{n})_{\rho^{\otimes n}}\right] \\
& \leq H(AC) - \lim_{n \to \infty} \left[ \max_{A^{n}-X-C^{n}} \frac{1}{n} H^{\ve - \delta}_{\min}(A^{n}C^{n}|X)_{\rho} \right ] \ ,
\end{align*}
where we remind the reader $C^{F,\ve}_{\max}$ is the smooth flipped max Wyner common information, Eq.~\eqref{eq:smooth-flipped-MCI}.
\end{proposition}
\begin{proof}
The proof method is the same as the previous proposition.
\end{proof}
Note that we don't particularly care about the upper bound as our achievability result for $C^{\ve}_{\max}(A \mcol C)$ held for all $\ve \in (0,1)$ (Lemma \ref{lem:max-wyner-common-info-achievability}). Therefore, our interest is in if the lower bound in the previous proposition can upper bound the common information in the regularized limit. To get Proposition \ref{prop:open-problem}, note that by a standard chain rule,
\begin{align*} 
C(A\mcol C)_{\rho} =& \min_{A-X-C} I(AC\mcol X)_{\rho} \\
 =& H(AC) - \max_{A-X-C} H(AC|X)_{\rho} \ ,
\end{align*} 
so to acquire the asymptotic bound, it would suffice that 
\begin{align*}
    & H(AC) - \max_{A-X-C} H(AC|X)_{\rho} \\
    \leq & H(AC)\\
    &\hspace{5mm} - \lim_{n \to \infty} \left[\frac{1}{n} \max_{\wt{\rho}\in\Bvec{\ve+\delta(\ol{\ve})}(\rho^{\otimes n})} \, \max_{A^{n}-X-C^{n}}  H_{\min}(A^{n}C^{n}|X) \right ] \ ,
\end{align*}
which by canceling and multiplying by negative one gets us the term in Proposition \ref{prop:open-problem}.

\paragraph{Difficulty of Applying Entropy Accumulation to establish a Strong Converse} As noted in the main text, one can decompose $A^{n}-X-C^{n}$ into a sequence of Markov chain conditions, $A_{i}-A^{i-1}_{1}XC^{i-1}_{1}-C_{i}$ for all $i \in [n]$ \cite{Hayashi-2006a}. This might suggest that it satisfies the constraints to apply some form of the entropy accumulation theorem (EAT) \cite{Dupuis-2020a,Metger-2022a}. First, we note that this is conceptually unlikely a good strategy as the Markov chain conditions given above do not satisfy the locality constraints of distributed source simulation. Nonetheless, as the EAT is about effective maps, this does not preclude its application. However, it will need access to $A^{i-1}_{1}C^{i-1}_{1}$ as side-information for generating the next round as well as being the previous entropy generating registers. It is not obvious that this can be done in the separable state case. This is because while we know $A^{n}-X-C^{n}$ is an extension of $\rho^{\otimes n}_{AC}$, it is not obvious from the structure of the Petz recovery map \cite{Wilde-2011a} nor the general structure of the recovery map for Markov chains specifically \cite{Sutter-2018a} that the recovery map $\cR_{C^{i-1}_{1}XA^{i-1}_{1} \to C^{i}_{1}XA^{i-1}_{1}}$ won't entangle some of the quantum systems. This may be viewed as the general difficulty of applying the EAT to this parallel setup.

We further note that we could restrict to a classical Markov chain problem, $X^{n}-Y-Z^{n}$, where registers can always be copied so we can properly define EAT maps $\cG_{i}$ per round. However, in this setting and by modifying the notation of some of the side-information registers for clarity, the generalised EAT \cite[Appendix A]{Metger-2022a} will be of the form
\begin{align*}
   H^{\ve}_{\max}(X^{n}Z^{n}|YE)_{\rho}  \leq \sum_{i} \max_{\ket{\omega}} H(X_{i}Z_{i}|X^{i-1}_{1}Z^{i-1}_{i}YE_{i})_{(\cG_{i}\otimes \id_{E_{i}})(\omega)} \ ,
\end{align*}
where $\omega$ is any purification of an input to $\cG_{i}$. The problem then is the LHS conditions on the quantum side-information register $E$ which we don't want to consider in our problem. Moreover, by strong subadditivity of smooth max-entropy, $H^{\ve}_{\max}(X^{n}Z^{n}|Y\wt{E}) \leq H^{\ve}_{\max}(X^{n}Z^{n}|Y)$, so this would require modifying the max-entropy version of the EAT to not include the quantum side-information. That is, it is not an immediate corollary of the EAT even in the fully classical setting.

\section*{On Variants of One-Shot Distributed Source Simulation}\label{sec:variants-of-DSS}
\begin{figure}
\centering
\begin{subfigure}[b]{0.95\columnwidth}
   \begin{center}
        \begin{tikzpicture}
            \tikzstyle{porte} = [draw=black!50, fill=black!20]
                \draw
                (0,0) node (qy) {$p_{X}$}
                ++(1.25,0) node[porte] (copy) {Copy}
                ++(1.5,0.75) node[porte] (Alice-Proc) {$\Phi_{X \to A}$}
                ++(1.5,0) node (A) {$A$}
                ++(0,-1.5) node (C) {$C$}
                ++(-1.5,0) node[porte] (Bob-Proc) {$\Psi_{X' \to C}$}
                ++(2.2,0.75) node (approx) {$\approx_{\ve}$}
                ++(0.8,0) node (pxz) {$\rho_{AC}$}
                ;
                \path[draw = black, ->] (qy) -- (copy); 
                \path[draw=black, ->] (copy) |- (Alice-Proc);
                \path[draw=black, ->] (Alice-Proc) -- (A);
                \path[draw=black, ->] (copy) |- (Bob-Proc);
                \path[draw=black, ->] (Bob-Proc) -- (C);
        \end{tikzpicture}
    \end{center}
   \caption{Correlation of Formation, $C^{\ve}_{F}$}
   \label{subfig:DSS} 
\end{subfigure}
\par\bigskip
\begin{subfigure}[b]{0.95\columnwidth}
\begin{center}
        \begin{tikzpicture}
            \tikzstyle{porte} = [draw=black!50, fill=black!20]
                \draw
                (0,0) node (px) {$p_{X}$}
                ++(1.75,0) node[porte] (prep) {$\Xi_{X \to ACX}$}
                ++(1.75,0.75) node (A) {$A$}
                ++(0,-1.5) node (C) {$C$}
                ++(0,0.75) node (X) {$X$}
                ++(0.75,0) node (approx) {$\approx_{\ve}$}
                ++(1,0) node (pxz) {$\rho_{A-X-C}$}
                ;
                \path[draw = black, ->] (px) -- (prep); 
                \draw[->] (prep) to[out=+90,in=+180] (A);
                \draw[->] (prep) to[out=-90,in=+180] (C);
                \path[draw = black, ->] (prep) -- (X);
        \end{tikzpicture}
    \end{center}
   \caption{EA Correlation of Formation, $\widehat{C}^{\ve}_{F}$}
   \label{subfig:EADSS}
\end{subfigure}
\par\bigskip	
\begin{subfigure}[b]{0.95\columnwidth}
   \begin{center}
        \begin{tikzpicture}
            \tikzstyle{porte} = [draw=black!50, fill=black!20]
                \draw
                (0,0) node (qy) {$p_{X}$}
                ++(1.25,0) node[porte] (copy) {Copy}
                ++(1.5,0.75) node[porte] (Alice-Proc) {$\Phi_{X \to A}$}
                ++(1.5,0) node (A) {$A$}
                ++(0,-1.5) node (C) {$C$}
                ++(-1.5,0) node[porte] (Bob-Proc) {$\Psi_{X' \to C}$}
                ++(2.2,0.75) node (approx) {$\approx_{\ve}$}
                ++(1,0) node (pxz) {$\rho_{A-X-C}$}
                ;
                \draw
                ++ (4.25,0) node (X) {$X$};
                \path[draw = black, ->] (qy) -- (copy); 
                \path[draw=black, ->] (copy) |- (Alice-Proc);
                \path[draw=black, ->] (Alice-Proc) -- (A);
                \path[draw=black, ->] (copy) |- (Bob-Proc);
                \path[draw=black, ->] (Bob-Proc) -- (C);
                \path[draw=black, ->] (copy) -- (X);
        \end{tikzpicture}
    \end{center}
   \caption{Private Correlation of Formation, $\wt{C}^{\ve}_{F}$}
   \label{subfig:privateDSS}
\end{subfigure}
\caption{The three correlation of formations and their corresponding tasks: (a) Correlation of formation captures the amount of randomness for distributed source simulation. (b) Entanglement-assisted correlation of formation captures the amount of broadcasted randomness needed such that there exists a set of (possibly entangled) states $\{\wt{\rho}_{AC}^{x}\}_{x \in \cX}$ to distribute so that the entire output is approximately indistinguishable from a distributed source simulation implementation of the target state. (c) Private correlation of formation measures the amount of randomness needed for a distributed source simulation protocol so that if the randomness were to be leaked it would be approximately indistinguishable from an exact distributed source simulation of the target state with leaked classical information.}
\label{fig:correlation-measure-depictions}
\end{figure}
In this section we introduce the variants of one-shot distributed source simulation that are better directly addressed by Shen et al. \cite{Shen-2022a}, establish their properties, provide the proof  of the one-shot rate (Proposition \ref{prop:one-shot-rate-for-EA-DSS}) and the asymptotic equipartition property needed for proving Theorem \ref{thm:first-order-asymptotic-equivalence}.

\subsection{Definitions and Equivalence}
Here we define two related operational quantities for separable states: ``private correlation of formation" and ``entanglement-assisted correlation of formation." The former has a stronger error criterion than one-shot DSS that captures the idea of the $X$ register also being public, and the latter allows for a type of entanglement-assistance in this stronger setting. One can think of this scenario where the $X$ register is public as a setting where an adversary wishes to verify the network structure is being respected. We then show these are in fact the same task--- i.e. entanglement-assistance does not help under this stronger error criterion. Given this result, for the rest of this appendix we will consider the variants of one-shot DSS, we focus on $\widehat{C}^{\ve}_{F}(A \mcol C)_{\rho}$ as it is easier to see the network structure. We note in the subsequent definitions we use trace distance.

\begin{definition}\label{def:private-DSS}
Let $\ve \in (0,1)$ and $\rho_{AC} \in \Density(A \otimes C)$. The one-shot private correlation of formation is defined as
\begin{equation}\label{eq:private-form-defn}
	\begin{aligned}
    &\wt{C}^{\ve}_{F}(A\mcol C)_{\rho} \\
     :=& \inf_{\wt{\rho}_{A-X-C}}\Bigg\{H_{0}(X)_{\wt{\rho}} : \exists \rho_{A-X-C} : \\
     & \hspace{20mm} \frac{1}{2}\|\wt{\rho}_{A-X-C}-\rho_{A-X-C}\|_{1} \leq \ve \Bigg\} \ .
	\end{aligned}
\end{equation} 
\end{definition}

\begin{definition}\label{def:EA-assisted-corr-of-form}
Let $\ve \in (0,1)$ and $\rho_{AC} \in \Density(A \otimes C)$. The one-shot entanglement-assisted correlation of formation is defined as
\begin{equation}\label{eq:EA-form-defn}
	\begin{aligned}
    &\widehat{C}^{\ve}_{F}(A\mcol C)_{\rho} \\ \coloneq&\inf_{\wt{\rho}_{AXC}} \Bigg\{H_{0}(X)_{\wt{\rho}} : \exists \rho_{A-X-C} : \\
     & \hspace{20mm} \frac{1}{2}\|\wt{\rho}_{AXC} - \rho_{A-X-C} \|_{1} \leq \ve  \Bigg\} \ .
     \end{aligned}
\end{equation}
\end{definition}
We remark that the sense in which both of these tasks are a `distributed' task are presented in Fig. \ref{fig:correlation-measure-depictions}. The private correlation of formation can be viewed either mathematically as a stronger error criterion than one-shot correlation of formation or operationally as a case where one is concerned the classical random variable $X$ will be leaked. That is, one imagines there is a distinguisher that not has access to the entire implementation. It is in this sense that we call it `private.' However, the motivation for entanglement-assistance would be to hope that a sender that makes use of entanglement may outperform one that cannot in this private setting. We now reduce these two quantities to being characterized by the same entropic quantity, which in particular implies entanglement assistance is not useful in this private setting. For this reason, we will focus on just the private correlation of formation variant, $\wt{C}^{\ve}_{F}$, for the rest of the work for clarity and notational simplicity.
\begin{lemma}\label{lem:characterization-of-variant-of-DSS}
For $\rho_{AC} \in \Density(A \otimes C)$,
$$ \wt{C}^{\ve}_{F}(A\mcol C)_{\rho}= \inf_{\rho_{A-X-C}} H^{\TD,\ve}_{0}(X)_{\rho}  = \widehat{C}^{\ve}_{F}(A \mcol C)_{\rho}  \ , $$
\end{lemma}
\noindent where $H^{\TD,\ve}_{0}(X)_{p} := \min \{H_{0}(X)_{q} : \frac{1}{2}\|p_{X} - q_{X}\|_{1} \leq \ve \, \& \, q \in \cP(\cX) \}$. In particular, this means these two settings are equivalent and the (sequence of) optimizer(s) is a (sequence of) Markov chain extension(s) $\wt{\rho}_{A-X-C}$.
\begin{proof}
	In the case no quantum Markov chain extension exists, by our convention, all optimizations take the value $+\infty$, so we focus on the case where there is an extension, which, by Lemma \ref{lem:sep-marg-of-QMC}, means $\rho_{AC} \in \Sep\Density(A \mcol C)$. We now prove the result for $\widehat{C}^{\ve}_{F}(A\mcol C)_{\rho}$ as the proof for $\wt{C}^{\ve}_{F}(A \mcol C)_{\rho}$ is effectively identical. Let $\wt{\rho}_{AXC}$ be feasible for $\wt{C}^{\ve}_{F}(A\mcol C)_{\rho}$. This means 
	$$\ve \geq \frac{1}{2}\|\wt{\rho}_{AXC} - \rho_{A-X-C}\|_{1}  \geq \frac{1}{2}\|\wt{\rho}_{X} - \rho_{X}\|_{1} \ , $$
where we have used the contractivity of the $L_{1}$-norm under CPTP maps, namely the partial trace. As this construction holds for any feasible point, this implies
	$$\inf_{\rho_{A-X-C}} H^{\Lone,\ve}_{0}(X)_{\rho} \leq \widehat{C}^{\ve}_{F}(A\mcol C)_{\rho} \ . $$ 

	Similarly, let $\widehat{\rho}_{X}$ such that there a QMC extension $\rho_{A-X-C}'$ of $\rho_{AC}$ such that $\frac{1}{2}\|\widehat{\rho}_{X} - \rho_{X}'\|_{1} \leq \ve$. This is then feasible for $\inf_{\rho_{A-X-C}} H^{\TD,\ve}_{0}(X)_{\rho}$. Identify $\cR,\ol{\cR}$ as the recovery maps of $\rho_{A-X-C}'$, which exist by Theorem \ref{thm:QMC-equivalences}. Define the preparation channels $\cE(\cdot) := (\Tr_{X} \circ \cR)(\cdot)$, $\cF(\cdot) := (\Tr_{X} \circ \ol{\cR}(\cdot))$. Then we may define $\widehat{\rho}_{A-X-C} := (\cE \otimes \cF)(\widehat{p}_{X})$. Thus, we have
	\begin{align*}
	 & \frac{1}{2}\|\widehat{\rho}_{A-X-C} - \rho_{A-X-C}' \|_{1} \\
	 =& \frac{1}{2}\sum_{x} \|\widehat{p}(x) {\rho'}^{x}_{A} \otimes {\rho'}^{x}_{C} - p'(x) {\rho'}^{x}_{A} \otimes {\rho'}^{x}_{C}\|_{1} \\
	 =& \frac{1}{2}\sum_{x} |\widehat{p}(x) - p'(x)| = \|\widehat{p}_{X} - p_{X}'\|_{1} \ ,
	\end{align*}
where we used in each case the conditional state is the same. Thus, we have 
	$$\inf_{\rho_{A-X-C}} H^{\Lone,\ve}_{0}(X)_{\rho} \geq \widehat{C}^{\ve}_{F}(A\mcol C)_{\rho} \ . $$ 
This completes the proof for $\widehat{C}^{\ve}_{F}$, and note that we never used the non-Markov structure of the optimizer $\wt{\rho}_{AXC}$ for  $\widehat{C}^{\ve}_{F}$, so it will also work for $\wt{C}^{\ve}_{F}$. 
\end{proof}

We stress the following point about the previous proposition: one would expect there are cases where the gap between $C^{\ve}_{F}$ and the other two quantities is large. In the case of $C^{\ve}_{U,F}$ this follows from approximating the non-uniform distribution as seen in Proposition \ref{prop:QMC-unif-close-enough}. In the case of $\wt{C}^{\ve}_{F}$, we expect this can be large because in $\wt{C}^{\ve}_{F}$ the distinguisher has access to the $X$ register. That is to say, in general one expects there to be a non-trivial difference between
$$  \min_{\wt{\rho}_{A-X-C}} \left\|\Tr_{X}(\wt{\rho}_{A-X-C}) - \rho_{AC} \right\|_{1}$$
and
$$ \min_{\substack{\wt{\rho}_{A-X-C}\\ \rho_{A-X-C}}} \sum_{x} \left\|\wt{p}(x)\wt{\rho}^{x}_{A} \otimes \wt{\rho}^{x}_{C}  - p(x)\rho_{A}^{x} \otimes \rho_{C}^{x} \right\|_{1} \ . $$ 

We end by noting that because $\wt{C}^{\ve}_{F}(A \mcol C)_{\rho}$ is only well-defined for separable states, so it takes a finite value under more restricted conditions than the correlation of formation and its uniform version that we considered in the main text.
\begin{proposition}\label{prop:distinction-bw-op-measures}
Let $\rho \in \Density(A \otimes C)$. 
\begin{enumerate}
\item For $\varepsilon \in [0,1]$, $\wt{C}^{\ve}_{F}(A\mcol C)$ are finite if and only if $\rho_{AC} \in \Sep\Density(A\mcol C)$. 
\item In contrast, $C^{\ve}_{F}(A\mcol C)_{\rho}$ is finite if and only if $E_{T}(A\mcol C)_{\rho} \leq \ve/2$. Likewise, $C^{\ve}_{U,F}(A\mcol C)_{\rho}$ is finite if and only if there exists $\delta \in (0,\ve)$ such that $E_{T}(A\mcol C)_{\rho} \leq (\ve - \delta)/2$.
\end{enumerate}
\end{proposition}
\begin{proof}
We begin with Item 1. The only if direction is immediate because if there exists an appropriate QMC extension, then the state is separable by Lemma \ref{lem:sep-marg-of-QMC}. Likewise, by Lemma \ref{lem:sep-marg-of-QMC}, if $\rho_{AC} \in \Sep\Density(A\mcol C)$, there exists a QMC extension of $\rho_{AC}$, $\rho_{A-X-C} = \sum_{x} p(x) \rho^{x}_{A} \otimes \dyad{x} \otimes \rho^{x}_{C}$ for some finite alphabet $\cX$. This state is then feasible for $\wt{C}^{\ve}_{F}$. Note none of this has relied on the choice of $\varepsilon$.

By definition, $C^{\ve}_{F}(A\mcol C)_{\rho} < +\infty$ only if there is a $\wt{\rho}_{AC} \approx_{\ve}^{L1} \rho_{AC}$ that admits a QMC extension. By Lemma \ref{lem:sep-marg-of-QMC}, this must be $\wt{\rho}_{AC} \in \Sep\Density(A\mcol C)$. Thus, it must be the case that there is $\wt{\rho}_{AC} \in \Sep\Density(A\mcol C)$ such that $\wt{\rho}_{AC} \approx_{\ve}^{L1} \rho_{AC}$. By definition of trace distance, this means it must be the case $E_{T}(A\mcol C)_{\rho} \leq \ve/2$. For the uniformity claim, note that if such a $\delta > 0$ exists then by Proposition \ref{prop:QMC-unif-close-enough} we may do the same argument again. This completes the proof.
\end{proof}
Therefore we see $C^{\ve}_{F}$ is a fundamentally distinct measure from $\wt{C}^{\ve}_{F}$ as when they are finite is not even in agreement.

\subsection{Hypothesis Testing Common Information}
Here we introduce the `hypothesis testing alternative common information.' As the name suggests, it is derived from the hypothesis testing divergence,
\begin{equation}\label{eq:hyp-test-div-defn}
\begin{aligned}
 \exp(-D^{\ve}_{h}(\rho\Vert\sigma)) 
 :=  \inf_{0 \leq \Lambda \leq \mbb{1}} \{ \Tr[\Lambda \sigma] : \Tr[\Lambda\rho] \geq 1 - \ve \ \} \ .
\end{aligned},
\end{equation}
and the mutual information quantity it induces
\begin{equation}\label{eq:hyp-test-MI}
	I_{h}^{\ve}(X \mcol B)_{\rho} := D_{h}^{\ve}(\rho_{XB} \Vert \rho_{X} \otimes \rho_{B}) \ . 
\end{equation}
We refer the reader to \cite{Khatri-2020a} for a review on the properties of the hypothesis testing divergence, but we do remark that unlike $I^{\uparrow,\ve}_{\max}(A\mcol B)$, the order of registers does not matter for hypothesis testing mutual information.

\begin{definition}\label{def:hyp-test-common-info}
Let $\rho_{AC} \in \Density(A \otimes C)$. The hypothesis testing common information is given by
\begin{equation}\label{eq:hyp-test-common-info} \wt{C}^{\ve}_{h}(A\mcol C)_{\rho} := \min_{A-X-C} I^{\ve}_{h}(AC\mcol X)_{\rho} \ ,
\end{equation}
where $|X| \leq |A|^{2}|C|^{2} + \edit{2}$.
\end{definition} 
The key difference between the max Wyner common informations (Definition \ref{def:max-Wyner-CIs}) is that the smoothing is not applied to the state. This results in taking the Markov chain extension in some sense `before' the smoothing, which is why it is only directly relevant for these variants of distributed source simulation. The cardinality bound in the definition follows from the same argument as Lemma \ref{lem:card-bound-for-doubly-max} as we now show.

\begin{proposition}\label{prop:card-bounds-for-hypothesis-testing-divergence}
	Let $\rho_{AC} \in \Sep\Density(A:C)$. Then
	$$ \inf_{\rho_{A-X-C}} I^{\ve}_{h}(AC:X) \geq \underset{\substack{\rho_{A-X-C} : \\
	|\cX| \leq |A|^{2}|C|^{2} \edit{+ 2}}}{\min} I^{\ve}_{h}(AC:X)_{\rho_{A-X-C}} \ . $$
\end{proposition}
\begin{proof}
The proof is similar to that of Lemma \ref{lem:card-bound-for-doubly-max}, so we just highlight the differences. Let $\rho_{A-X-C} = \sum_{x \in \cX} p(x) \sigma^{x}_{A} \otimes \dyad{x} \otimes \sigma^{x}_{C}$ be a QMC extension of $\rho_{AC}$. Let $0 \leq \Lambda \leq \mbb{1}$ be an arbitrary POVM element such that $\tr[\rho_{A-X-C}\Lambda]\geq 1-\ve$.  Note that since $X$ is classical, without loss of generality we can assume that $\Lambda=\sum_x\Lambda_{AC}^x\otimes \dyad{x}_X$.  Hence, for each fixed set of $\{\Lambda_{AC}^x\}$ with $0\leq \Lambda_{AC}^x\leq \mbb{I}$, fix the numbers
\begin{align}
k_1&=\sum_{x\in\mc{X}}p(x)\tr[\rho_{AC}\Lambda_{AC}^x]=\tr[\rho_{AC}\otimes\rho_X\Lambda]\notag\\
k_2&=\sum_{x\in\mc{X}}p(x)\tr[\sigma_A^x\otimes\sigma_C^x\Lambda^x]=\tr[\rho_{A-X-C}\Lambda],
\end{align}
and define the convex set
\begin{align*}
	\mbb{P}:= \left\{\begin{array}{ll}
	q(x) \geq 0:& \rho_{AC} = \sum_{x} q(x)\sigma^{x}_{A} \otimes \sigma^{x}_{C} \\
	  & k_1=\sum_{x\in\mc{X}}q(x)\tr[\rho_{AC}\Lambda_{AC}^x] \\
	 & k_2=\sum_{x\in\mc{X}}q(x)\tr[\sigma_A^x\otimes\sigma_C^x\Lambda^x] \end{array} \right\} \ .
\end{align*}
By the same argument as before, it follows that there are distributions in $\mbb{P}$ (namely its extreme points) having no more than $|A|^{2}|C|^{2} \edit{+ 2}$ nonzero elements.  The proposition then readily follows.
\end{proof}

\subsection{One-Shot Rate}
We now derive the one-shot rate (Proposition \ref{prop:one-shot-rate-for-EA-DSS}). The achievability is a near immediate application of Shen et al. \cite{Shen-2022a} and the converse follows a similar argument to what we did in the main text for correlation of formation proper, but we make use of Lemma \ref{lem:characterization-of-variant-of-DSS}, which will allow us to keep track of what approximations introduce looseness in our bound.
\begin{proof}[Proof of Proposition \ref{prop:one-shot-rate-for-EA-DSS}]
\textit{Achievability:} Using the definition of error in soft-covering from \cite{Shen-2022a}, which is in terms of trace distance, one can use the achievable rate given for soft-covering, $I^{1-\ve+3\delta}_{h}(X \mcol AC)_{\rho} + \log(\nu^{2}/\delta^{4})$ for any Markov chain extension of $\rho_{AC}$, $\rho_{A-X-C}$, which exists as we assumed the state was separable. Then minimizing over the Markov chain extension and using Definition \ref{def:hyp-test-common-info} completes the achievability proof. \\

\textit{Converse:} As Lemma \ref{lem:characterization-of-variant-of-DSS} includes an infimization, we start from an arbitrary feasible Markov chain extension $\rho_{A-X-C}$ and will use that in Lemma \ref{lem:characterization-of-variant-of-DSS} we showed the optimizer of $H_{0}^{\Lone}(X)_{\rho_{A-X-C}}$ is a Markov chain $\wt{\rho}_{A-X-C}$ defined using the preparation maps induced by the recoverability maps of $\rho_{A-X-C}$, $\cR,\ol{\cR}$.
\begin{align}
	& H_{0}^{\TD}(X)_{\rho} \notag \\
	=& \min \{H_{0}(X)_{\wt{\rho}} : \frac{1}{2}\|\wt{p}_{X} - p_{X}\|_{1} \leq \ve \, \& \, \wt{p}_{X} \in \cP(\cX) \} \notag \\
	=& \min \left\{\begin{array}{ll} I_{\max}(X'X'':X)_{\chi^{\wt{p}}} :& \frac{1}{2}\|\wt{p}_{X} - p_{X}\|_{1} \leq \ve \\
	& \wt{p}_{X} \in \cP(\cX) \end{array} \right\} \notag \\
	\geq& \min \left\{\begin{array}{ll} I_{\max}(AC:X)_{(\cR \otimes \ol{\cR})(\chi^{\wt{p}})} :& \frac{1}{2}\|\wt{p}_{X} - p_{X}\|_{1} \leq \ve \\
	& \wt{p}_{X} \in \cP(\cX) \end{array} \right\} \notag \\
	\geq& \min \{I_{\max}(AC:X)_{(\cR \otimes \ol{\cR})(\chi^{\wt{p}})} :  P(\wt{\rho}_{AXC},\rho_{A-X-C}) \leq \sqrt{2\ve} \} \notag \\
	=& I^{\uparrow,\sqrt{2\ve}}_{\max}(AC:X)_{\rho} \ , \label{eq:alternative-max-converse-bound}
\end{align}
where the first equality is by definition, the second is using Proposition \ref{prop:Cmax-of-perfectly-correlated}, the first inequality is by data processing. The second inequality is using $\frac{1}{2}\|\wt{p}_{X} - p_{X}\|_{1} \leq \ve$ implies $P(\wt{p}_{X},p_{X}) \leq \sqrt{2\ve}$, and, by the Uhlmann's theorem for generalized fidelity \cite{Tomamichel-2015a}, there exists an extension of $\wt{p}_{X}$, $\wt{\rho}_{AXC}$ such that $P(\wt{\rho}_{AXC},\rho_{A-X-C}) \leq \sqrt{\ve}$. As we started by an arbitrary QMC extension, we may infimize over QMC extensions, but by Lemma \ref{lem:cardinality-bound-for-SMCI}, we know the there is a cardinality bound. Finally, to get the hypothesis testing bound, one defines $\sqrt{1-\ol{\ve}} = \sqrt{2\ve}$ and applies \cite[Eqn. 22]{Tomamichel-2013a}.
\end{proof}

We note the above result seems to not be tight because of the need to convert between purified distance and trace distance. If one were to add a restriction that the $X$ register of the strategy was uniform, one would be able to more directly relate to hypothesis testing divergence as the following result shows.
\begin{proposition}\label{prop:Hartley-MI}
    For perfectly correlated state $\chi^{|p}_{XX}$, 
    $$ I_{0}(X:X')_{\chi^{|p}} = -\log(\|p\|^{2}_{2}) \ .$$
  	Moreover, if $p$ is the uniform distribution, then $I_{0}(X:X')_{\chi^{|\pi}} = H_{0}(X)_{\chi^{|\pi}}$.
\end{proposition}
\begin{proof}
First,
    \begin{align*}
    &\exp(-I^{\ve}_{h}(X:X)_{\chi^{|p}}) \\ =& \inf\{\Tr[\Lambda \rho_{X}^{\otimes 2}] : \Tr[\Lambda \chi^{|p}_{X}] = 1-\ve \} \\
    =& \inf_{\vec{0} \leq q \leq \vec{1}} \big\{ \sum_{x \in \cX} q(x)p(x)^{2} : \sum_{x} q(x)p(x) = 1-\ve \big\} \ ,
    \end{align*}
    where in the second equality we have used that the only elements relevant are $\dyad{x}^{\otimes 2}$ for the constraint, so we may as well restrict our POVM to those and the constraints just mean $q$ is restricted between the all zeros and all ones vectors. 
    
    Setting $\ve = 0$, we obtain $\exp(-I_{h}^{0}(X:X)_{\chi^{|p}}) = \sum_{x} p(x)^{2} = \|p\|_{2}^{2} \ ,$ where $\|\cdot \|_{2}$ is the $2$-norm. In the case that $p=\pi$, i.e. is the uniform distribution, a direct calculation will verify $I^{0}_{h}(X\mcol X')_{\chi^{|\pi}} = \log(|\cX|) = H_{0}(X)_{\chi^{|\pi}}$.
\end{proof}

\subsection{On Asymptotics}
As stated in the main text, we are able to establish equivalence to first-order of our variants and distributed source simulation proper (Theorem \ref{thm:first-order-asymptotic-equivalence}).

This follows from deriving a weak asymptotic equipartition property for $C_{h}^{\ve}(A \mcol C)_{\rho}$. Unlike for smooth max Wyner common information, the achievability is very easy in this case. The converse proof is very similar to that in the main text. We provide both for completeness.
\begin{lemma}
Let $\rho_{AC} \in \Sep\Density(A\mcol C)$. Then,
\begin{align*}
	\lim_{n \to \infty} \left[\frac{1}{n} C^{\ve}_{h}(A^{n} \mcol C^{n})_{\rho^{\otimes n}} \right] = C(A\mcol C)_{\rho} \ .
\end{align*}
\end{lemma}
\begin{proof}
	\textit{Achievability} Let $\rho_{A-X-C}$ be the Markov chain extension that obtains the Wyner common information \edit{\eqref{eq:Wyner-common-information}, which is obtained by Proposition \ref{prop:card-bound-common-info}.} Then
	\begin{align*}
		C^{\ve}_{h}(A^{n} \mcol C^{n})_{\rho^{\otimes n}} \leq& I^{\ve}_{h}(X^{n} \mcol A^{n}C^{n})_{\rho^{\otimes n}_{A-X-C}} \\
		=& nI(X:AC)_{\rho_{A-X-C}} + o(n) \\
		=& C(A \mcol C)_{\rho} + o(n) \ ,
	\end{align*}
	where the inequality is our choice of a Markov chain extension, the first equality is the asymptotic expansion of hypothesis testing divergence on an i.i.d. inputs \cite{Tomamichel-2013a}, and the second equality is definition of Wyner common information. Dividing by $n$ and taking the limit completes this direction.
	
	\textit{Converse} The proof is similar to that of Lemma \ref{lem:SMCI-converse}, so we only note the major differences. Note the bound is trivial whenever $\rho_{AC}$ is not separable by definition of $\wt{C}^{\ve}_{\max}$. Thus, we assume $\rho_{AC} \in \Sep\Density(A\mcol C)$. Next, note that \eqref{eq:alternative-max-converse-bound} tells us that we could have also gotten a one-shot converse in terms of $\wt{C}^{\sqrt{2\ve}}_{\max}(A\mcol C)_{\rho} := \min_{A-X-C} I^{\sqrt{2\ve}}_{\max}(A \mcol C)_{\rho}$. For simplicity, this is the quantity we use which we can then immediately lower bound the max mutual information with the mutual information as in Lemma \ref{lem:SMCI-converse}. Now, the state $\sigma_{A^{n}XC^{n}}$ that optimizes $\wt{C}^{\sqrt{2\ve}}_{\max}(A^{n}\mcol C^{n})_{\rho^{\otimes n}}$ is not necessarily a Markov chain, though it is classical on the auxiliary register as follows from data-processing. This follows as the smoothing is done on the choice of Markov chain. Nonetheless, using Lemma \ref{lem:general-conditional-chain-rule}, we establish 
\begin{equation}\label{eq:alt-cmax-converse-step-1}
\begin{aligned}
\frac{1}{n} I(A^{n}C^{n}\mcol X)_{\sigma} \geq& \frac{1}{n} H(A^{n}C^{n})_{\sigma}
    - \frac{1}{n} \sum_{i=1}^{n} H(A_{i}C_{i})_{\sigma} \\
    & + \frac{1}{n} \sum_{i=1}^{n} I(A_{i}C_{i}:X)_{\sigma} \ ,
\end{aligned}
\end{equation}
where the inequality is because we no longer have that $\sigma$ is a Markov chain.

For the second step, there is more to change as now we don't know if $\sigma_{A_{i}XC_{i}}$ is ever a Markov chain. Let $\widehat{\rho}_{A^{n}-X-C^{n}}$ be a (in case it is not unique) Markov chain that corresponds to the optimizer $\sigma$ in the sense that $\sigma \in \Bve(\widehat{\rho})$ where $\widehat{\rho}_{A^{n}-X-C^{n}}$ is a QMC extension of $\rho_{AC}^{\otimes n}$. This state must exist by definition of $\wt{C}^{\ve}_{\max}$ and that we are focusing on $\rho_{AC}\in  \Sep\Density(A\mcol C)$. This means $\widehat{\rho}_{A^{n}C^{n}} = \rho_{AC}^{\otimes n}$, so $\widehat{\rho}_{A_{i}C_{i}} = \rho_{AC}$ for all $i \in [n]$. It also means $\widehat{\rho}_{A_{i}XC_{i}}$ is a $A_{i} - X - C_{i}$ Markov chain for all $i \in [n]$. It follows $\tau_{AXIC} = \frac{1}{n} \sum_{i \in [n]} \dyad{i}_{I} \otimes \widehat{\rho}_{A_{i}XC_{i}}$ is a $A-IX-C$ Markov chain as conditioned on $I$, the remaining state is a Markov chain conditioned on $X$. Moreover, $\tau_{AC} = \frac{1}{n} \sum_{i \in [n]} \Tr_{X} \widehat{\rho}_{A_{i}XC_{i}} = \frac{1}{n} \sum_{i \in [n]} \rho_{AC} = \rho_{AC}$. Thus, $\tau$ is a Markov chain extension of $\rho_{AC}$.
Therefore, we have
\begin{align*}
    C(A:C)_{\rho} 
    \leq & I(AC:XI)_{\tau} \\
    = & I(AC:I)_{\tau} + I(AC:X|I)_{\tau} \\
    =& I(AC:I)_{\tau} + \frac{1}{n} \sum_{i=1}^{n} I(A_{i}C_{i}:X)_{\widehat{\rho}} \\
    =& H(AC)_{\tau} - H(AC|I)_{\tau} + \frac{1}{n} \sum_{i=1}^{n} I(A_{i}C_{i}:X)_{\widehat{\rho}} \\
    =& H(AC)_{\rho} - \frac{1}{n} \sum_{i=1}^{n} H(A_{i}C_{i})_{\widehat{\rho}} + \frac{1}{n} \sum_{i=1}^{n} I(A_{i}C_{i}\mcol X)_{\widehat{\rho}} \ ,
\end{align*}
which implies
\begin{align*}
    - \frac{1}{n} \sum_{i=1}^{n} H(A_{i}C_{i})_{\widehat{\rho}} \geq&  C(A:C)_{\rho} - H(AC)_{\rho} -\frac{1}{n} \sum_{i=1}^{n} I(A_{i}C_{i}\mcol X)_{\widehat{\rho}} \ .
\end{align*}
The issue is now our first bound was in terms of $\sigma$ and this new bound is in terms of $\widehat{\rho}$. However, we can use the Alicki-Fannes-Winter (AFW) inequalities in the following manner. As $\widehat{\rho} \approx_{\ve} \sigma$, we have $\widehat{\rho}_{A_{i}BC_{i}} \approx_{\ve} \sigma_{A_{i}BC_{i}}$ for all $i \in [n]$ and likewise if we trace off $B$. Therefore using the AFW inequalities (for both unconditional entropy and mutual information \cite{Wilde-2011a}),
\begin{align*}
    &- \frac{1}{n}\sum_{i=1}^{n}H(A_{i}C_{i})_{\sigma} \\
     \geq& C(A\mcol C)_{\rho} - H(AC)_{\rho} - \frac{1}{n} \sum_{i=1}^{n} I(A_{i}C_{i}\mcol X)_{\sigma} \\
    & - 4\ve \log|AC|- h_{2}(\ve) - 2(\ve+1)\log(\ve+1) + \ve\log(\ve)  \ .
\end{align*}
Then we can plug this into \eqref{eq:alt-cmax-converse-step-1} and use Fannes-Audenaert inequality on $H(A^{n}C^{n})_{\sigma}$ in the same fashion as the previous proof to obtain
\begin{align*}
\frac{1}{n}\wt{C}^{\ve}_{\max}(A^{n}:C^{n})_{\rho^{\otimes n}_{AC}} 
\geq & C(A:C)_{\rho} - 3\ve\log(|AC|) \\
& - 2(\ve+1)\log(\ve+1)+\ve \log(\ve) \ .
\end{align*} 
Taking the limit $n \to \infty$ followed by letting $\ve \to 0$ completes the proof where we use $0\log(0) = 0$.
\end{proof}

\end{document}